\documentstyle[namedreferences,psfig,float]{kluwer}

\def\cm3{{cm$^{-3}$}\/} 
\def\ergs.cm2.s{{ergs~cm$^{-2}$~s$^{-1}$}\/}
 
\def\etal{{ et~al.}\/ }
\def\ergs.s{{ergs~s$^{-1}$}\/} 
\def\H2O{{H$_2$}O\/}
\def\Lsun{\hbox{$L_{\odot}$}} %% DON'T USE IN MATH MODE
\def\Msun{\hbox{$M_{\odot}$}} %% DON'T USE IN MATH MODE
\def\Mdot{M\kern-.6em\raise2.ex\hbox{.}~} % for use in math mode

\def\Sec{\hbox{${}^{\prime\prime}$\llap{.}}}

\def\deg{\hbox{${}^\circ$}} 

\def\sec{\hbox{${}^{\prime\prime}$}} 
\def\kms{km s$^{-1}$}

\def\hst{{\it HST}~}
\def\hstnsp{{\it HST}}
\def\rl{{R_{BLR}-L_{\lambda}}}  
\def\mh{{M_{\bullet}}}

\def\msun{{M$_{\odot}$}} 

\def\ms{{\mh-\sigma}} 
\def\mm{{\mh-M_{DM}}}

 \def\vc{{\textsl v_c}} 

\def\vs{{\vc-\sigma_c}}   
\def\ml{{\mh-L}}
\def\mlb{{\mh-L_B}} 
\def\mlk{{\mh-L_K}} 
\def\mc{{\mh-C}}
\def\lae{\mathrel{<\kern-1.0em\lower0.9ex\hbox{$\sim$}}}
\def\gae{\mathrel{>\kern-1.0em\lower0.9ex\hbox{$\sim$}}}

\def\lesssim{\mathrel{\hbox{\rlap{\hbox{\lower4pt\hbox{$\scriptscriptstyle\sim$}}}
\hbox{$\scriptscriptstyle<$}}}}

\def\gtrsim{\mathrel{\hbox{\rlap{\hbox{\lower4pt\hbox{$\scriptscriptstyle\sim$}}}
\hbox{$\scriptscriptstyle>$}}}}

\begin{opening}

\title{SUPERMASSIVE BLACK HOLES IN GALACTIC NUCLEI: PAST, PRESENT AND
FUTURE RESEARCH}

\author{Laura \surname{Ferrarese}} \institute{Herzberg Institute of Astrophysics, National Research Council Canada\\ Victoria, BC, V9EÊ2E7
Canada\\ and\\
Department of Physics
and Astronomy, Rutgers University\\ Piscataway, NJ 08854 USA}
\author{Holland \surname{Ford}} \institute{Department of Physics and
Astronomy, Johns Hopkins University\\ Baltimore, Maryland 21218 USA }

\date{\today}

\end{opening}
\runningtitle{SUPERMASSIVE BLACK HOLES IN GALACTIC NUCLEI}
\begin{document}

\begin{abstract}

This review discusses the current status of supermassive black hole
research, as seen from a purely observational standpoint. Since the early
'90s, rapid technological advances, most notably the launch of the
Hubble Space Telescope, the commissioning of the VLBA and
improvements in near-infrared speckle imaging techniques, have not only
given us incontrovertible proof of the existence of supermassive black
holes, but have unveiled fundamental connections between the mass
of the central singularity and the global properties of the host galaxy. It
is thanks to these observations that we are now, for the first time,
in a position to understand the origin, evolution and cosmic relevance
of these fascinating objects.

\end{abstract}

\section{INTRODUCTION}
\label{sec:intro}

The menagerie of Active Galactic Nuclei (AGNs) is as eclectic as could
be imagined. Quasars, radio galaxies, Seyfert nuclei, Blazars, LINERS,
BL-Lac objects, to name a few, are set apart from each other both by
the detailed character of the activity which takes place in the
nuclei, and by the traits of the galaxies which host them. Underneath
this apparent diversity, however, lie three revealing common
properties. First, AGNs are extremely compact.  Flux variability -- a
staple of all AGNs -- confines AGNs to within the distance light can
travel in a typical variability timescale. In many cases, X-ray
variability is observed on time scales of less than a day, and flares
on time scales of minutes (e.g.  MCG 6-30-15, McHardy 1988). Second,
the spectral energy distribution is decisively non-stellar: roughly
speaking, AGNs' power per unit logarithmic frequency interval is
constant over seven decades in frequency, while stars emit nearly all
of their power in a frequency range a mere factor three wide.  Third,
AGNs must be very massive, a conclusion supported by two independent
arguments.  AGNs' bolometric luminosities are astoundingly large: at
least comparable, and often several orders of magnitude larger than
the luminosity of the entire surrounding galaxy. Masses in excess of
$\sim 10^6 M_{\odot}$ are needed for an AGN not to become unbound by
its own outpouring of energy. Furthermore, according to our best
estimates, AGNs remain active for upward of $10^7$ years: during this
period, an enormous amount of material, well over a million solar
masses, must be consumed to sustain their luminosity, even assuming a
very high efficiency of energy production.

Taken together, these considerations lead to the inescapable conclusion
that the source of the nuclear activity is accretion onto a central,
supermassive black hole (SBH; Rees 1984). Indeed, evidence of a
relativistic regime is betrayed, at least in some AGNs, by
superluminal motions of the radio jets, and by the broadening of low
excitation X-ray emission lines (see \S~\ref{sec:kalpha}).  In our
standard picture, the accreted matter is thought to be confined in an
accretion disk, or more generally optically thick plasma, glowing
brightly at ultraviolet (UV) and perhaps soft X-ray
wavelengths. Medium and hard X-ray emission is produced by inverse
Compton scattering in a corona of optically thin plasma which might
surround or ``sandwich'' the disk. Clouds of
line-emitting gas move at high velocity around this complex core and
are in turn surrounded by an obscuring torus or warped disk of gas and
dust, with a sea of electrons permeating the volume within and above
the torus. What is commonly referred to as the `AGN paradigm' states
that  the detailed character of the nuclear activity can always be
reproduced by finely tweaking, rather than completely revising, this
basic picture. Changes in the angle at which the AGN is observed,  in
the spin and/or mass of the black hole, in the accretion rate, and in
the modalities with which the surrounding interstellar medium
interacts with the emerging AGN flux, account for the varied types
found in the AGN zoo.

Although the black hole paradigm originated and evolved exclusively
within the AGN context, modern SBH searches have targeted almost
exclusively quiescent or weakly active nearby galaxies - and it's on
these galaxies that this review will mainly focus. There are two good
reasons for this. First, ``dormant'' SBHs are {\it expected} to be
found in the nuclei of quiescent galaxies. The cumulative SBH mass
density needed to explain the energetics of high redshift powerful
quasars falls short, by at least two orders of magnitudes, to the one
required to power local AGNs (Padovani, Burg \& Edelson 1990;
Ferrarese 2002a). The unaccounted SBHs must therefore reside in local,
quiescent galaxies. Second, the telltale Keplerian dynamical signature
imprinted by a central compact object on the motion of the surrounding
gas and stars can only be resolved in the most nearby galactic centers
and, unfortunately, most nearby galaxies are not powerful AGNs. In
passing, it must be mentioned that although it is now accepted that
SBH  are present in the nuclei of quiescent galaxies, we still do not
completely understand how the two coexist; in view of the abundant
supply of gas and dust in galactic centers, preventing a SBH from
accreting and immediately producing an AGN is not a simple task (Fabian
\& Canizares 1988). Indeed, a definitive answer to this dilemma has
yet to be found (Rees \etal 1982; Narayan \& Yi 1994; Blandford \&
Begelman 1999; di Matteo \etal 1999).

The most recent review on SBHs was published in 1995 (Kormendy \&
Richstone 1995). Since then, progress in this field has been so rapid
that any attempt to summarize it was destined to be outdated by
publication time. The number of local SBH detections has gone from a
few in 1995 to almost three dozens in 2004.  Strong connections
between SBHs and their host galaxies have emerged. Formation and
evolutionary scenarios have become more tightly constrained. After
such feverish activity, we are now at a turning point, when progress
is once again slowing down as observational facilities are been
exploited to their limit.

This review will be concerned exclusively with supermassive black
holes. There is controversial evidence that ``intermediate'' mass
black holes (IBHs), bridging the gap between the stellar mass (a few
to a few tens of solar masses) and supermassive (over a million solar
masses) varieties, might exist in the off-nuclear regions of some
star-forming galaxies and perhaps at the centers of globular
clusters. An excellent review of intermediate mass black holes is
given by Miller \& Colbert (2004), and we will not discuss the issue
any further. Some useful formalism will be introduced in
\S~\ref{sec:forma}. We will then present a brief historical overview
of the subject (\S~\ref{sec:past}). Although most reviews dispense
with it, the history of SBHs is a fascinating example of the long
trail of tentative steps, missed clues, and heterogeneous research
areas which ultimately need to congeal for seemingly unforeseen and
revolutionary ideas to emerge. Readers who are familiar with this
history are invited to skip to \S~\ref{sec:resstudies}; readers
desiring a comprehensive history of the theoretical developments
should refer to Thorne's book {\it Black Holes and Time Warps,
Einstein's Outrageous Legacy} (1994).  We will then move on to discuss
the several methods which can be used to measure SBH masses
(\S~\ref{sec:resstudies}-~\ref{sec:agn}), with particular emphasis on
resolved stellar and gas dynamical studies carried out with the Hubble
Space Telescope. Scaling relation, linking SBH masses to the overall
properties of the host galaxies, are discussed in
\S~\ref{sec:demo}. SBH demographics, from high redshift quasars to
local galaxies, is discussed in \S~\ref{sec:demo}. 
Finally, in \S~\ref{sec:future} we will
discuss the most pressing open questions and the issues on which
future progress is most likely to be made.

\section{SOME USEFUL FORMALISM}
\label{sec:forma}

For convenience, we present in this section some terminology and
equations which will recur in the remainder of this review.

An important measure of the accretion rate onto a BH of mass $\mh$ is
provided by the Eddington luminosity, {\it i.e.} the
luminosity at which radiation pressure on free electrons  balances
the force of gravity. Because the force due to radiation ,ure has
exactly the same inverse square dependence on distance as gravity, but
does not depend on mass, $L_E$ is independent of  distance but depends 
on $\mh$:

\begin{equation}%2 
L_E = \frac{4 \pi G M m_p c}{\sigma_T} \sim 1.3 \times 10^{46} \left(
\frac{\mh}{10^8{\rm M_{\odot}}} \right) {\rm erg~s^{-1}}\end{equation}

\noindent where $m_p$ is the proton rest mass and $\sigma_T$ is the
Thomson cross section. Above the Eddington luminosity, the
source is unable to maintain steady spherical accretion (although
the presence of magnetic fields can considerably complicate the
picture, Begelman 2001).

Related to the Eddington luminosity is the  Salpeter time

\begin{equation} t_S = \frac{\sigma_T c}{4 \pi G m_p} \sim 4 \times 10^8
~\epsilon {\rm ~yr}\end{equation}

\noindent $t_S$ can be interpreted in two, equivalent ways. It would
take a black hole radiating at the Eddington luminosity a time $t_S$
to dissipate its entire rest mass. Also, the luminosity (and mass) of
a black hole accreting at the Eddington rate with constant
$\dot{M}/M$ will increase exponentially, with e-folding time $t_S$.
$\epsilon$ is the efficiency of conversion of mass into energy, and
depends on the spin of the black hole, varying between 6\% if the
black hole is not spinning, and 42\% if the black hole is maximally
spinning.

The ``boundary'' of a (non-rotating) black hole of mass $\mh$ is a
spherical surface called the event horizon, the radius of which is
given by the Schwarzschild (or gravitational) radius:

\begin{equation}r_{Sch} = \frac{2G\mh}{c^2} \sim 3 \times 10^{13} \left(
\frac{\mh}{10^8{\rm M_{\odot}}} \right) {\rm cm} \sim 2
\left(\frac{\mh}{10^8{\rm M_{\odot}}}\right) {\rm A.U.}\end{equation}

At the Schwarzschild radius the gravitational time dilation goes to
infinity and lengths are contracted to zero.

The radius $r_{st}$ of the last stable orbit, inside which material
plunges into the black hole, depends on the black hole angular
momentum, being smaller for spinning Kerr black holes. For a non
rotating Schwarzschild black hole:

\begin{equation}r_{st} = \frac{6G\mh}{c^2} = 3 r_{Sch}\end{equation}

The photon sphere, of radius $1.5 r_{Sch}$, is defined as the surface
at which gravity bends the path of photons to such an extent that
light orbits the hole circularly.

For a Kerr (rotating) black hole there are two relevant surfaces, the
event horizon, and the static surface, which completely encloses
it. At the static surface, space-time is flowing at the speed of
light, meaning that a particle would need to move at the speed of
light in a direction opposite to the rotation of the hole in
order to be stationary.  In the region of space within the static
surface and the event horizon, called the ergosphere, the rotating
black hole drags space around with it (frame dragging) in such a way
that all objects must corrotate with the black hole. For a maximally
rotating black hole, the radius of the last stable orbit is

\begin{equation}r_{st} = \frac{1.2G\mh}{c^2}\end{equation}

Because of the dependence of $r_{st}$ on the black hole spin, the
latter can be inferred provided a measure of the former, and an
estimate of the black hole mass, are available, for instance from
rapid flux variability (see also \S~\ref{sec:revmap}).

In the case of supermassive black holes inhabiting galactic nuclei, 
the ``sphere of influence''
is defined as the region of space within which the gravitational
potential of the SBH dominates over that of the surrounding stars. 
Its radius is given by:

\begin{equation}%2 
r_h \sim G~\mh /{\sigma}^2 \sim 11.2~(\mh/10^8 {\rm M_{\odot}}) /
(\sigma/ 200~ {\rm km~ s^{-1}})^2~ {\rm pc}.
\end{equation}

\noindent where $\sigma$ is the velocity dispersion of the surrounding
stellar population. Beyond a few thousand Schwarzschild radii from the
central SBH, but within the sphere of influence, the motion of stars
and gas is predominantly Keplerian (relativistic effects are minimal),
with a component due to the combined gravitational potential of stars,
dust, gas, dark matter, and anything else  contributing mass to within
that region. Beyond the sphere of influence, the gravitational
dominance of the SBH quickly vanishes.

\section{A BRIEF HISTORICAL OVERVIEW}
\label{sec:past}

There is perhaps no better way to describe the long chain of events
that culminated, in the mid 1960s, in postulating the existence of
black holes than the famous quote from Albert Szent-Gyorgyi:
``Research is to see what everybody else has seen, and to think what
nobody else has thought.''

Although Newton theorized that gravity acts on light, it was the
British natural philosopher Reverend John Mitchell who pursued the
implications of this idea. In a paper delivered to the Royal Society
in London in 1783, Mitchell envisioned the existence of  ``dark
stars''. He realized that the escape velocity would become larger if
the star radius were to be increased while maintaining its density
constant. Eventually, the escape velocity would exceed the velocity of
light: such star would become invisible to a distant observer, since
the ``corpuscles'' of light, after librating some distance above the
star's surface in their attempt to escape, would inevitably be pulled
back. Thirteen years later, and with no mention of Mitchell's work,
Pierre Laplace published a very similar argument in {\it Exposition du
Systeme du Monde}, only to drop it in the  3rd edition of the
book. The reason for the omission is not clear, but it might have been
prompted by the gaining popularity of Christian Huygen's ondulatory
theory of light -- in view of Young and Fresnel's experiments -- and
the lack of a physical understanding of how gravity and waves
intermingle.

Such understanding had to wait until the day, in 1915, when Albert
Einstein delivered a lecture on his theory of general relativity to
the German Academy of Science in Berlin.  Within a month of the
publication of Einstein's work, Karl Schwarzschild, while serving in
the German Army on the Russian front, solved Einstein's field
equations for a non-rotating spherical star.  His solution (1916a,b)
for the spacetime geometry, now known as the Schwarzschild metric,
enabled him to calculate, for a star of a given mass, the critical radius
at which light emitted from the surface would have an infinite
gravitational redshift, and thereby infinite time dilation. Such star,
Schwarzschild concluded, would be undetectable by an external observer
at {\it any} distance from the star -- a proposition received with
considerable skepticism by most theorist of the time, including
Einstein himself.

Whether a real star could ever reach this critical radius was
addressed quite serendipitously in the 1930s. During the voyage from
Madras, India, to England to begin graduate study at Cambridge
University, nineteen year old Subrahmanyan Chandrasekhar set himself
the task of deriving the structure of white dwarfs, whose existence
had been known since the very early days of the Hertzprung-Russell
diagram. By using the theory of stellar polytropes (Eddington, 1930)
in combination with Fowler's (1926) recently published equation of
state for a non-relativistic degenerate electron gas, Chandrasekhar
demonstrated that the density (or radius) of a white dwarf is a very
simple function  of its mass. After calculating the central density
for Sirius B, however, Chandrasekhar realized that white dwarfs of
mass 1$~\Msun$~reach high enough densities in their cores for the
electron gas to becomes fully relativistic, invalidating Fowler's
equation of state (cf Thorne 1994 for a thorough discussion).  By
allowing for the relativistic increase in the electrons' momentum,
Chandrasekhar was able to deduce, though not prove, that the
dependence of pressure on electron density softens from Fowler's
$N_e^{5/3}$ to $N_e^{4/3}$. This change has drastic consequences -- so
bizarre, in fact, to be publicly ridiculed by the most influential
physicist of the time, sir Arthur Eddington. A fully relativistic
polytropic gas can only be in equilibrium for a mass of precisely .91
$M_{\odot}$\footnotemark -- no matter what its radius or density.

\footnotetext{Subsequent improved calculations have shown this mass to
be ~1.4 $M_{\odot}$, e.g. Harwit (1998).}

In his 1931 paper, Chandrasekhar interpreted this value -- now known
as the ``Chandrasekhar limit'' -- as the maximum mass attainable by a
white dwarf as it approaches fully relativistic degenerate
conditions. In 1932, the russian physicist Lev Davidovich Landau --
who had reached a conclusion analogous to Chandrasekhar's in the context of
degenerate neutron stars --  went a step further. He clearly stated
that addition of matter over the critical limit would lead to
unavoidable collapse: the star would shrink in free fall to a
point. This remarkable result has a simple physical explanation. For a
star of small enough mass -- whether it is composed of a degenerate
gas of electrons or neutrons --  the quantum pressure due to Pauli's
exclusion principle can always be brought into balance with gravity by
increasing the density via contraction (according to Fowler's
equation). As the star becomes more massive, however, further
contractions will eventually lead to high enough core densities to
bring the kinetic energy at the top of the Fermi sea to levels
comparable to the rest energy of the particles; the gas then becomes
relativistic. Because of the softening in the equation of state that
follows the transition into the relativistic regime, the increase in
quantum pressure which derives from further contraction always fall
short of balancing gravity: the contraction cannot be halted.

Following these pioneering works, a detailed study of the equilibrium
configuration was published by Chandrasekhar (1935) for white dwarfs,
and Oppenheimer \& Volkoff (1938) for neutron stars. In 1939,
Oppenheimer \& Snyder wrote what was destined to become the foundation
for the new field of physics concerned with gravitational
singularities. In  ``On Continued Gravitational Contraction'', they
provided a fully analytical solution for the collapse of a cloud of
gas,  and drew attention to how collapsed objects are a unique testbed
for a fully relativistic theory of gravitation.  By applying the
general relativistic field equations to a sphere of cold neutrons,
Oppenheimer \& Snyder concluded that ``when all thermonuclear sources
of energy are exhausted a sufficiently heavy star will
collapse. Unless fission due to rotation, the radiation of mass, or
the blowing off of mass by radiation, reduce the star's mass to the
order of that of the Sun, this contraction will continue indefinitely
[...] The total time of collapse for an object comoving with the
stellar matter is finite [..]  The star thus tends to close itself off
from any communication with a distant observer; only its gravitational
field persists''. Black holes -- a term coined by Princeton theorist
John Wheeler only in 1960 -- were effectively born.

\begin{figure}[t]
\begin{minipage}[t]{4.0cm}
\psfig{figure=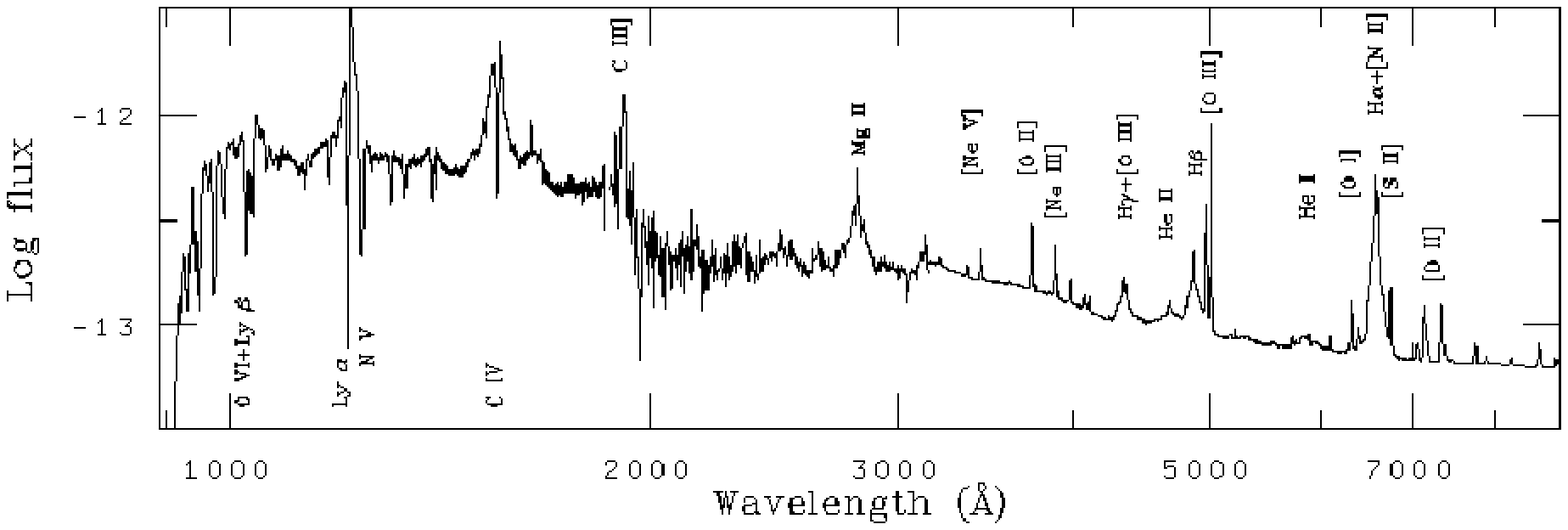,height=4.0 cm}
\end{minipage}
\begin{minipage}[t]{4.0cm}
\psfig{figure=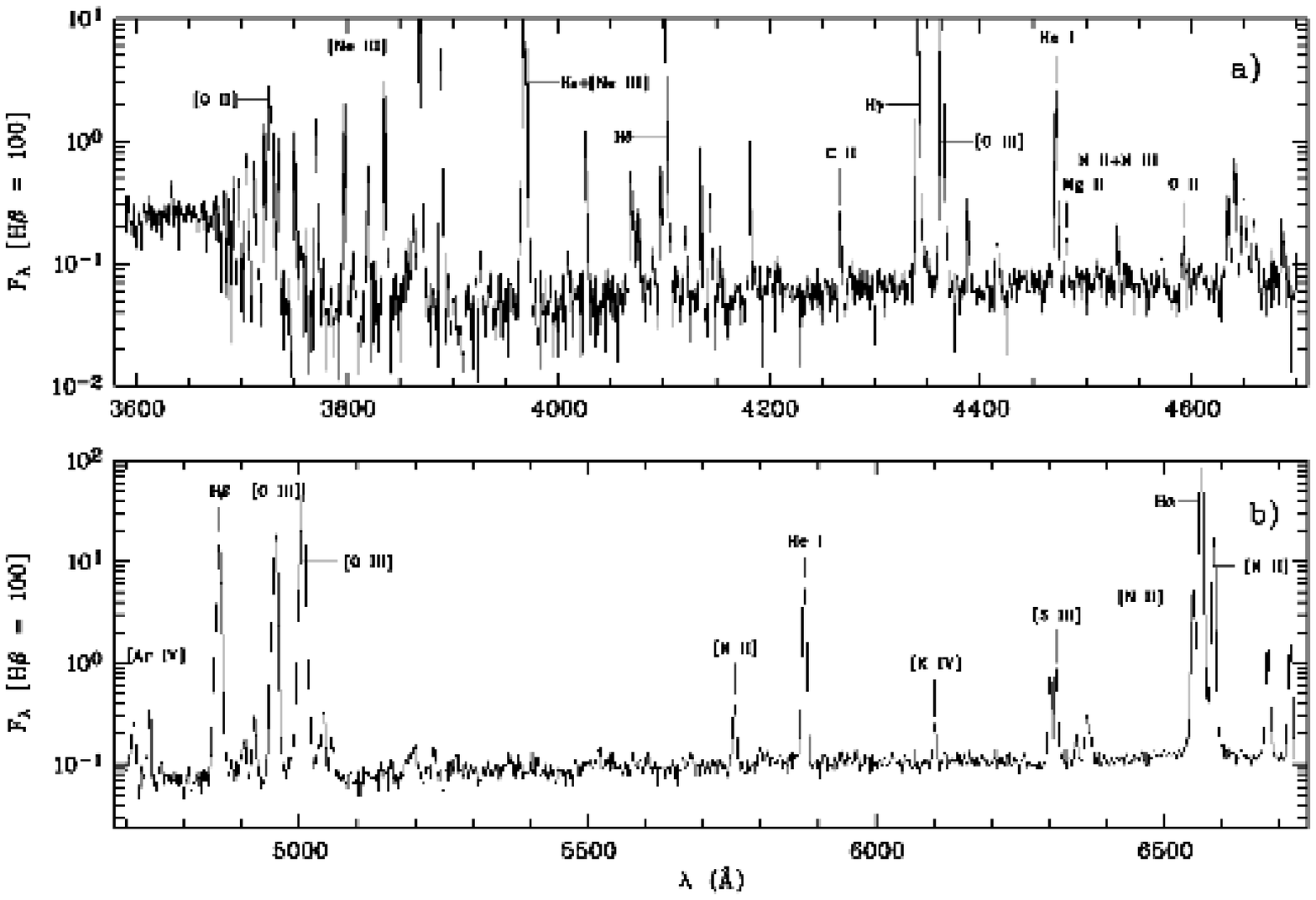,height=8.0 cm}
\end{minipage}
\caption{A spectrum of the nucleus of NGC 4151 (top panel), compared to the
spectrum of the planetary nebula M2-24 (middle and lower panel), illustrating the similarities
and differences (broad lines and strong continuum in N4151).  NGC 4151
spectrum courtesy of W. C. Keel; M2-24 spectrum courtesy of Y. Zhang
\& X.-W. Liu.}
\label{Seyfert2}
\end{figure}

Meanwhile, observational work on black holes was lagging behind the
theory by at least a decade. Although unrealized at the time, the very
first data supporting the existence of black holes -- not a few, but
millions of solar masses -- had been accumulating since the early
'40s.  In a 1943 paper titled ``Nuclear Emission in Spiral Nebulae'',
Carl Seyfert identified 12 galaxies with highly unusual nuclei.  
Unlike the centers of normal galaxies which
contain only old stars, these nuclei had ``high excitation nuclear
emission lines superposed  on a normal G-type spectrum''
(Figure~\ref{Seyfert2}). Seyfert remarked that ``profiles of the
emission lines show that the lines are broadened, presumably by
Doppler motion, by amounts up to 8500 \kms~ for the total widths of
the hydrogen line in NGC 3516 and NGC 7469 [...] The lines of the
other ions show no evidence of wide wings." In our galaxy, only
supernovae explosions were known to generate mass motions of this
magnitude. Finally, Seyfert noted that these unusual galaxies have
``bright nuclei, scarcely distinguishable from stars."  For instance,
the nucleus of NGC 4151 in the near ultraviolet is as bright as
$\sim$70\% of the entire ultraviolet luminosity of M 31.  Later
observations of variability timescales showed that many Seyfert nuclei
were smaller than a few light minutes in size.

The significance of Seyfert's work was not fully appreciated until the
1950s, when studies of the newly discovered radio sources made it
abundantly clear that  new and extraordinary physical processes were
at play in the nuclei of galaxies.  The most unequivocal evidence came
from M87, the dominant galaxy in the Virgo cluster (Figure
~\ref{M87-1}). In 1954, Baade \& Minkowski associated the bright radio
source Vir A with the seemingly unremarkable giant elliptical galaxy.
Two facts suggested that M87 was the source of the  radio
emitting plasma, and specifically that the radio source had
originated in the nucleus.  First,   a bright narrow optical jet 1 kpc
long emanated from the nucleus (Curtis 1918). Second,  spectra showed
that the nucleus contains ionized gas with unusual line ratios and
line widths of several hundred km s$^{-1}$. Interestingly, forty years
later, M87 was the first galaxy for which the presence of a
supermassive black hole was firmly established using the Hubble Space
Telescope  (see \S~\ref{sec:gasdyn}; Ford \etal 1994; Harms \etal
1994).

\begin{figure}[t]
\centerline{\parbox[l][8cm]{9.2cm}{\vskip 3cm THIS FIGURE IS INCLUDED IN THE FULL VERSION OF THE
MANUSCRIPT AVAILABLE AT http://www.physics.rutgers.edu/~lff/publications.html.}}
%\centerline{\psfig{figure=f2.eps,height=8.0 cm}}
\caption{VLA (top left), \hst (top right), and VLBI (bottom) images of
M87.   The images show a bright radio and optical source at the
center of the galaxy, and a helical jet emanating from the nucleus.
The radio and optical emission is synchrotron radiation. Courtesy of
NASA, National Radio Astronomy Observatory/National Science
Foundation, John Biretta (STScI/JHU), and Associated Universities,
Inc.}
\label{M87-1} 
\end{figure}
  
It soon became clear that M87 was far from being an isolated case. In
1953, Jennison \& Das Gupta showed that the radio source Cygnus A was
double (Figure~\ref{CygA}).  The following  year Baade and Minkowski
(1954)  identified the radio source with a galaxy at the center of a
rich cluster at z=0.057 (D $\sim$ 250 Mpc for $H_0=70$ \kms
Mpc$^{-1}$), implying that Cyg A was one of the brightest radio
sources in the sky. The radio emission originated from two lobes on
either side of the visible object, suggesting that it might be due to
relativistic particles ejected in opposite direction from the
nucleus. Like M87, the nucleus of the visible galaxy associated with
Cyg A showed strong extended emission with unusual line strengths. The
case for galaxy nuclei being the sites of violent activity tightened
even further when interferometric observations showed that at least
some of the radio sources could not be resolved even at 1 arcsec
resolution (Allen \etal 1962).

\begin{figure}[t]
\centerline{\parbox[l][5cm]{9.2cm}{\vskip 3cm THIS FIGURE IS INCLUDED IN THE FULL VERSION OF THE
MANUSCRIPT AVAILABLE AT http://www.physics.rutgers.edu/~lff/publications.html.}}
%\centerline{\psfig{figure=f3.eps,height=5.0 cm}}
\caption{VLA 1.4 6cm image of the bright radio source Cygnus A.   The
jets transport energy from the nucleus to the radio lobes at distance
of $\sim$220 Kpc, before  being stopped by the intergalactic medium
surrounding  the galaxy. Image courtesy of NRAO/AUI.}
\label{CygA}
\end{figure}

By 1955 astronomers had made a connection between the optical and
radio emission from the Crab Nebula (the remnant of a supernova
observed on July 4 1054 by Chinese astronomers) and M87's polarized
optical and radio emission from the jet (Baade 1955). The theory of
synchrotron radiation from a relativistic plasma newly developed by
Shklovski (1954) allowed estimates of  the total energy required to
power up a radio source. Using the observed power law spectrum,
luminosity, and volume of Cyg A, the  minimum energy associated with
the relativistic electrons and magnetic  field was calculated to be $4
\times 10^{58}~{\rm ergs}$,  equivalent to the rest mass energy of
11,000 M$_{\odot}$.  If the contribution of the relativistic protons
is accounted for, these estimates can be plausibly multiplied by
factors of $\sim$100. Such requirements are clearly difficult to
reconcile with any classical power source, once the  small  physical
size of the nuclei is folded in.

With the discovery of quasi stellar objects (QSOs) in the early '60s,
this energy crisis could no longer be ignored.  Lunar
occultations carried out with the 250 ft telescope at Jodrell Bank
showed that the radio source 3C273 consisted of two components, one of
which had a small angular size and flat spectrum (Hazard, MacKay \&
Shimmins 1965). Inspection of photographic plates taken at the Palomar
200 inch Hale telescope showed that the second source coincided with a
13 mag star, while the more extended radio source matched a faint
optical nebulosity which seemed to protrude from the star (Figure~\ref{3c273}).  When the
same year Caltech's astronomer Martin Schmidt identified the broad lines
from the ``star'' as Balmer emission at a redshift $z = 0.158$, he
opened the door to a field of astronomy that has led to observations
of the most energetic phenomena and the most distant objects in the
Universe.

\begin{figure}[t]
\centerline{\parbox[l][8cm]{9.2cm}{\vskip 3cm THIS FIGURE IS INCLUDED IN THE FULL VERSION OF THE
MANUSCRIPT AVAILABLE AT http://www.physics.rutgers.edu/~lff/publications.html.}}
%\centerline{\psfig{figure=f4.eps,height=8.0 cm}}
\caption{ A pair of Hubble Space Telescope images of the quasar  3C 273, taken with the
Wide Field and Planetary Camera 2 (left), and the Advanced Camera
for Surveys (right). In the latter image, the bright nucleus was
placed behind an occulting finger, to reveal the light from the host galaxy. The WFPC2
image clearly shows the optical synchrotron jet
extending 50 Kpc from the nucleus.  For comparison, the jet in M87 is
only 2 kpc long. Credit: NASA, the ACS Science Team and ESA}
\label{3c273}
\end{figure}

\begin{figure}[t]
\centerline{\psfig{figure=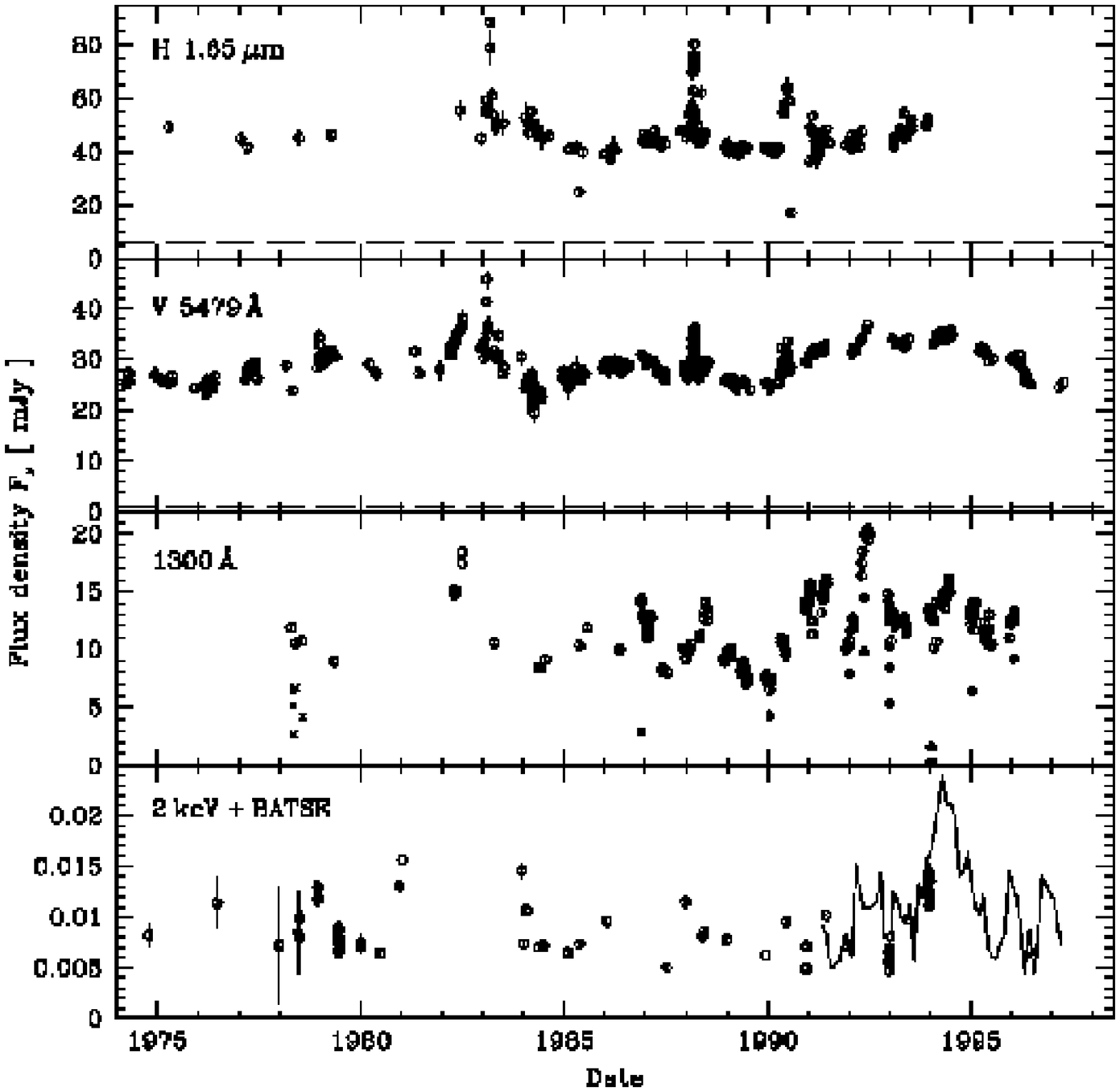,height=9.0 cm}}
\caption{The infrared-to-X-ray continuum variability in 3C273, from 1974 to 1998. 
The dashed line indicates the contribution from the underlying galaxy.
From Turler \etal (1999).  }
\label{xray-variation}
\end{figure}

Within two years, it was recognized that almost a third of high
latitude radio sources were QSOs at large redshifts, the most luminous
of which  were found to be  $\sim$1000 times brighter than the
Andromeda galaxy. The discovery that some QSOs showed variability on
timescales of the order of 1 year (Smith \& Hoffleit 1963; Sandage
1964; Greenstein \& Schmidt 1964) proved that the light was emitted
from a region less than 1 pc in size. Orbiting X-ray telescopes
launched in the 1970s set even tighter constraints on the sizes of the
energy sources by establishing that active galactic nuclei (AGNs) are
luminous in X-rays, and that the luminosity can  change by  factors
$\sim$2 with timescales of days, hours, and even minutes, setting a
corresponding upper limit to the size of the central engine
(Figure~\ref{xray-variation}).  If the  engines are massive black
holes,  the relationship  between the Schwarzschild radius and the
black hole mass opens the possibility of setting an upper limit to the
latter from the variation timescale.

Since the early 1960s there was much speculation about how the observed
luminosities could be produced within such a small
region. Thermonuclear reactions, which have efficiencies of 0.7\% at
best (in the case of fusion of H into He), were quickly eliminated.
Speculations concentrated on runaway  explosions triggered by
supernovae outbursts in a cluster of tightly packed stars (Burbridge
1962),  the collapse and disintegration of a $10^5 - 10^8$ $M_{\odot}$
star (Hoyle and Fowler 1963; the authors note that ``The concept of
stellar-type objects with masses up to $10^8 M_{\odot}$ is of course
strange, but the very nature of the case demands an unusual physical
situation''), rapid star formation in a newly born galaxy (Field
1964), and energy generated in galaxy collisions (Harrower 1960). By
the 1963 Texas symposium (Robinson \etal 1965), much pondering was
given to the idea that the energy source was gravitational. For
example, Wheeler envisioned a gravitational singularity at the center
of a galaxy, converting into energy much of the matter falling onto
it.  Zel'dovich \& Novikov (1964) and Salpeter (1964) further
described the growth of a massive object at the center of a galaxy
through accretion, and the accompanying release of energy. Lynden-Bell
(1969) made an attempt  to explain the phenomenology observed in QSOs
and Seyfert galaxies directly in terms of a black hole formalism.

In spite of these promising developments, it is fair to say that the
single most influential event contributing to the acceptance of black
holes was the 1967 discovery of pulsars by graduate student Jocelyn
Bell. The clear evidence of the existence of neutron stars --  which
had been viewed with much skepticism until then -- combined with the
presence of a critical mass above which stability cannot be achieved,
made the existence of stellar-mass back holes inescapable.  The first detection of a solar mass black hole came when the mass
of the rapidly variable X-ray source Cygnus X-1 was proven to be above
the maximum allowed for a neutron star (Brucato \& Kristian 1972;
Bolton 1972; Mauder 1973;
Rhoades \& Ruffini 1974). The first secure detection of a supermassive
black hole in a galactic nucleus trailed twenty years behind. How it
came about, and what SBHs tell us about galaxy formation and
evolution, will be the subject of the remainder of this review.

\section{CURRENT STATUS OF SUPERMASSIVE BLACK HOLE SEARCHES: 
AN OVERVIEW}
\label{sec:resstudies}

A SBH which forms or  grows in a galactic nucleus will produce a cusp
in the stellar density (Peebles 1972; Young 1980; Quinlan, Hernquist,
\& Sigurdsson 1995; van der Marel 1999). Unfortunately, as
demonstrated very effectively by Kormendy \& Richstone (1995), the growth,
or even the presence of a SBH is not a necessary condition for
a ``light cusps'' to form.  Moreover, even when originally present, central density cusps
can be destroyed during galaxy mergers, as a consequence of the hardening of the SBH binary which 
forms at the center of the merger product (Milosavljevic \& Merritt
2001). 

\begin{figure}[t]
\centerline{\psfig{figure=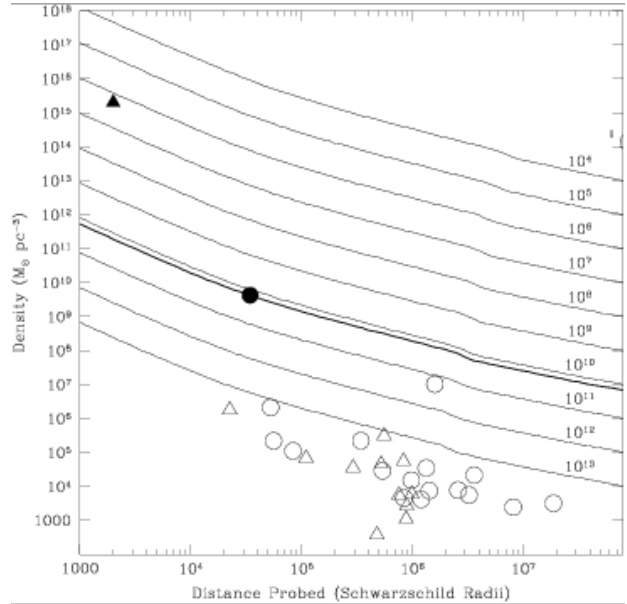,height=8.0 cm}}
\caption{The innermost radius (normalized to the Schwarzschild radius)
probed by current experiments which have led to the detection of SBHs
in nearby galactic nuclei, plotted against the inferred mass density
within that region.  The detection in the Milky Way (\S~\ref{sec:mw}) is shown as the
solid triangle, water maser detections (\S~\ref{sec:maser}) are shown as solid circles,
detections based on stellar (\S~\ref{sec:intsd}) and gas (\S~\ref{sec:gasdyn}) dynamics are shown as open circles
and triangles respectively. The solid curves  show the maximum
lifetime of a dark cluster against collision and evaporation, using
the prescription in Maoz (1998), from $10^4$ to $10^{13}$ years. The
thick solid line represents a lifetime of 15 billion years.}
\label{sbhdensity}
\end{figure}

The dynamical signature imprinted by a SBH on the motion of
surrounding matter is, however, unique. Within the sphere of
influence, a Keplerian  rotation or velocity dispersion of stars or
gas is unambiguous proof of the existence of a central mass
concentration. The ultimate test as to its nature (a singularity or a
dense star cluster?) can only reside in the detection of relativistic
velocities within a few Schwarzschild radii. Only observations of the
Fe K$\alpha$  emission line in Type 1 AGNs might give us a change of
peering within the relativistic regime of a SBH
(Table~\ref{tab:methods} and \S~\ref{sec:kalpha}),  although this is
still considered to be a controversial issue.

\begin{table}[t] 
\scriptsize
\caption{Probing the Centers of Galaxies}
\begin{tabular}{lcccc} 
\hline 	       Method \& & Scale & No. of SBH & $\mh$ Range & Typical
Densities \\  Telescope & ($R_S$) & Detections & (M$_{\odot}$)
&(M$_{\odot}$ pc$^{-3}$)  \\  
\hline 
Fe K$\alpha$ line     & 3-10 & 0 & N/A & N/A \\  
(XEUS, ConX) & & & & \\  
& & & &\\  
Reverberation Mapping & 600  & 36 & $10^6 - 4\times10^8$ & $\gae 10^{10}$\\ 
(Ground based optical)\\  
& & & &\\  
Stellar Proper Motion & 1000 & 1  & $4\times10^6$& $4\times 10^{16}$ \\ 
(Keck, NTT, VLT) & & & &\\ 
& & & &\\  
\H2O Megamasers       & $10^4$  & 1  & $4\times10^7$ & $4\times10^{9}$ \\ 
(VLBI) & & & &\\  
& & & &\\  
Gas Dynamics (optical)& $10^6$  & 11 & $7\times10^7 - 4\times10^9$ & $\sim10^5$ \\
(Mostly \hstnsp) & & & &\\  
& & & &\\  
Stellar Dynamics      & $10^6$  & 17 & $10^7 - 3\times10^9$ & $\sim10^5$ \\ 
(Mostly \hstnsp) & & & &\\ 
\hline
\end{tabular}
The columns give all methods which can (or, in the case of the Fe
K$\alpha$ line emission, might) be used to estimate SBH masses, and
the telescopes needed for the observations;  the typical distance from
the singularity of the material probed by each method; the number of
SBH detections claimed based on that method; the range in the detected
SBH masses; and the corresponding implied central mass density.
\label{tab:methods}
\end{table}

In the absence of the ultimate, relativistic signature, the case for the detected 
masses to be confined within a singularity becomes stronger as the
corresponding mass-to-light ratio and mass density increase.  Maoz
(1998) provided rough calculations of the lifetime of a dark cluster
against evaporation and collapse. For any choice of the cluster's mass
and density (or, equivalently, radius and density) such lifetime
depends on the cluster's composition. Maoz considered the case of
clusters of brown dwarfs (with masses down to $3\times 10^{-3}$
\msun), white dwarfs, neutron stars and stellar black holes. His
results are reproduced in  Figure~\ref{sbhdensity}, where the maximum
lifetime attainable by any such cluster is plotted as a function of
the cluster's density and radius (an upper limit to which is given
by the scale probed by the data), the latter normalized to the Schwarzschild
radius. The case for a SBH is tight, according to this simple
argument, when the observations imply densities and masses for which
the dark cluster lifetime is short compared to the age of the
galaxy. When all SBH detections claimed to date are considered
(Table~\ref{tab:allmasses}), this condition is verified only in the
case of the Milky Way, NGC 4258 and Circinus. For all other galaxies,
although we will tacitly assume for the rest of this paper that the
detected masses are indeed SBHs, dark clusters can provide viable
explanations.

Table~\ref{tab:methods} summarizes typical central mass densities 
inferred for each of the methods
which are  (or might become) available to measure SBH masses in
galactic nuclei. Observations of the Fe K$\alpha$ emission line will
be discussed in \S~\ref{sec:kalpha}. The line, which seems to be an
almost ubiquitous feature in the X-ray spectra of Seyfert 1 galaxies,
is thought to arise from material within a few Schwarzschild radii of
the central SBH  (Nandra \etal 1997; Reynolds 1997;  but see also
Done \etal 2000; Gondoin \etal 2001a\&b). As such, it provides a
powerful testbed of the properties of spacetime in strong
gravitational fields.  Future generations of X-ray satellites (most
notably the European mission XEUS and NASA's Constellation X), will
reveal whether the line responds to flares in the X-ray continuum (a
point which is debated, Nandra \etal 1997; Nandra \etal 1999; Wang
\etal 2001; Takahashi \etal 2002). If it does, the mass of the central
SBH can conceivably be estimated using reverberation mapping
techniques (\S~\ref{sec:revmap}).

In its current application, reverberation mapping
(\S~\ref{sec:revmap}) targets the Broad Line Region (BLR) of Type 1
AGNs (Blandford \& McKee 1982; Peterson 1993; Peterson 2002). It is
currently the least secure, but potentially most powerful of the
methods which we will discuss, probing material within only a few
hundred Schwarzschild radii from the singularity, a factor $10^3$
closer than can be reached by stellar and gas dynamical studies using
\hstnsp. As discussed in detail in \S~\ref{sec:revmap},  although
there is compelling, albeit indirect observational evidence that the
BLR motion is Keplerian, the kinematics and morphology of the BLR have
not yet been mapped directly.   In the Keplerian hypothesis, the SBH
mass is derived from the observed ``average'' size and velocity of the
BLR. If the method does indeed measure masses (and the evidence  that
this is indeed the case is growing stronger by the day), the inferred
central densities leave no doubt that such masses are indeed confined
in a singularity.

To date, the only secure detections of SBHs (as opposed to dense
clusters of stars or exotic particles) come from stellar proper motion
in the Galactic center and the \H2O megamaser study of the nearby
Seyfert 2 galaxy NGC 4258 (\S~\ref{sec:mw} and \S~\ref{sec:maser}
respectively). The applicability of either method is however limited
(to one galaxy, the Milky Way, in the case of proper motion
studies). The most prolific methods are based on optical stellar and
gas dynamical studies, generally carried out using the Hubble Space
Telescope (\hstnsp, \S~\ref{sec:intsd} and \S~\ref{sec:gasdyn}).  On the
downside, these methods can rarely reach closer than several million
Schwarzschild radii from the singularity; the implied central
densities are always far lower than needed to conclude that the mass
is indeed collapsed into a SBH.

Table~\ref{tab:allmasses} lists all galaxies for which a SBH detection
has been claimed based on stellar proper motion, \H2O megamasers, or
optical stellar and gas  dynamical studies (reverberation mapping
detections will be listed in \S~\ref{sec:revmap}). Cases for which the
analysis did not lead to a successful determination of the SBH mass 
(according to the original investigators) are grouped at the end of the table. For
each galaxy, we list the Hubble type, distance (mostly from Tonry et
al. 2001), SBH mass and reference, central bulge velocity dispersion,
total and extinction corrected blue magnitude (from the RC3, de
Vaucouleurs \etal 1991, corrected for Galactic extinction using the
reddening maps of Schlegel \etal 1998), and fraction of the total
light judged to be in the hot stellar component (from Fukugita et
al. 1998). In the last column, the ratio $r_h/r_{res}$ between the
radius of the SBH sphere of influence (equation 6) and the spatial
resolution of the data is given as a rough indicator of the quality of
the SBH mass estimate. All studies which have addressed the issue
(Ferrarese \& Merritt 2000; Merritt \& Ferrarese 2001b\&c;
Graham \etal 2001; Ferrarese 2002a; Marconi \& Hunt 2003; Valluri et
al. 2004) have concluded that resolving  the sphere of influence  is
an important (although not sufficient)  factor: not
resolving $r_h$ can lead to systematic  errors on $\mh$ or even spurious
detections.

This rather intuitive fact explains the dominance of \hst in this
field. As an example, consider our close neighbor, the Andromeda
galaxy. At a fiducial stellar  velocity dispersion of $\sigma \sim
160$ km s$^{-1}$, and assuming a SBH with mass $\sim 3 \times 10^7$
\msun~(unfortunately, the well known presence of a double nucleus does
not allow for an accurate determination of the SBH mass - Bacon et
al. 2001), the radius of the sphere of influence is 5.2 pc, or 1\Sec4
at a distance of 770 kpc. This is resolvable from the ground. However,
if Andromeda were just a factor two or three further, ground based observations would 
be unable to address the question of whether its nucleus hosts a SBH. 
In the absence of maser clouds or an active nucleus, \hst data would offer the only viable option. 
At the

\begin{table}[t] 
\scriptsize
\caption{Complete List of SBH Mass Detection Based on Resolved
Dynamical Studies}
\begin{tabular}{llccccccc} 
\hline 	     Object & Hubble & Distance & $\mh$ & $\mh$ Ref. \& &
$\sigma$ & $M_{B,T}^0$ & $L_{B,bulge}/$ & $r_{h}/$\\ 
& Type & (Mpc) &
($10^8$\msun) & Method & (km s$^{-1})$ & (mag) & $L_{B,total}$ &
$r_{res}$ \\ 
\hline      
MW     & SbI-II   & 0.008 & $0.040^{+0.003}_{-0.003}$ &  1,PM & 100$\pm$20 & -20.08$\pm$0.50 & 0.34 & 1700\\ 
N4258  & SAB(s)bc & 7.2   & $0.390^{+0.034}_{-0.034}$ &  2,MM & 138$\pm$18 & -20.76$\pm$0.15 & 0.16 & 880\\ 
N4486  & E0pec    & 16.1  & $35.7^{+10.2}_{-10.2}$    &  3,GD & 345$\pm$45 & -21.54$\pm$0.16 &  1.0 & 34.6\\ 
N3115  & S0       & 9.7   & $9.2^{+3.0}_{-3.0} $      &  4,SD & 278$\pm$36 & -20.19$\pm$0.20 & 0.64 & 22.8\\ 
I1459  & E3       & 29.2  & $26.0^{+11.0}_{-11.0}$    &  5,SD & 312$\pm$41 & -21.50$\pm$0.32 & 1.0  & 17.0\\ 
N4374  & E1       & 18.7  & $17^{+12}_{-6.7}     $    &  6,GD & 286$\pm$37 & -21.31$\pm$0.13 &  1.0 & 10.3\\ 
N4697  & E6       & 11.7  & $1.7^{+0.2}_{-0.3}  $     &  7,SD & 163$\pm$21 & -20.34$\pm$0.18 & 1.0  & 10.2\\ 
N4649  & E2       & 16.8  & $20.0^{+4.0}_{-6.0} $     &  7,SD & 331$\pm$43 & -21.43$\pm$0.16 &  1.0 & 10.1\\ 
N221   & cE2      & 0.8   & $0.025^{+0.005}_{-0.005}$ &  8,SD  & 76$\pm$10  & -15.76$\pm$0.18 &  1.0 & 10.1\\ 
N5128  & S0pec    & 4.2   & $2.0^{+3.0}_{-1.4}  $     &  9,GD & 145$\pm$25 & -20.78$\pm$0.15 & 0.64 & 8.41\\ 
M81    & SA(s)ab  & 3.9   & $0.70^{+0.2}_{-0.1}  $    & 10,GD & 174$\pm$17 & -20.42$\pm$0.26 & 0.33 & 5.50\\ 
N4261  & E2       & 31.6  & $5.4^{+1.2}_{-1.2}    $   & 11,GD & 290$\pm$38 & -21.14$\pm$0.20 &  1.0 & 3.77\\ 
N4564  & E6       & 15.0  & $0.56^{+0.03}_{-0.08}$    &  7,SD & 153$\pm$20 & -19.00$\pm$0.18 &  1.0 & 2.96\\ 
CygA   & E	  & 240   & $25.0^{+7.0}_{-7.0} $     & 12,GD & 270$\pm$87 & -20.03$\pm$0.27 & 1.0  & 2.65\\ 
N2787  & SB(r)0   & 7.5   & $0.90^{+6.89}_{-0.69}$    & 13,GD & 210$\pm$23 & -18.12$\pm$0.39 & 0.64 & 2.53\\ 
N3379  & E1       & 10.6  & $1.35^{+0.73}_{-0.73}$    & 14,SD & 201$\pm$26 & -19.94$\pm$0.20 &  1.0 & 2.34\\ 
N5845  & E*       & 25.9  & $2.4^{+0.4}_{-1.4}  $     &  7,SD & 275$\pm$36 & -18.80$\pm$0.25 & 1.0  & 2.28\\ 
N3245  & SB(s)b   & 20.9  & $2.1^{+0.5}_{-0.5} $      & 15,GD & 211$\pm$19 & -20.01$\pm$0.25 & 0.33 & 2.10\\ 
N4473  & E5       & 15.7  & $1.1^{+0.5}_{-0.8} $      &  7,SD & 188$\pm$25 & -19.94$\pm$0.14 &  1.0 & 1.84\\ 
N3608  & E2       & 22.9  & $1.9^{+1.0}_{-0.6} $      &  7,SD & 206$\pm$27 & -20.11$\pm$0.17 & 1.0  & 1.82\\ 
N4342  & S0       & 16.7  & $3.3^{+1.9}_{-1.1} $      & 16,GD & 261$\pm$34 & -17.74$\pm$0.20 & 0.64 & 1.79\\ 
N7052  & E        & 66.1  & $3.7^{+2.6}_{-1.5}  $     & 17,GD & 261$\pm$34 & -21.33$\pm$0.38 &  1.0 & 1.53\\ 
N4291  & E3       & 26.2  & $3.1^{+0.8}_{-2.3} $      &  7,SD & 269$\pm$35 & -19.82$\pm$0.35 & 1.0  & 1.52\\ 
N6251  & E        & 104   & $5.9^{+2.0}_{-2.0}  $     & 18,GD & 297$\pm$39 & -21.94$\pm$0.28 & 1.0  & 1.19\\ 
N3384  & SB(s)0-  & 11.6  & $0.16^{+0.01}_{-0.02}$    &  7,SD & 151$\pm$20 & -19.59$\pm$0.15 & 0.64 & 1.12\\ 
N7457  & SA(rs)0- & 13.2  & $0.035^{+0.011}_{-0.014}$ &  7,SD & 73$\pm$10  & -18.74$\pm$0.24 & 0.64 & 0.92\\ 
N1023  & S0       & 11.4  & $0.44^{+0.06}_{-0.06}$    &  7,SD & 201$\pm$14 & -20.20$\pm$0.17 & 0.64 & 0.89\\ 
N821   & E6       & 24.1  & $0.37^{+0.24}_{-0.08}$    &  7,SD & 196$\pm$26 & -20.50$\pm$0.21 &  1.0 & 0.74\\ 
N3377  & E5       & 11.2  & $1.00^{+0.9}_{-0.1}$      &  7,SD & 131$\pm$17 & -19.16$\pm$0.13 &  1.0 & 0.74\\ 
N2778  & E        & 22.9  & $0.14^{+0.08}_{-0.09}$    &  7,SD & 171$\pm$22 & -18.54$\pm$0.33 &  1.0 & 0.39\\ 
\hline \hline
\multicolumn{9}{c}{Galaxies for which the dynamical models might be in error.}\\ 
Object & Hubble & Distance & $\mh$ & $\mh$ Ref. \& & \multicolumn{4}{c}{Notes}\\ 
& Type & Mpc & $10^8$\msun & Method & & & &  \\ 
\hline 
Circinus &SA(s)b & 4.2   & $0.017^{+0.003}_{-0.003} $  & 19,MM & \multicolumn{4}{c}{disk inclination angle not constrained}\\ 
N4945  & SB(s)cd &  3.7  & $0.014^{+0.007}_{-0.005}$   & 20,MM & \multicolumn{4}{c}{no 2-D velocity field}  \\ 
N1068  & Sb & 23.6 &  $0.17^{+0.13}_{-0.07}$           & 21,MM & \multicolumn{4}{c}{maser disk is self gravitating} \\ 
N4459  & SA(r)0+ &  16.1 & $0.70^{+0.13}_{-0.13}$      & 13,GD & \multicolumn{4}{c}{disk inclination angle not constrained}\\ 
N4596  & SB(r)0+ &  16.8 & $0.8^{+0.4}_{-0.4} $        & 13,GD & \multicolumn{4}{c}{disk inclination angle not constrained}\\ 
N4594  & SA(s)a  &  9.8  & $10.0^{+10.0}_{-7.0}$       & 22,SD & \multicolumn{4}{c}{no 3-integral models}  \\ 
N224  &  Sb      & 0.77 & $0.35^{-0.25}_{+0.25}$       & 23,SD & \multicolumn{4}{c}{double nucleus}  \\ 
N4041&	Sbc     & 16.4      & $<$ 0.2		       & 24,GD & \multicolumn{4}{c}{disk might be
dynamically decoupled}\\ 
\hline
\label{tab:allmasses}
\end{tabular}
\end{table}

\clearpage
\scriptsize
\noindent Notes to Table II: The columns give the galaxy's Hubble type; distance (from Tonry et
al. 2001 whenever available; derived from the heliocentric systemic
velocity  and $H_0 = 75$ \kms~Mpc$^{-1}$ in all other cases); black
hole mass,  reference (coded below) and method of detection (PM =
stellar proper motion, GD= gas dynamics, SD = stellar dynamics, MM =
\H2O megamasers); central bulge velocity dispersion; total, extinction
corrected blue magnitude (from the RC3, de Vaucouleurs \etal 1991,
corrected for Galactic extinction using the reddening maps of Schlegel
\etal 1998); fraction of the total light judged to be in the hot
stellar component (from Fukugita \etal 1998); the ratio of the
diameter of the SBH sphere of influence to the spatial resolution of
the data.  References:  
1. Ghez \etal 2003 --  
2. Miyoshi \etal 1995 -- 
3. Macchetto \etal 1997 --  
4. Emsellem \etal 1999 --
5. Cappellari \etal 2002 -- 
6. Bower \etal 1998 --
7. Gebhardt \etal 2003 --
8. Verolme \etal 2002 --  
9. Marconi \etal 2001 --
10. Devereux \etal 2003 --  
11. Ferrarese \etal 1996 -- 
12. Tadhunter \etal 2003 -- 
13. Sarzi \etal 2001 --  
14. Gebhardt \etal 2000a --  
15. Barth \etal 2001 --  
16. Cretton \& van den Bosch 1999--  
17. van der Marel \& van den Bosch 1998 --  
18. Ferrarese \& Ford 1999 -- 
19. Greenhill \etal 2003b --
20. Greenhill, Moran \& Herrnstein 1997 -- 
21. Greenhill \etal 1996 -- 
22. Kormendy \etal 1988 -- 
23. Bacon \etal 2001 --  
24. Marconi \etal 2003.
\normalsize

\vskip .2in
\noindent distance of the Virgo cluster, 15 Mpc, the
sphere of influence of a $\sim 3 \times 10^7$ \msun~SBH would shrink to a projected 
radius of 0\Sec07, not only well beyond the reach of any ground based telescope, 
beyond even \hst capabilities. Overall, as will be discussed in
\S~\ref{sec:future}, the number of galaxies for which the SBH sphere of
influence can be resolved with ground-based optical observations can
be counted on the fingers of one hand. \hst has enabled 
that number to be increased by well over an order of magnitude.

\section {STELLAR DYNAMICAL STUDIES}
\label{sec:sdyn}

\subsection {A Special Case: The Galactic Center}
\label{sec:mw}

The case for a massive object at the Galactic Center has been building
since the 1970's detection of strong radio emission originating from
the innermost 1-pc (Balick \& Brown 1974; Ekers \etal 1975). Not only
is the source, dubbed Sgr A$^*$, extremely compact (VLBI observations
at 86 GHz set a tight upper limit of 1 A.U. to its size, Doeleman et
al. 2001), but the  absence of any appreciable proper motion implies that
it must also be very massive. In the most recent study on the subject,
Reid \etal (2003) argue that Sgr A$^*$ must be in excess of
$4\times10^5$ \Msun, thus excluding that it might consist of a compact
cluster of stellar objects.  Because of its proximity ($8.0 \pm 0.4$
kpc, Eisenhauer \etal 2003), the Galactic Center can be studied at a
level of detail unimaginable in any other galaxy. Proper motions of the
star cluster surrounding Sgr A$^*$ can be detected using near infrared
speckle imaging techniques. On-going monitoring studies, conducted for
the past ten years at the ESO NTT and Keck Telescope first, and at the
ESO VLT more recently, have reached a  staggering 0\Sec003 (0.1 mpc)
astrometric accuracy in the stellar positions (Eckart \etal 1993;
Ghez \etal 1998; Ghez \etal 2000; Sch\"{o}del \etal 2003; Ghez et
al. 2003): proper motion has been measured for over 40 stars within
1\Sec2 of Sgr A$^*$; deviations from linear motion has been detected
for eight stars and four stars in particular have passed the
pericenter of their orbits since monitoring began (Ghez \etal 2003;
Sch\"{o}del \etal 2003). In the three cases for which accurate orbits can be
traced, the stars orbit Sgr A$^*$ with periods
between 15 and 71 years, reaching as close as 87 A.U. from the central
source.  Using a simultaneous multi-orbital solution, Ghez et
al. (2003) derive a best fit central mass of $(4.0 \pm 0.3)\times10^6$
\Msun. The implied central mass density of $4\times 10^{16}$
\Msun~Mpc$^{-3}$, provides virtually incontrovertible evidence that the
mass is indeed in the form of a singularity.

An excellent review of the nature and observations of Sgr A$^*$ can be found in
Melia \& Falcke (2001), to which we refer the reader for
a detailed discussion.

\subsection {Integrated Stellar Dynamics}
\label{sec:intsd}

Modeling the kinematics of stars in galactic nuclei has historically
been the method of choice to constrain the central potential, and for
good reasons: stars are always present, and their motion is always
gravitational. But, as for every method, downsides also exists.
Stellar absorption lines are faint and the central surface brightness,
especially in bright ellipticals, is low (Crane \etal 1993; Ferrarese
\etal 1994; Lauer \etal 1995; Rest \etal 2001).  Acquiring stellar
kinematical data therefore often entails walking a fine line between
the need for high spatial resolution and the need for high spectral
signal-to-noise: the latter benefits from the large collective area of
ground based telescopes, while the former demands the use of \hst in
all but a handful of cases.  Theoretical challenges arise from the
fact that the stellar orbital  structure is unknown and difficult to
recover from the observables. Although dynamical models have reached a
high degree of sophistication (Verolme \etal 2002; Gebhardt et
al. 2003; van de Ven \etal 2003), the biases and systematics which
might affect them have not been fully investigated and could be severe
(Valluri, Merritt \& Emsellem 2004).  For the rest of this section, we
will explore in detail some of the issues related to stellar dynamical
modeling.

Unlike globular clusters (e.g. Binney \& Tremaine 1987), and with the
possible exception of unusually dense galactic nuclei -- M32 and
NGC205  could be such cases -- stars in a galaxy have not had enough
time to become aware of each other's individual existence. Putting it
more rigorously, both the characteristic crossing time $t_c$, and the
galaxy age, are much shorter (typically by a factor $10^8$ and $10^6$
respectively) than the  relaxation time, $t_{rel}$, defined as the characteristic timescale over
which, due to the cumulative effects of stellar encounters, a typical
star acquires a transverse velocity equal to its initial
velocity.  To very good approximation, therefore, galaxies can be
treated as collisionless
stellar systems: each star can be thought of as moving in the combined
gravitational potential $\Phi(\vec{x},t)$ of all other stars. This makes it possible
to describe the system analytically, while the dynamics of globular clusters, for instance, 
must be studied with the aid of numerical simulations. The distribution
function $f(\vec{x},\vec{v},t)$ (DF), defined as the number of stars
which occupy a given infinitesimal volume in phase-space, obeys a
continuity equation, i.e. the rate of change of the number of stars
within a given phase space volume is equal to the amount of inflow
minus the amount of outflow:

\begin{equation}%2 
{{\partial f} \over {\partial t}} + \vec{v} \cdot \vec{\nabla} f -
\vec{\nabla} \Phi(\vec{x},t) \cdot {{\partial f} \over {\partial \vec{v}}}
= 0
\end{equation}

Equation 12 is known as the Collisionless Boltzmann Equation
(CBE). $\Phi(\vec{x},t)$ is linked to the total mass density $\rho$ by
the Poisson equation:

\begin{equation}%2 
\nabla^2 \Phi(\vec{x},t) = 4 \pi G \rho(\vec{x},t)
\end{equation}

In the above equation, $\rho$ comprises all mass present  within the
system, including not only stars, but also SBHs  and dark matter, if
present.

Modulo a multiplicative factor (the stellar mass-to-light ratio), the
stellar mass density and the six components of the streaming velocity
and velocity dispersion involve integrals of the DF in velocity
space. This implies that the DF can be reconstructed given the stellar
mass density and velocity tensor. Once the DF is known, the total
gravitational potential follows from the CBE, and the total mass
density (and hence the SBH mass, if the stellar mass density is known)
from Poisson's equation.

Unfortunately, not all seven variables can be extracted from
observational data: galaxies' images and spectra  contain information
on the projected surface brightness profile and the integrated, line
of sight velocity and velocity dispersion only. It is only at the cost
of making further simplification that the CBE and Poisson's equations
can be solved analytically. For instance, if there are grounds to
believe that the system is in a steady state (i.e. all time derivatives are
null), spherically symmetric, and isotropic, the first velocity moment
of the CBE becomes:

\begin{equation}%2 
GM(r) = -r \sigma_r^2 \left[{{{\rm d ln}\nu}\over{{\rm d ln}r}} + {{{\rm d
ln}\sigma_r^2}\over{{\rm d ln}r}}\right]
\end{equation}

\noindent where $\sigma_r^2$ is the velocity dispersion, and $\nu$ is
the stellar mass density. Both are related to the surface brightness
profile and the line of sight velocity dispersion through Abel
integrals which can be easily inverted; therefore equation 9 can be
solved analytically to derive $M(r)$.

\begin{figure}[t]
\begin{minipage}[t]{6.0cm}
\psfig{figure=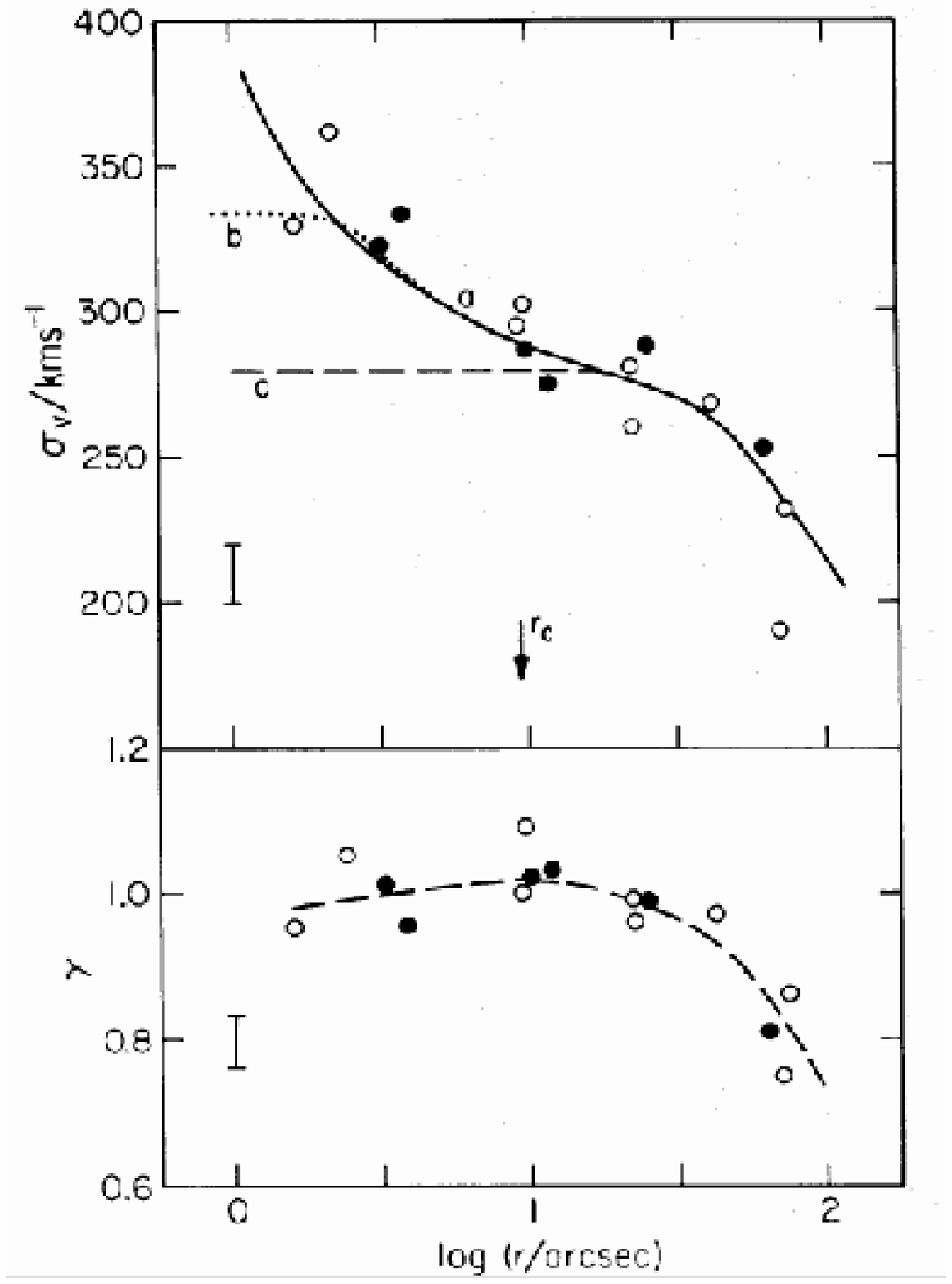,height=9.0 cm}
\end{minipage}
\begin{minipage}[t]{6.0cm}
\psfig{figure=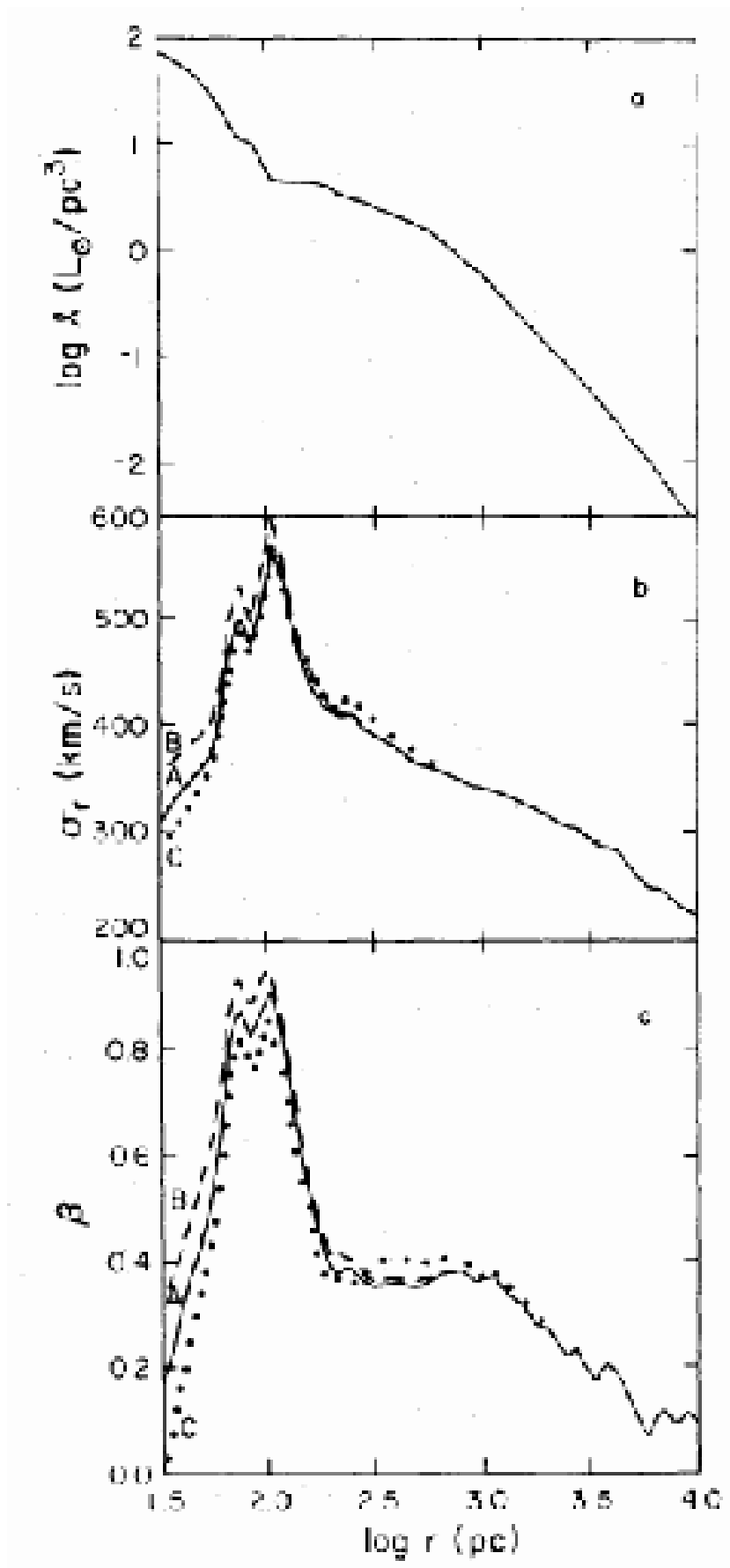,height=9.0 cm}
\end{minipage}
\caption{Left: The velocity dispersion for M87, as measured by Sargent
\etal (1978). The open and solid symbols show points W and E of the
nucleus respectively. The solid, dotted and dashed lines correspond to
models including a $5\times10^9$ \msun~SBH, the same model  convolved
with the seeing disk, and a spherically symmetric, isotropic model
with no black hole. From Sargent \etal 1978.  Right: The results of
Binney \& Mamon dynamical modeling of Sargent \etal (1978) data for
M87. The mass to light ratio is assumed to be constant, but the level
of anisotropy in the system is allowed to vary as a function of
radius. From top to bottom, the panels show the computed luminosity
density, radial velocity dispersion and anisotropy parameter $\beta =
(\sigma_r^2 - \sigma_{\theta}^2) / \sigma_r^2$ predicted by the
model. The models fit the observables without requiring a central mass
concentration -- in particular they reproduce the sharp rise in
velocity dispersion measured by Sargent \etal within 2$''$ from the
center. The three separate curves correspond to three different
extrapolations of the observed velocity dispersion within 2$''$ ($\sim 150$ pc). From
Binney \& Mamon (1982).}
\label{m87_1}
\end{figure}

The first attempts to use morphological and dynamical data to
constrain the presence of a central mass were indeed based on Equation
9: assuming an isotropic, spherically symmetric system, Sargent et
al. (1978)  ``detected'' a central $\sim 5 \times 10^9$ \msun~dark
mass within the inner 110 pc of M87 (Figure~\ref{m87_1}). 
But how much does this result
depend on the assumptions of sphericity and isotropy? Like all giant
ellipticals, M87 does not appreciably rotate, and is well described
(at least in the inner parts) by a spherical system. However, the
assumption of isotropy is unfounded. The ratio between velocity
dispersion and circular velocity in massive ellipticals is far higher
than predicted for an isotropic model (Illingworth 1977), implying
that these systems are supported by an anisotropic velocity
tensor. Relaxing the assumption of isotropy drastically alters the
dynamical models. The Sargent \etal data were reanalyzed by Binney \&
Mamon (1982), and later Richstone \& Tremaine (1985). Both groups
lifted the assumption of isotropy. To still be able to solve the CBE
analytically, if isotropy is not assumed, some other restrictions must
be imposed on the system. In the specific case of M87, the mass to light
ratio was constrained to be constant throughout the galaxy. The model
which is able to fit the data has a mass to light ratio of 7.6 (in the
$V-$band), plausible for a stellar system, and a highly anisotropic
velocity dispersion in the inner 300 pc  (Figure~\ref{m87_1}). This is easy to understand: a
highly radial velocity dispersion will mimic an increase in the
observed line of sight velocity dispersion in the central regions (as
the presence of a central dark object would) since, compared to the
isotropic case, the space-volume sampled by the data includes more
radially directed stars moving closely along the line of sight.

\begin{figure}[t]
\centerline{\psfig{figure=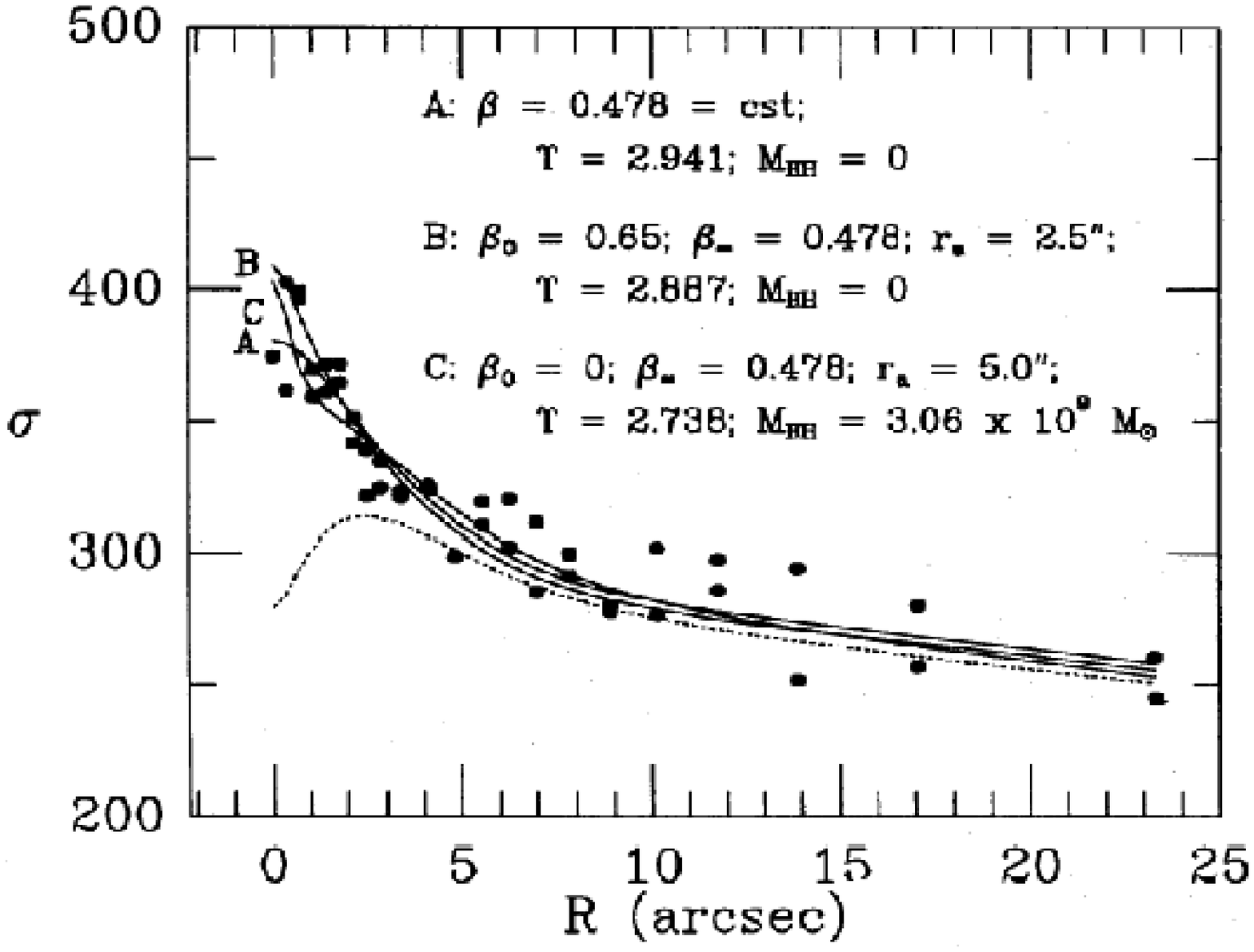,height=9.0 cm}}
\caption{The figure, from van der Marel (1994), shows the best stellar
dynamical data obtained to date for M87. The velocity dispersion is
seen to increase towards the center, in agreement with earlier
observations by Sargent \etal (1978). Models A and B are radially
anisotropic and have no central black hole, while model C is isotropic
towards the center and requires a central black hole to fit the data
(the dotted line represent the same model when the black hole is
removed).}
\label{M87_3}
\end{figure}

Although the Binney \& Mamon model has not been tested for dynamical
stability -- indeed, {\it no} anisotropic dynamical models of galactic
nuclei have -- the example serves to illustrate the difficulty in
modeling stellar kinematical data. It is important to realize that,
especially for non rotating, giant ellipticals, such difficulty cannot
be rooted out given the observables (see also the illuminating
discussion at the beginning of \S 4 of Kormendy \& Richstone
1995). Indeed, for M87 the ambiguity remained even when more recent,
state of the art data (van den Marel 1994) were used (Figure~\ref{M87_3}). A potentially
powerful way to break the degeneracy between a varying mass-to-light
ratio, and velocity anisotropy comes from analyzing the Line of Sight
Velocity Distribution (LOSVD), defined as:

\begin{equation}%2 
LOSVD(v_z,x,y) = {{1}\over{\mu}} \int\!\!\int\!\!\int
f(\vec{x},\vec{v})dv_x dv_y dz
\end{equation}

\noindent where $\mu$ is the projected surface brightness profile at
position $(x,y)$, measured in the plane of the sky. The LOSVD, which is reflected in the shape of the
absorption line profiles, is routinely expressed in terms of
Gauss-Hermite moments, which arise from an expansion of the profile
shape in terms of orthogonal functions. The second moment in
particular depends on the level of anisotropy of the system (van der
Marel \& Franx 1993; Gerhard 1993). For instance, tangentially
(radially) anisotropic DFs produce LOSVDs which are more flat-topped
(peaked) compared to the isotropic case. The presence of a SBH further
influences the LOSVD by stretching its wings, due to  high velocity
stars orbiting in the Keplerian potential (van der Marel 1994).  In
practice, a comprehensive study of the LOSVD requires higher S/N than
available (or obtainable) for the vast majority of galaxies. It
therefore remains generally true that stellar dynamical models are
best applied to rapidly rotating systems (most faint ellipticals are
in this class, e.g. Kormendy \& Richstone 1992). Not only are these
galaxies more accurately described as isotropic systems
(Illingworth 1977) but also, as the streaming velocity  becomes the
dominant kinematical component, terms involving the velocity
dispersion play only a second order effect in the CBE.  The most
extreme  cases of such systems are galaxies hosting small nuclear
stellar disks, a not uncommon occurrence especially in late type
ellipticals (Kormendy \& Richstone 1992; Scorza \& Bender 1995; van
den Bosch \& de Zeeuw 1996; van den Bosch, Jaffe \& van der Marel
1998; Cretton \& van den Bosch 1999). Van den Bosch \& de Zeeuw (1996)
find that if such a galaxy hosts a sufficiently massive SBH, the wing
of the bulge-dominated LOSVD will show a clear signature of the fast
rotation of the disk. Unfortunately, for this to happen, the mass of
the SBH needs to be a few percent of the mass of the hot stellar
component, at least a factor 10 larger than has been measured
(\S~\ref{sec:demo}).

Since the 1978 work of Sargent \etal, the level of sophistication of
the dynamical models applied to stellar kinematics has been improving
steadily. The complexity of the models is reflected in the complexity
of the distribution function needed to describe the data. For
instance, the DF of a spherical, isotropic system depends only on one
integral of motion, the total energy of the system. The assumption of
isotropy makes it possible to have a one-to-one correspondence between
mass density $\rho(r)$ and distribution function: given $\rho(r)$
(which can be determined from the data once a stellar-mass-to-light
ratio $\gamma_{star}$ and, for our application, a central dark mass
$M_{\bullet}$ are assumed), it is always possible to find analytically
a DF which self-consistently generates  $\rho(r)$ (and
vice-versa). Once the DF is known, the velocity dispersion can be
predicted uniquely. It follows that it is always possible to determine
the values of $\gamma_{star}$ and $M_{\bullet}$ which best fit the
observables. As mentioned earlier, whether these values actually
describe the real galaxy hinges critically on whether the assumptions
do, which should not be taken for granted.

If the velocity dispersion is anisotropic, the DF must depend on at
least two integrals of motion; in the simplest case $f=f(E,L_z)$,
where $L_z$ is the vertical component of the angular momentum. 
Such a DF can be fully constrained
provided that the mass density and the tangential component of the
streaming motion $v_{\theta}$ are known. As in the spherical isotropic
case, therefore, the mean square velocity (velocity dispersion and
streaming motion) can be uniquely predicted in any given gravitational
potential. Two-integral (2I) models are handled through the Jeans
equations, which relate the second order of the velocity moments to
the potential and density of the stellar system. The procedure 
follows the steps below:

\begin{enumerate}

\item Deproject the observed 2-D surface brightness profile to derive
the underlying luminosity density. The deprojection is not unique, and
an inclination angle $i$ must be assumed.

\item Translate the luminosity density to a mass density by assuming a
(generally constant with radius) $\gamma_{star}$ and a central point
mass $M_{\bullet}$. Compute the gravitational potential $\Phi$
corresponding to the mass density thus derived.

\item Solve the Jeans equation for the mean square velocities.

\item Project the mean velocities onto the plane of the sky to get the
line of sight velocity and velocity dispersion.

\item Compare the predicted and observed velocities.

\item Return to point 1, and iterate until the values  $i$,
$\gamma_{star}$ and $M_{\bullet}$ which produce the best fit to the
data are found. In practice, the data are never of high enough quality
for all three parameters to be constrained, and an inclination angle
$i$ is always assumed a priori.

\end{enumerate}

As in the case of spherical isotropic models, 2I models can produce
good fits to the data, but the solution might be severely in error if
the galaxy under study is not well approximated by the assumption that
$f=f(E,L_z)$.  Unfortunately, there is ample evidence that this
condition is in fact not generally obeyed.  Not only do 2I models predict
major-axis velocity dispersions which are larger than observed, they
also require identical velocity dispersions in the radial and vertical
direction, a condition not verified, for instance, in the solar
neighborhood. Furthermore, numerical simulations show that most orbits
are not completely described by two integrals of motion: a third
integral must be admitted, although an analytical description of such an
integral is not known. Dynamical models in which the distribution
function depends on three integrals of motion are referred to as 3I
models. State of the art 3I models are now routinely applied to
stellar kinematics (Verolme \etal 2002; Gebhardt \etal 2003).
Schwarzschild (1979) is credited with devising a way to construct
galaxy models without an explicit knowledge of the integrals of
motions:

\begin{enumerate}

\item Start with a choice of potential, as for the 2I models (items
  1. and 2.)

\item For a grid of cells in position space, choose initial conditions
for a set of orbits. For each orbit, the equations of motion are
integrated over many orbital periods, and a tally is kept of how much
time each orbit spends in each cell. This provides a measure of how
much mass is contributed to each cell by each orbit.

\item Determine non-negative weights for each orbit such that the
summed mass and velocity structure in each cell, when integrated along
the line of sight, reproduce the observed surface brightness and
kinematical constraints.

\end{enumerate}
 
With the introduction of 3I models, any kind of stellar system can in
principle be modeled in the most general and unconstrained way. In
practice, however, some assumptions must still be made. With one
exception (Verolme \etal 2002), the inclination angle of the galaxy
is always assumed a priori (e.g. Gebhardt \etal 2003); not doing so
would introduce an extra degree of freedom which cannot be
constrained given the observables.  Furthermore, although steps toward
a formalism for triaxial systems have recently been taken (van de Ven
\etal 2003), all current models assume axisymmetry.

These problems notwithstanding, systematic uncertainties could be
hidden in the method itself (if a lesson is to be learned from the M87 example
discussed above, it is that stellar dynamical studies can be fallible!). 
It is therefore rather surprising that to date there
has been only one study specifically aimed at investigating the
incidence of systematics in 3I models (Valluri, Merritt \& Emsellem
2004). This study reaches the rather bleak conclusion that even in the
case of the best observational datasets, 3I models admit too many
degrees of freedom to constrain $\mh$. Valluri \etal identify a
fundamental reason for this: since in the 3I case there is an infinite
number of distribution functions corresponding to a given $\rho(r)$
and $v_{\theta}$ (e.g. Binney \& Tremaine 1987), a change in the
gravitational potential (for instance due to the introduction of a
central point mass) can always be compensated with a change in the DF,
so to leave the goodness of the fit of the model to the data
unchanged. Valluri \etal further argue that this intrinsic
indeterminacy becomes apparent only when the number of orbits used in
the simulations (step 2 above) is large (at least a factor 10)
compared to the number of observational constraints. When only a few
orbits are used (relative to the number of constraints), the system is
artificially restricted to such an extent that an apparently  well
constrained (but very likely biased) solution is found.  Although it
seems unlikely that 3I models are severely flawed (after all, SBH
masses determined using these models agree, in a statistical sense,
with those measured using completely independent methods, such as
dynamical studies of gas kinematics, and reverberation mapping,
\S~\ref{sec:scale}) quantifying possible systematics is a must. Consequently, it is imperative that more studies address this
issue in the future.

The 17 claimed SBH detections that are based on 3I models are listed in
Table~\ref{tab:allmasses}, ranked from highest to lowest resolution of the 
SBH sphere of influence. With no exceptions, they all required the
use of \hst data. Even then, the sphere of influence was not resolved
in five cases, and barely resolved in several others.

\section {GAS DYNAMICS FROM WATER MASER CLOUDS}
\label{sec:maser}

The 1995 announcement of the discovery of a massive black hole in NGC
4258 (Miyoshi \etal 1995) marked the beginning of an unexpected and
powerful new way to measure central masses in AGNs.  At a distance of
7.2 Mpc (Herrnstein \etal 1999), and given the low bulge velocity
dispersion of the galaxy (Heraudean \& Simien 1998) the Hubble Space
Telescope would be able to resolve the sphere of influence only of a
SBH more massive than a few $10^8$ \msun~(equation 6).  Scaling
relations (\S~\ref{sec:scale}) predict a SBH mass an order of
magnitude smaller. NGC 4258, however, is blessed by the presence of
water masers clouds, confined to a thin, regular disk extending only a
fraction of a parsec from the central source. Emitting at 22 GHz, the
masers can be studied with the VLBA at spatial resolutions a factor
$\sim 200$  higher than can be achieved using the Hubble Space
Telescope, instantly pushing the ability to detect a SBH to
correspondingly smaller masses.

The history of extragalactic water masers is a recent one, although
Galactic water masers, which are likely the result of collisional
excitation of warm interstellar gas (Neufeld \& Melnick 1991) were
first detected in  1968 (Cheung \etal 1969) and subsequently
identified in a number of star forming regions and late-type stellar
envelopes. The first extraordinarily luminous, extragalactic example
was found in the Seyfert 2 galaxy NGC 4945 (dos Santos and Lepine
1979).  In the following five years very luminous ($L > 100 $\Lsun)
\H2O masers were detected in four additional active galaxies: Circinus
(Gardner \& Whiteoak 1982), NGC 4258, NGC 1068 (Claussen, Heiligman,
\& Lo 1984), and NGC 3079 (Henkel \etal 1984).  The  22 GHz water
maser emission line luminosity in NGC 1068 is $L_{22\rm GHz} = 350~
\Lsun$, exceeding the luminosity of a typical Galactic water maser by
a factor of $\sim 3.5 \times 10^5$, and exceeding the brightest
Galactic source, W49 (Walker, Matsakis, and Garcia-Barreto 1982) by a
factor of 350.

VLA observations by Clausen \& Lo (1986) were unable to resolve the maser sources
in NGC 1068 and NGC 4258, which must therefore be confined within
3.5 and 1.3 pc of the nucleus.  Based on marked variability of the two
sources and preliminary VLBI observations, Clausen \& Lo concluded
that the true sizes of the sources were 10-100 times smaller than the
VLA upper limits. This made it unlikely that the masers could be a
superposition of Galactic-type water masers, which are known to be
powered by massive young stars.  To further strengthen the lack of
association between the nuclear and Galactic masers, they noted that
two nearby starburst galaxies, M82 and NGC 253, are not luminous water
maser sources, and that the infrared luminosity that would be required
to excite the \H2O maser in NGC 4258 exceeded the observed IR
luminosity of the galaxy by a factor of nearly 300. The logical
conclusion was that the excitation energy was produced in the active
nucleus.  In an inspired insight, Clausen \& Lo suggested that the
masing in NGC 1068 might arise in the postulated -- but never directly
observed -- pc-size obscuring torus which is at the heart of the
Seyfert 1/2 unification scheme (Antonucci and Miller 1985).

Neufeld, Maloney, \& Conger (1994) show that luminous water masers can
be produced when X-rays, presumably  generated by the innermost
accretion disk around the central SBH, illuminate and heat a torus of
dense circumnuclear gas and dust.  Their model leads to ``a sandwich
structure in which the high-pressure gas closest to the midplane is
molecular but the lower-pressure gas above and below the plane is
atomic.''  When the X-ray heating is large enough to raise the
temperature in gas and dust above 400 K, but small enough to permit
the existence of H$_2$, a reaction network O + H$_2 \rightarrow $ OH +
H and OH + H$_2 \rightarrow $ H$_2$O + H is very efficient at
producing water with an abundance ratio of water to hydrogen  of
several $10^{-4}$.  Over a range of X-ray fluxes a stable two phase
structure is possible with an atomic phase at $T\sim 6000$ to 8000 K
and  a molecular phase at $T\sim 600$ to 2500 K.  Warping in the maser
disk is argued to be critical: if the disk were flat, the path through
the  disk would be optically thick to X-rays and the X-ray opacity
would become too high for the heating required to excite the
masers. This prediction is  supported by observations of NGC 4258,
which indeed show the maser disk to be mildly warped
(\S~\ref{sec:n4258})

Unfortunately, \H2O masers are not common. The first large survey was
conducted by Braatz, Wilson, \& Henkel (1994, 1996). Targeting 354
galaxies mostly within 10 Mpc, including Seyfert galaxies, LINERS, and
radio galaxies, they found a 7\% detection rate (13 sources) among
Seyfert 2 and LINERS, but no detections in Seyfert 1 nuclei. This is
not surprising, since to detect water masers the observer must see a
long path length through the torus. Therefore, when the geometry  is
favorable to observe masing, the Seyfert 1 nucleus is hidden by the
optically thick torus. The Greenhill \etal (1997) survey of 26 AGNs with
the 70m antenna of the NASA Deep Space Network produced only one
detection, NGC 3735. One additional detection stemmed from a survey
of 131 AGNs with the Parkes Observatory (Greenhill \etal 2002). In
most of the sources the masers were detected only at the systemic
velocity of the galaxy.  In a few cases, however, NGC 4258 being the
most noticeable (Nakai \etal 1995), high velocity ($v \sim 1000$ km
s$^{-1}$), maser clouds were also detected. It is these high velocity
clouds which, as we will discuss in more detail below, open the
possibility of constraining the central potential through the study of
the masers' velocity field.

\begin{figure}[t]
\centerline{\psfig{figure=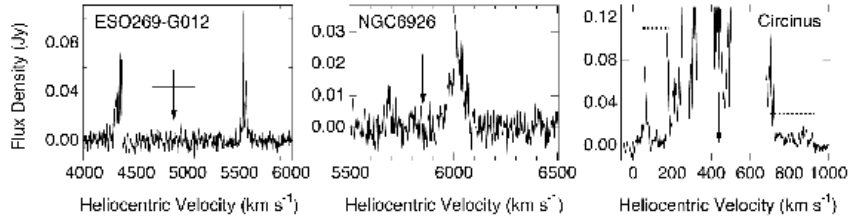,height=3.0 cm}}
\caption{The NASA Deep Space Antenna spectra of Circinus and two newly
discovered sources showing \H2O high velocity maser emission, obtained by 
Greenhill et al (2003a).  The arrows indicate the galaxies systemic
velocity, assuming the radio definition of Doppler shift.
The velocity range of the newly discovered emission in Circinus 
is indicated by the heavy horizontal dotted lines.
These sources are excellent candidates for VLBI followup, in the hope
that the spatial and kinematical structure of the clouds can be resolved.}
\label{masers}
\end{figure}

The most recent, and most promising results come from Greenhill et
al. (2003a). Benefiting from  a larger sensitivity and wider wavelength
coverage than previous surveys, this study led to the discovery of
seven new sources  among 160 nearby ($cz < 8100$ km s$^{-1}$) AGNs
surveyed with the NASA Deep Space Antenna. More exciting still, two of
the sources exhibit high velocity masers and are promising targets for
VLBI follow-ups (Figure~\ref{masers}).

Besides the fact that no \H2O maser emission is detected in Seyfert 1
galaxies, no strong correlations have been found between maser
emission and the global properties of the host galaxies, although
where X-ray measurements are available, all known masers lie in
galaxies with large X-ray obscuring column densities (Braatz et
al. 1997). All galaxies in which \H2O nuclear masers have been
detected are listed in Table~\ref{masertable}.

\begin{table}[t] 
\scriptsize
\caption{Luminous 22 GHz H$_2$0 Masers in the Nuclei of Active
Galaxies}
\begin{tabular}{lllllll} 
\hline Galaxy & Type	& Distance	&Reference &  High v &  Rot. &
SBH Mass\\ & 	& (Mp\rlap{c)}	& &   Masers?&  Curve? & (\msun)\\
\hline     M51   & S2/L              & 9.6 & Hagiwara \etal 2001a & yes
& no &  \\ NGC 1052  & L      & 20 & Braatz \etal 1996    & no    &
no &  \\ NGC 1068  & S2    & 16 & Greenhill \etal 1996 & yes   & yes&
$\sim 1.5 \times 10^7$\\ NGC 1386   & S2   & 12 & Braatz \etal 1996 &
no    & no &  \\ NGC 2639   & S2   & 44 & Braatz \etal 1996 & no    &
no &  \\ NGC 2824    & S?   & 36 & Greenhill \etal 2003a & no & no &
\\ NGC 2979    & S2   & 36 & Greenhill \etal 2003a    & no    & no &
\\ NGC 3079    & S2   & 16 & Trotter \etal 1998       & yes   & yes &
$\sim 10^6$\\ NGC 3735& S2  & 36 & Greenhill \etal 1997 & no    & no
&  \\ NGC 4258& S2  & 7.2 & Greenhill \etal 1995     & yes   & yes&
$(3.9\pm0.34) \times 10^7$ \\ NGC 4945& S2  & 3.7 & Greenhill et
al. 1997     & yes   & yes& $\sim 10^6$\\ NGC 5347  & S2 & 32 & Braatz
\etal 1996        & no    & no &  \\ NGC 5506  & S2 & 24 & Braatz et
al. 1996        & no    & no &  \\ NGC 5643   & S2 & 16 & Greenhill et
al. 2003a     & no    & no &  \\ NGC 5793   & S2 & 50 & Hagiwara et
al. 2001b     & yes   & yes& $\sim 10^7$\\ NGC 6240 & L      & 98 &
Hagiwara \etal 2002    & yes    & no &  \\ NGC 6300 & S2      & 15 &
Greenhill \etal 2003a    & no    & no &  \\ NGC 6929 & S2      & 78 &
Greenhill \etal 2003a    & yes   & no &  \\ IC1481  & L        & 83 &
Braatz \etal 1996    & no    & no &  \\ IC2560   & S2 & 38 & Ishihara
\etal 2001  & yes   & no & $2.8\times10^6$\\ Mrk1 & S2      & 65 &
Braatz \etal 1996    & no    & no &  \\ Mrk348     & S2      & 63 &
Falcke \etal 2000    & yes    & no &  \\ Mrk1210  & S2 & 54 & Braatz
\etal 1996    & no    & no &  \\ Mrk1419 & S2 & 66 & Henkel et
al. 2002 & yes   & no & $\sim 10^7$\\ Circinus   & S2 & 4 & Greenhill
\etal 2003a & yes   & yes& $(1.7 \pm 0.3) \times 10^6$\\ ESO269-G012&
S2      & 65 & Greenhill \etal 2003a & yes   & no &  \\
IRASF18333-6528& S2      & 57 & Braatz \etal 1996    & no    & no &
\\ IRASF22265-1826& S2     &100& Braatz \etal 1996 & no    & no &  \\
IRASF19370-0131& S2     & 81 & Greenhill \etal 2003a & no    & no &
\\ IRASF01063-8034& S2     & 53 & Greenhill \etal 2003a & no & no &
\\ \hline
\end{tabular}
The columns give the galaxy name; AGN type (L = LINER, S2 = Seyfert 2);
distance (from the heliocentric velocity, assuming $H_0 = 75$
\kms~Mpc$^{-1}$); whether high velocity emission has been detected;
whether the emission has been spatially resolved; and the estimated
SBH mass.
\label{masertable}
\end{table}

\subsection{ The Small Masing Disk in NGC 4258}
\label{sec:n4258}

NGC 4258 is a fascinating object. It was one of the 12 galaxies
identified by Seyfert (1943) as having peculiar nuclei, but the
peculiarities do not end there. In addition to two well-defined spiral
arms delineated by HII regions and bright OB associations, there are
two anomalous arms that are visible only in emission lines at optical
wavelengths (van der Kruit et al. 1972) (Figure~\ref{n4258}).   The
anomalous arms emit bright synchrotron radiation at radio wavelengths
(Turner \& Ho 1994). The arms have both a morphological (Ford \etal 1986)
and kinematical (Dettmar \& Koribalski 1990; Cecil, Wilson, \& Tully 1992) twist; the latter has
the same sense as the rotation of the molecular disk discussed below.

\begin{figure}
\centerline{\parbox[l][5.9cm]{9.2cm}{\vskip 3cm THIS FIGURE IS INCLUDED IN THE FULL VERSION OF THE
MANUSCRIPT AVAILABLE AT http://www.physics.rutgers.edu/~lff/publications.html.}}
%\begin{minipage}[t]{5.6cm}
%\psfig{figure=f10a.eps,height=5.9 cm}
%\end{minipage}
%\begin{minipage}[t]{5.6cm}
%\psfig{figure=f10b.eps,height=5.9 cm}
%\end{minipage}
\caption{Left: an H$\alpha$ on-band minus off-band image of NGC4258,
taken with the Kitt Peak No. 1 0.9-m telescope (Ford et al. 1986).
The image clearly shows the ``anomalous arms", extending almost
vertically from the center.  Ford et al. (1986) and Cecil et al. (2000) argued that
i) the anomalous arms may be the result of a precessing jet and ii), the sharp turn in the SE 
anomalous arms may be where they break out of the disk. Right: an HST
WFPC H$\alpha$ on-band minus off-band image of the nucleus taken by H. Ford. Cecil et al.
(2000) suggest that the ``pincher" shaped features are a complete ring formed from the projection of the edges of a bubble of hot gas inflated by the northern jet and ionized by
radiation from the central engine.}
\label{n4258}
\end{figure} 

The arms have long been thought to mark the location of a
bi-directional jet which is moving through the disk of NGC 4258. For
this to be the case, the axis of the central engine and the galaxy's
rotational axis must be misaligned by $\sim $90\deg, a condition that
might also apply to M51 and NGC 1068 (Ford \etal 1986).  If this
interpretation is correct, then the central accretion disk and
molecular torus, the axes of which should align with that of the SBH,
should be seen nearly edge on. This is precisely what the maser
observations suggest.

\begin{figure}[t]
\centerline{\psfig{figure=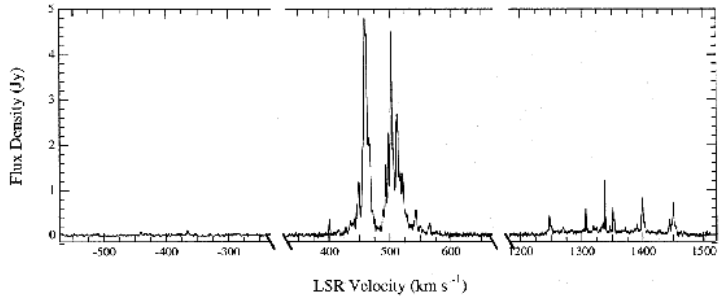,height=4.0 cm}}
\caption{The Nobeyama spectrum of NGC 4258 that first revealed the 
presence of high velocity \H2O maser emission (From Nakai \etal 1993)}
\label{n4258-spectrum}
\end{figure}

Nakai \etal (1993) used the Nobeyama 45-m telescope to discover high
velocity \H2O maser lines at $\sim \pm 1000$ \kms~relative to NGC
4258's systemic velocity (Figure~\ref{n4258-spectrum}).  They combined
very long baseline interferometric observations from the 45-m telescope and the Kashima 34-m
telescope to show that the strong features near systemic velocity are
within $\pm$0\Sec05 ($\pm 1.6$~pc) of the nucleus.   Nakai \etal
proposed three possible explanations for the observations: a
circumnuclear molecular torus in Keplerian orbit around a $\sim
10^8$~\Msun~central object, bi-directional outflow, and stimulated
Raman scattering of the emission features near systemic velocity.

Nakai et al.'s results precipitated a great deal of theoretical and
observational activity. Almost simultaneously, three independent teams
(Watson \& Wallin 1994; Haschick \etal 1994; Greenhill \etal 1995a)
arrived at the conclusion that the maser emission in NGC 4258 must
originate in a rapidly rotating Keplerian disk viewed nearly edge-on.
The incriminating piece of evidence, beside the existence of the high
velocity features, resided in the fact that the systemic velocities
showed both a secular and spatial variation, ${\rm d}(\Delta\bar v)/{\rm d}t =
6$~km~s$^{-1}$ yr$^{-1}$  and  ${\rm d}(\Delta\bar v)/{\rm d}{\alpha} =
280$~km~s$^{-1}$ mas$^{-1}$ respectively (Haschick \etal 1994;
Greenhill \etal 1995a\&b), exactly as  expected if
the masers originate in an edge on disk. In this case, the high
velocity lines arise from masing along the long lines of sight through
the two opposing tangent points in the disk, while the clouds which project
along the line of sight to the nucleus should have velocities close to the systemic velocity. The velocity of individual masing clouds which
project against the nucleus should therefore change with time (secular
change) as the clouds move along their trajectory, while clouds
projecting at slightly different locations should have slightly
different velocity (spatial change), as observed\footnotemark:

\footnotetext{According to this model, there should be no frequency
drift in the high velocity features, again in agreement with the
observations.}
  
\begin{equation}
{{{\rm d}(\Delta \bar v)} \over{{\rm d}t} }\simeq { {v^2_0}\over{r_0}},
\end{equation}

\begin{equation}
{{{\rm d}(\Delta \bar v)} \over{{\rm d}(D\alpha)} } = { {v_0}\over{r_0}}.
\end{equation}

Here, $v_0$ is the circular velocity at the outer radius ($r_0$) of
the disk, $D$ is the distance to the galaxy,  and $\alpha$ is the
angular displacement along the arc.

Watson \& Wallin modeled the hypothetical edge-on disk as a thin
unsaturated annulus surrounding a transparent gap in the disk and a
small central opaque region, which was included to account for the
splitting of the emission near systemic velocity into two sets of
lines. Assuming a distance $D = 6.6$ Mpc\footnotemark~ Watson \&
Wallin used equations 11) and 12) and advance publication results from
Greenhill \etal (1995b) to find $v_0 \sim 700$ \kms~ and $r_0 =
0.1$~pc.  The agreement of their derived $v_0$ with the observed
values of the high velocity satellite lines (740 to 980 \kms~and -760
to -920 \kms) lent strong support to the Keplerian disk hypothesis.
Knowing $v_0$ and $r_0$, they calculated the central mass as $10^7$~
\Msun.

\footnotetext{Subsequent high resolution VLBA observations, combined
with measurements of the angular size of the disk, the central mass,
and the observed temporal and spatial variation of the systemic lines
(equations 11 and 12) enabled Herrnstein \etal (1999) to derive a geometric distance to NGC 4258, $D = 7.2 \pm0.3$~Mpc.}

\begin{figure}[t]
\centerline{\psfig{figure=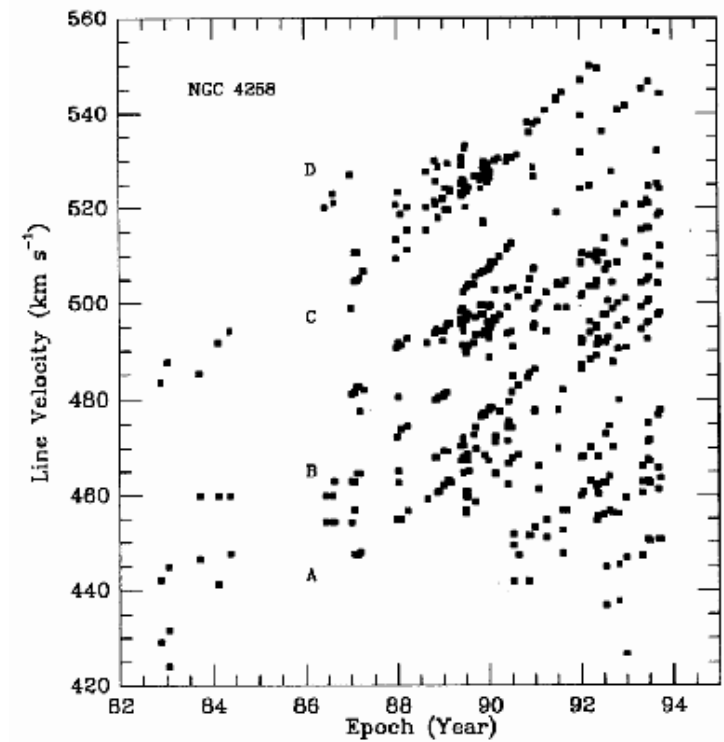,height=10.0 cm}}
\caption{The time dependence of the radial velocities of the systemic \H2O 
maser lines in NGC 4258.  The ``bands" show that the clouds drift
with a mean rate of 7.5 \kms yr$^{-1}$. (Courtesy of Haschick \etal
1994)}
\label{Haystack}
\end{figure}

\begin{figure}[t]
\centerline{\psfig{figure=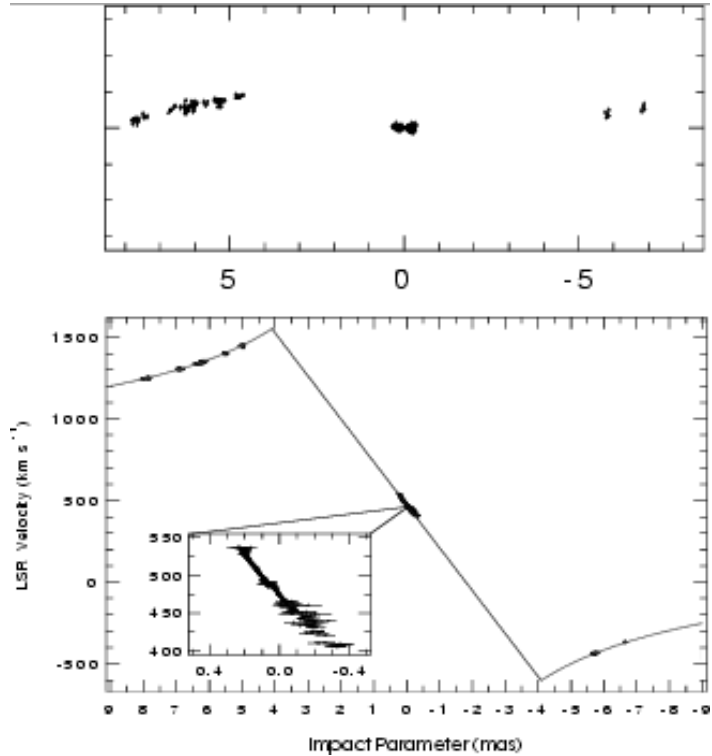,height=10.0 cm}}
\caption{The top panel shows the spatial distribution of the \H2O
megamasers in NGC 4258. Units are milliarcseconds, with 1 mas
corresponding to 0.035 pc at the distance of the galaxy. The
positional errors for the systemic and high velocity features are
$\sim 0.05$ mas and  $\sim 1$ mas respectively. The bottom panel shows
the rotation curve traced by the maser clouds. (From Moran \etal\
1999) }
\label{Miyoshi3}
\end{figure}

Similar conclusions were reached by Haschick \etal  ($r_0 = 0.11$~pc,
$v_0 = 900$~\kms, $\mh = 2.2 \times 10^7$~\Msun) by combining
Greenhill et al.'s (1995b) observed linear gradient in velocity, with
the results of a seven year monitoring program showing an acceleration
for the systemic \H2O maser lines of 7.5 \kms~yr$^{-1}$
(Figure~\ref{Haystack})\footnotemark. \footnotetext{They also noted
the previously reported fact (Haschick \& Baan 1990) that the flux
density of the systemic lines showed strong variability between 1988.6
and 1989.6.  The strongest feature ($v \sim 465$~\kms) varied with an
apparent 85 day period.  The periodicity, which has disappeared in
subsequent observations, remains unexplained.}  Interestingly, neither
Haschick \etal nor Watson \& Wallin suggested that the central mass
might reside in a massive black hole. The final proof had to await
until the day when Miyoshi's \etal (1995)  high resolution ($0.6
\times 0.3$~mas at a position angle of $\sim 7$\deg) VLBI observations
could spatially resolved the disk's structure. The masing clouds,
which are confined in an annulus with inner and outer radii of 0.14
and 0.28 pc respectively,  display the unmistakable $v^2 = GM/r$
signature of Keplerian motion around  $3.9 \times 10^7$ solar
masses\footnotemark. Miyoshi's \etal paper, combined with the \hstnsp
measurement of the mass of the SBH in M87, marked a turning point in
the debate about the existence of massive black holes.
Figure~\ref{Miyoshi3} shows the VLBA velocities plotted versus
distance along the disk major axis.  The high velocity lines on
opposing sides of the disk (which is very thin, Figure~\ref{Miyoshi3})
and the central lines are an almost perfect fit to a Keplerian disk
inclined to the plane of the sky by 86\deg.  Moran et al. (1999)
summarizes the properties of the disk and central mass.

\footnotetext{Values have been corrected to a distance of $7.2
\pm0.3$~Mpc (Herrnstein \etal 1999).}

The absence of measurable perturbations in the $r^{-0.5}$ Keplerian
dependence of the high velocity lines requires $M_{disk} < 4 \times
10^6$~\Msun.  Taking the 4.1 mas inner radius as an upper limit to the
size of the central object, Miyoshi \etal (1995) found that the mass density is $\rho >
4 \times 10^9$~\Msun~pc$^{-3}$, larger than the  density in the
densest known star clusters, globular clusters, where $\rho <
10^5$~\Msun~pc$^{-3}$, and several  orders of magnitude larger than
measured in any other galactic nucleus for which a SBH has been
claimed, with the exception of the Milky Way
(Figure~\ref{sbhdensity}). Stronger constraints still can be obtained
by assuming that the central source has a size less than the angular
extent of the systemic lines on the sky ($\sim 0.6$ mas = 0.02 pc).
This requires $\rho > 1.4 \times 10^{12}$~\Msun~pc$^{-3}$, leaving
virtually no doubt that the detected mass belongs to a central black
hole, with the masers orbiting as close as $\sim$40,000 Schwarzschild
radii. At this distance, relativistic corrections are small but not
negligible; the transverse Doppler shift is $\sim 2$~\kms, the
gravitational redshift is $\sim 4$ \kms, while the gravitational
deflection of the maser positions is only $0.1~\mu$as (Moran \etal
1995).

\begin{figure}[t]
\centerline{\psfig{figure=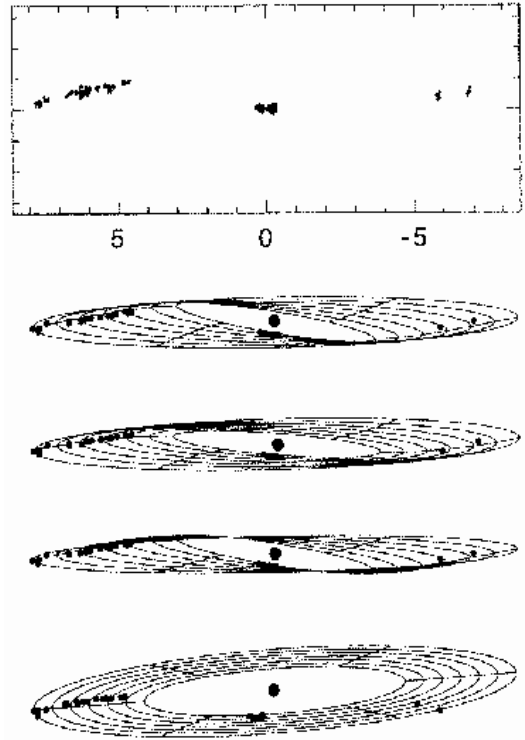,height=10.0 cm}}
\caption{The spatial distribution of water masers in NGC 4258 (top
panel, showing the east-west and north-south offsets, in
milliarcseconds), compared to  different models for the warping. 
The two top-most models allow only for a change of the position
angle of the disk with radius, while the third model allows for a
change in both position angle and inclination. The last model
is the best fitting flat disk, and clearly does a poor job at reproducing
the observed distribution of the masers. From Herrnstein \etal (1996).}
\label{Herrnstein}
\end{figure}

The VLBI observations of NGC 4258 are so exquisite to allow a most
detailed study of the physical properties of the molecular disk (Moran
\etal 1999; Neufeld \& Maloney 1995; Neufeld, Maloney \& Conger
1994). Herrnstein \etal (1996) convincingly argue that the disk is
warped; the degree of warping can be constrained given the relative
positions of the clouds on the sky, the line of sight velocities and
the accelerations for each of the maser clouds. The warping can be
represented by concentric rings with varying inclination and a
progression in the line of nodes; Herrnstein \etal conclude that a
model in which the line of nodes progresses by several (5 to 10)
degrees from the inner to the outer edge of the disk fits the
observations best. A few degree change in the position angle of the
disk cannot be excluded from the observations, because of the limited
coverage of the maser clouds (see Figure~\ref{Herrnstein}).  The warp
in the disk naturally explains the lower intensity of the blueshifted
masers compared to the redshifted emission and might be a necessary
condition for the maser emission to ensue (Neufeld \& Maloney 1995).

The disk is found to be at most $H =$ 0.01 mas (0.0003 pc) thick,
giving it a height to radius ratio of $H/R \leq 0.0025$. The disk is
then likely close to hydrostatic equilibrium, so that $H/R = c_s/v$
($v$ is the Keplerian rotational velocity). Knowing the sound speed
allows one to calculate Toomre's parameter $Q= {c_s \Omega}/{\pi G
\Sigma}$ for the stability of a disk (cf Binney \& Tremaine, 1987),
where $\Omega$ is the angular rotation speed. $Q$ is found to decrease
from the inside out, finally dropping below one (i.e. the disk becomes
unstable) at the outer edge of the disk.  Instability beyond this
radius may be what sets the outer radius of the masing portion of the
disk. The existence of the inner radius might result if the disk
becomes geometrically flat at that radius (Neufeld \& Maloney 1995).

Neufeld \& Maloney (1995)  estimate the mass accretion rate to be
$\Mdot = 7\times10^{-5} \alpha$ \msun yr$^{-1}$, where $\alpha$ is a dimensionless parameter 
that conventionally characterizes the disk viscosity. They find that if the accretion rate has been
constant over the $4 \times 10^6$ years required for material to
transit the disk from $R = 0.25$ pc into the nucleus, the efficiency
for conversion of rest mass into 2-10 Kev X-rays is
$0.01{\alpha}^{-1}$.  Since X-rays typically account for ~10\% of the
total luminosity of AGNs (Mushotzky, Done \& Pounds 1993), the
required efficiency for generation of the bolometric luminosity is
$\sim 0.1 {\alpha}^{-1}$, in agreement with expectations from standard
accretion disk models.

\subsection{The Maser Emission in NGC 1068}
\label{sec:n1068}

Because of NGC 1068 key role in AGN physics, we will discuss the \H2O
maser observations in this galaxy in some detail. The prototypical
Seyfert 2, NGC 1068 is the galaxy  for which the existence of a
parsec-scale, dense, dusty torus surrounding the nucleus was
postulated to  reconcile the spectral properties of Seyfert 1 and
Seyfert 2 nuclei  (Antonucci and Miller 1985). As mentioned at the
beginning of this section,  Clausen \& Lo (1986)  suggested that \H2O
maser emission in NGC 1068 might in fact arise from such an obscuring,
molecular torus.

\begin{figure}[t]
\centerline{\psfig{figure=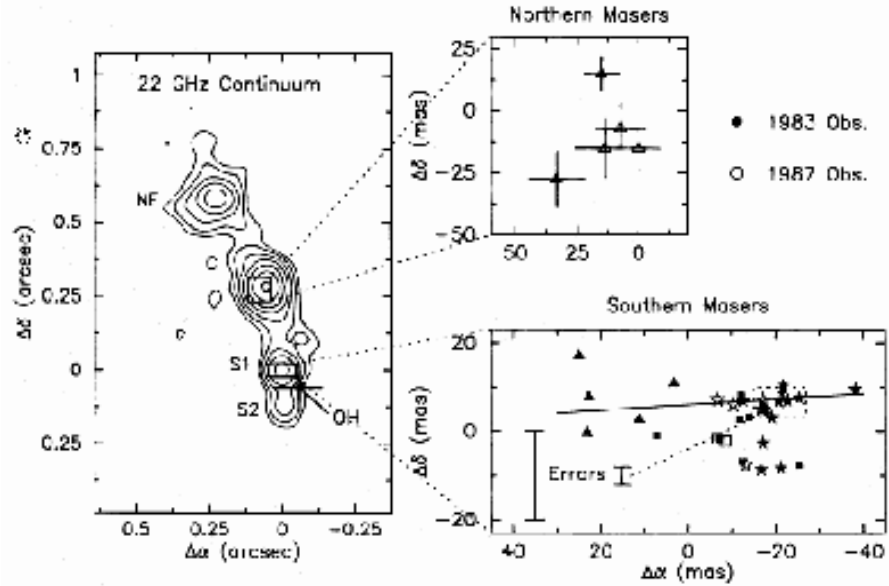,height=8.0 cm}}
\caption{The left inset shows the 22 GHz continuum in
the nucleus of NGC 1068.  The right hand insets show the positions
of the \H2O masers in the nucleus (S1) and source C. All data are
taken with the VLA (Gallimore \etal 1996b)}
\label{Gallimore1}
\end{figure}

The maser emission, which spans a 300 km s$^{-1}$ range from systemic
velocity, was partially resolved into a 40 mas structure by Gallimore
\etal (1996a,b\&c) using VLA observations.
Figure~\ref{Gallimore1}  shows the VLA positions of the masers
relative to a 22 GHz continuum map.

On the basis of its inverted radio spectrum, Gallimore
\etal (1996a,b\&c) identified the S1
feature shown in Figure~\ref{Gallimore1} with the  nucleus of the
galaxy. This conclusion was subsequently reinforced by Thatte et
al. (1997) using speckle and adaptive optics $K-$band and stellar 2.29
$\mu$m CO absorption bandhead observations. To within the astrometric
errors, the position of S1 coincides with that of the unresolved ($\le
2$~pc), heavily extincted $K-$band nucleus, and of the resolved
($\sim45$~pc), relatively young ($< 1.3 \times 10^9$~years) nuclear
star cluster revealed by the CO observations. Thatte \etal (1997)
speculate that the $K-$band thermal emission from the point source
arises from dust that is at the sublimation temperature because of
heating by the central engine.  The sublimation is most likely
occurring at the inner edge of the torus that hides the Seyfert 1
nucleus.

Also based on the radio spectra, Gallimore
\etal (1996a,b\&c) identify the  radio knots C and
NE (Figure~\ref{Gallimore1}) as knots in a jet that emanates from S1.
They conclude that the jet hits a molecular cloud at C and is
deflected by 20\deg~toward the NE knot.  They attribute the excitation
of the \H2O masers in knot C to a shock at the interface between the
radio jet and a dense molecular cloud.  In a subsequent paper
Gallimore \etal (1997) used VLBI continuum measurements to resolve S1
into a clumpy structure with dimensions $\sim$ 1 pc in length and
0.2pc wide.  S1 is approximately perpendicular to the direction of the
local radio jet.  They conclude that the radio continuum emission
comes from a hot ($T_e \geq 10^6$~K) dense ($n_e \geq
10^6$~\cm3)~plasma that is most likely heated and ablated from the
inner edge of the molecular disk due to intense heating by the central
AGN.

\begin{figure}[t]
\centerline{\psfig{figure=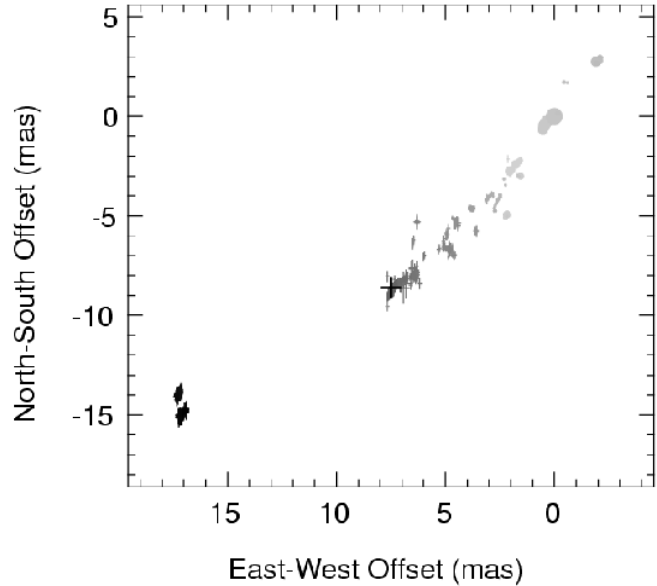,height=8.0 cm}}
\caption{The VLBI spatial distribution of \H2O masers in NGC 1068.
The gray scale shows the line-of-sight velocity, such that more redshifted
velocities are shown in lighter color.  The cross marks the emission
at the systemic velocity of the galaxy. (From Greenhill \& Gwinn 1997) }
\label{Greenhill2}
\end{figure}

Further insight to the morphology of the innermost regions in NGC 1068
came from VLBI observations of  Greenhill \etal (1996) and Greenhill
\& Gwinn (1997) who spatially resolved the high velocity maser
emission (Figure~\ref{Greenhill2}).

\begin{figure}[t]
\centerline{\psfig{figure=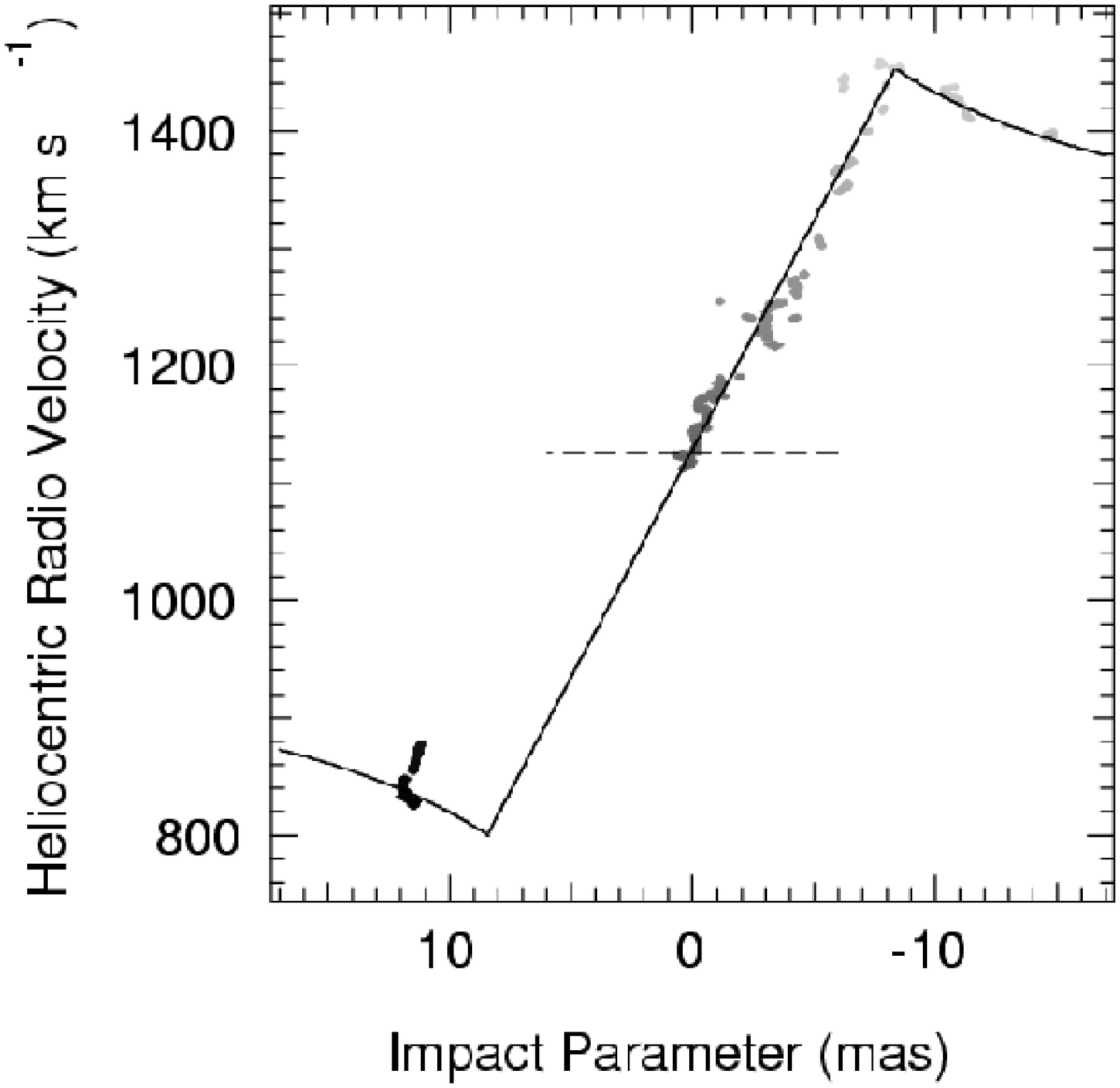,height=8.0 cm}}
\caption{The position-velocity diagram for the redshifted and
blueshifted \H2O masers in NGC 1068 (From Greenhill \& Gwinn 1997).}
\label{Greenhill3}
\end{figure}

Greenhill \& Gwinn (1997) plausibly interpret the nearly linear loci
of \H2O masers as a thin, flat, edge-on disk with rotation velocities
of 330 \kms~at 0.65 pc. The rotation axis of the disk does not
correspond to the axis of the jet seen on larger scales. 
Assuming the disk is Keplerian (although they note that the rotation
curve falls more gradually than predicted by Kepler's law, as shown in 
Figure~\ref{Greenhill3}),  Greenhill \& Gwinn estimate the mass
enclosed within 0.65 pc to be $\sim 1.5 \times 10^7$~\Msun.

Although the NGC 1068 observations provide only partial evidence for
the presence of a massive black hole (since Keplerian motion needs to
be assumed), they do give us perhaps the clearest picture of the
center of an AGN to date.  The active nucleus produces a radio jet
that is roughly perpendicular to the molecular disk. The central
accretion disk is surround by a flattened, high temperature, high
density ``hot zone'' that is likely heated by X-ray photoionization
from the innermost accretion disk.  The hot zone appears to be
interior to the molecular disk.  A luminous thermal infrared source
composed of hot dust at the sublimation temperature is centered on the
active nucleus.  The hot dust is likely ablating from the inner radius
of the molecular disk due to intense heating by the central source.
The molecular disk and central black hole are at the center of a
compact, relatively young nuclear star cluster that provides at least
10\% of the nuclear bolometric luminosity.

\subsection{\H2O Masers in Other Galaxies}
\label{sec:othermas}

For the rest of the section, we will briefly touch upon additional
\H2O maser studies.

{\bf M51}: Hagiwara \etal (2001a) used Effelsberg 100 m telescope and VLA
observations of this Seyfert 2/LINER galaxy to find low-luminosity, high velocity (up to 120 km s$^{-1}$) \H2O maser
clouds, distributed asymmetrically with respect to the systemic
features. The maser emission is spatially unresolved, but based on 
slight offsets between the nuclear and maser positions, Hagiwara et
al. conclude that the masers in M51 are most likely associated with
the nuclear jet, although they cannot completely rule out the
possibility that the emission originates from a circumnuclear disk.

{\bf Mrk 1419}: Henkel \etal (2002) used the Effelsberg 100m telescope to
detect \H2O emission from the nucleus of this LINER galaxy.  Both
systemic and high velocity features are seen; for the former,  a
secular velocity drift of $2.8 \pm 0.5$ km s$^{-1}$ yr$^{-1}$ is
detected.  As for the case of M51, no spatially resolved observations
of the maser clouds  exist, but assuming that the emission originates
in an edge-on Keplerian disk, the data implies a binding mass of $\sim
10^7$ \msun~within 0.13 pc of the nucleus, corresponding to a mass
density of  $10^9$ \msun~pc$^{-3}$.

{\bf NGC 5793}: VLA observations (Hagiwara \etal 2001b) detected both
systemic  and high velocity ($\sim 250$ km s$^{-1}$ relative to
systemic) \H2O maser clouds in the nucleus of this Seyfert 2 galaxy.
The spatial distribution of the blueshifted clouds was resolved by the
VLBI in two clumps with a separation of 0.16 pc but, unfortunately,
both the systemic and redshifted features (which are variable) were
too weak to be detected at the time of the VLBI observations. Assuming
that the blueshifted features (which do not show any evidence of a
velocity drift) arise from the edge of a Keplerian disk, Hagiwara et
al. estimate a  binding mass  of the order of $\sim 10^7$ \msun~
within 0.13 pc of the nucleus, corresponding to a mass density of
$2\times10^6$ \msun pc$^{-3}$.  VLBI imaging of the redshifted
features will be necessary to exclude different scenarios, for
instance one in which the highly blueshifted velocity of the maser
features might trace outflowing gas surrounding the active nucleus.

{\bf IC 2560}: \H2O emission at the systemic velocity in this Seyfert 2
galaxy was detected by Braatz \etal (1996). Using single dish VLBI
observations, Ishihara \etal (2001) detected high velocity clouds
symmetrically distributed up to a 418 \kms~  on either side of the
systemic velocity.  Furthermore, they detected a secular velocity
drift of the systemic features. The high velocity clouds are not
spatially resolved; however, by assuming that the maser clouds
lie in a  Keplerian disk inclined 90\deg~to the line of sight,
Ishihara \etal measure a SBH mass of $2.8 \times 10^6$ \msun~within
0.068 pc,  giving a density in excess of 2.1 $\times 10^9$ \msun
pc$^{-3}$.

{\bf NGC 4945}: Greenhill, Moran \& Herrnstein (1997) reported on VLBA observations of
this starburst/Seyfert 2 galaxy. \H2O emission at $\sim 150$ km
s$^{-1}$  is found symmetrically distributed on both sides of the
systemic features.  The rotation curve, however, is not as cleanly
Keplerian as in the case of NGC 4258. Greenhill, Moran \& Herrnstein note that while
on the blue side the  the velocities roughly decline with angular
distance from the systemic emission, ``the position errors are
significant, the redshifted emission does not appear to mirror this
structure, the decline in velocity for the blueshifted emission is
faster than what is characteristic for a Keplerian rotation curve, and
the non-point-source emission that seems to exist at some systemic and
blueshifted velocities could not be mapped.'' If the maser clouds are
distributed in a  Keplerian disk, the implied binding mass would be
$\sim 10^6$ \msun~within a volume of radius $\sim 0.3$ pc.

{\bf NGC 3079}: The high velocity ($\sim 500$ km s$^{-1}$ relative to
systemic) water maser features in NGC 3079 were studied with the VLBA
by Trotter \etal (1998). The galaxy, which has a LINER nucleus,
features a large scale ($\sim 700$ pc), rotating molecular ring
(Young, Claussen, \& Scoville 1988; Sofue \& Irwin 1992). The \H2O
maser clouds are confined within 0.5 pc of the center, but their
distribution and  velocity structure is consistent with them belonging
to an ``inward'' extension of the molecular disk. In a simple
Keplerian approximation, the implied binding mass is $\sim 10^6$
\msun, although Trotter \etal note that ``the velocity dispersion
of clumps within the distribution and the suggestion of internal
velocity gradients within the clumps indicate that non rotational,
possibly turbulent motions are significant. The masers may arise in
shocks driven by a wide opening-angle nuclear wind.''

\begin{figure}[t]
\centerline{\psfig{figure=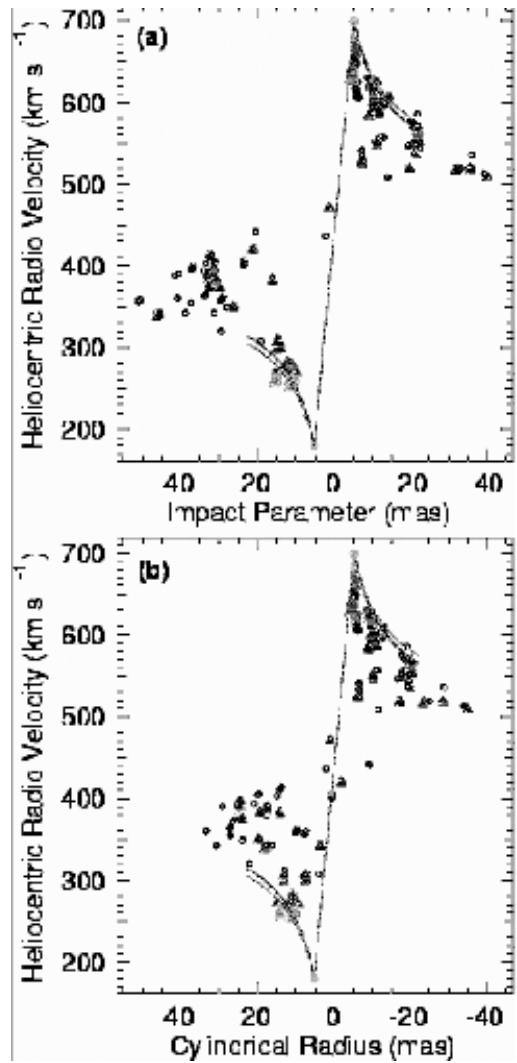,height=14.0 cm}}
\caption{Rotation curve for the \H2O maser emission in the Circinus
Galaxy. The upper panel shows velocity against impact parameter
measured from the estimated position of the central engine. The bottom
panel shows velocity against cylindrical radius measured from the
rotation axis of the innermost orbit of the proposed model disk. The
solid line represent a Keplerian rotation curve, while the dashed line
shows a rotation curve following $v \propto r^{-0.45}$  (From
Greenhill \etal 2003b)}
\label{Circinus}
\end{figure}

{\bf Circinus:} Greenhill \etal (2003b) used VLBA observations to resolve
the maser clouds into  two distinct morphological and kinematical
components. Some of the clouds trace a warped accretion disk extending
from 0.1 pc to 0.4 pc from the center of this galaxy, with a peak
rotational speed of 260 km s$^{-1}$. A second population of clouds
defines a wide-angle bipolar outflow up to 1 pc from the center.  The
clouds belonging to the disk follow a Keplerian rotation curve rather
closely (Figure~\ref{Circinus}), implying a central mass of $(1.7 \pm
0.3) \times 10^6$ \msun, corresponding to a  mass density of $(3.2 \pm
0.9) \times 10^8$ \msun pc$^{-3}$. Greenhill \etal  note that the
inclination of the disk could not be measured directly, and therefore
``formally, the mass and density estimates depend on the inclination
and are lower limits''.  Higher velocity (up to 460 km s$^{-1}$)
emission subsequently detected by Greenhill \etal (2003b), but for
which no VLBI observations exist, might help to further define the
velocity structure in this galaxy.

\clearpage

\section {GAS DYNAMICS OF NUCLEAR DUST/GAS DISK}
\label{sec:gasdyn}

In their seminal 1978  paper discussing stellar kinematics in the
nucleus of M87, Sargent \etal note that the spectrum of the galaxy
shows an  [OII]$\lambda$3727 emission doublet with a broadened,
unresolved component, plus a narrower, asymmetric feature. Although
they ``appreciate the dangers of associating the broad lines with gas
clouds swirling around a massive object, especially in view of the
possibility of ejection or infall'', Sargent \etal point out that the
width of the narrow line, 600 km s$^{-1}$, corresponds to the stellar
velocity dispersion measured at the same radius. Interpreting the
broadening of the unresolved component as due to Keplerian rotation,
they notice that the implied mass is of the same order as that derived
from the stellar dynamical modeling.

In the Galactic Center, gas kinematics hinted at the presence of a
central mass concentration well before the spectacularly conclusive
proper motion studies described in \S~\ref{sec:mw}. The sharp increase
in the gas velocity from $\sim 100$ \kms~ at 1.7 pc, to 700 \kms~ at
0.1 pc implies a virial mass of a few $10^6$ \msun~ within this radius
(Wollman \etal 1977; Lacy \etal 1980 and 1982; Crawford \etal 1985;
Serabyn \& Lacy 1985; Mezger \& Wink 1986). Although often interpreted
as evidence for a central black hole,  concerns over the  possibility
of gas outflows or inflows ultimately limited the credibility of such
claims (e.g. Genzel \& Townes 1987).

\begin{figure}[t]
\centerline{\psfig{figure=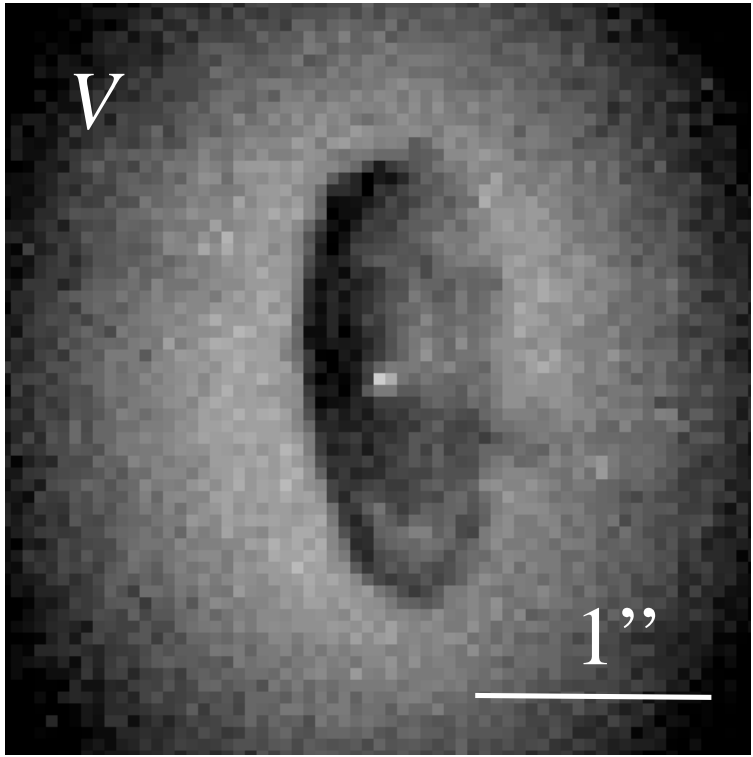,height=10.0 cm}}
\caption{An \hstnsp/WFPC2 image of the nuclear dust disk in NGC 4261,
taken with the F555W filter ($\sim$ Johnsons' $V$). The disk, which
surrounds the bright, non-thermal, unresolved nucleus, is 230 pc
across, and roughly perpendicular to the galaxy's radio
jets. H$\alpha$+[NII] narrow band images reveal an ionized gas
component associated with the inner parts of the disk. Adapted from
Ferrarese, Ford \& Jaffe (1996).}
\label{n4261disk}
\end{figure}

Indeed gas, unlike stars, can easily be accelerated by non
gravitational forces, and dynamical studies based on gas kinematics
have often been quickly -- and sometimes unjustly -- dismissed by the
establishment. For instance, as late as 1995, Kormendy and Richstone
remark that although the \H2O rotation curve of NGC 4258 ``looks
Keplerian'', ``as usual it is not certain that gas velocities measure
mass''. This concern is of course unfounded: NGC 4258, with the Milky Way,
has given us the only undeniable proof of the existence of a
SBH in a galactic nucleus, and the most reliable and elegant
determination of its mass (\S~\ref{sec:n4258}). This frame of mind is
no longer popular; the turning point, in our opinion,  can be traced
back to the early '90s \hst observations of NGC 4261 (Jaffe et
al. 1993) and M87 (Harms \etal 1994; Ford \etal 1994). NGC 4261 was
one of the first elliptical galaxies to be imaged with the Hubble
Space Telescope, as part of a magnitude limited sample of 12 early
type galaxies in the Virgo cluster. It was later realized that NGC
4261 should not have been included in the sample, since it in fact
belongs to a small group of galaxies about twice as far as
Virgo. Its inclusion turned out to be a lucky mistake: the image 
of the 270 pc, perfectly defined
nuclear dust disk surrounding the bright non-thermal nucleus of the
galaxy (Figure~\ref{n4261disk}), has since gone on to become one of
the most popular of all Hubble's images.

NGC 4261 is a well studied FR-I radio galaxy (e.g. Jones \etal 2000),
and the disk, which is roughly perpendicular to the radio jets,
immediately evokes textbook sketches of the inner molecular torus
which is at the heart of the AGN paradigm (Antonucci \& Miller
1985). The moment the \hst image of NGC 4261 was downloaded, it was
realized that the regularity of the dust disk must reflect the
signature of material in cold, circular rotation -- if verified by
kinematical data, the major objection to using gas to trace gravity
would be lifted. \hst spectra of NGC 4261 had not yet been taken, but
they were available for Virgo's cD galaxy, M87. Narrow band
H$\alpha$+[NII] images of M87  showed the presence of a disk-like
structure, although not as regular as the one seen in NGC 4261 (Ford
\etal 1994, Figure~\ref{M87_4}). Spectra taken with the 0\Sec26 Faint
Object Spectrograph (FOS) at five different locations (including one centered on
the nucleus) showed that the gas velocity reached  $\pm 500$ km
s$^{-1}$ 12 pc on either side of the nucleus (Harms \etal 1994).  If
interpreted as due to Keplerian motion, this implies a central mass of
$(2.4 \pm 0.7) \times 10^9$ \msun. The limited amount of kinematical
information available to Harms \etal did not allow for very
sophisticated modeling; not only was Keplerian motion assumed, but also the
inclination angle of the disk was fixed to the value estimated by Ford
\etal (1994) from the analysis of the narrow band images. The
correctness of these assumptions was verified a few years later by a
much more extensive kinematical survey of the disk, using the \hst Faint
Object Camera (FOC) (Macchetto \etal 1997). The 0\Sec062
wide FOC slit was positioned at the nuclear location and on either
side, which sampled the disk at over 30 (roughly)
independent positions within 0\Sec5 of the nucleus, at a spatial
resolution of 3 pc. This allowed for very detailed dynamical modeling
to be performed, leaving virtually no doubt as to the presence of a
$(3.2 \pm 0.9) \times 10^9$ \msun~SBH (Figure~\ref{M87-velocity}).

\begin{figure}[t]
\centerline{\parbox[l][7cm]{9.2cm}{\vskip 3cm THIS FIGURE IS INCLUDED IN THE FULL VERSION OF THE
MANUSCRIPT AVAILABLE AT http://www.physics.rutgers.edu/~lff/publications.html.}}
%\centerline{\psfig{figure=f20.eps,height=7.0 cm}}
\caption{A WFPC H$\alpha$ map of M87. The inset to the left shows details
of the nuclear dust disk. From Ford \etal (1994).}
\label{M87_4}
\end{figure}

\begin{figure}[t]
\centerline{\psfig{figure=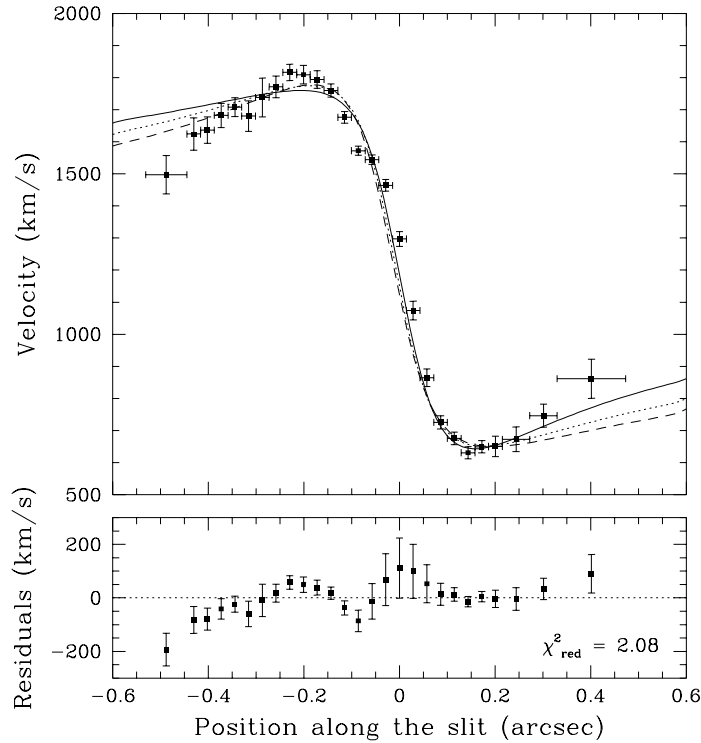,height=10.0 cm}}
\caption{The \hstnsp/FOC rotation curve of the ionized gas disk in M87 (from
Macchetto \etal 1997). The different lines correspond to different
model fits to the data; all require the presence of a central black
hole of a few $10^9$ \msun, but differ slightly in the other
parameters, most notably the position angle of the disk's line of
nodes. }
\label{M87-velocity}
\end{figure}

A kinematical study of NGC 4261 followed in 1996 (Ferrarese et al.),
claiming a $(4.9 \pm 1.0) \times 10^8$ \msun~SBH. In the meantime, it
started to become apparent that nuclear dust/gas disks are not
uncommon in early type galaxies. The most recent and statistically
complete study (Tran \etal 2001) finds NGC4261-type disks in 18\% of
early type galaxies, and almost 40\% of early type galaxies with IRAS
100 $\mu$m emission. Figure~\ref{dustdisks} shows some examples of
such disks. Tran \etal (2001) find that all galaxies hosting dust
disks show signs of nuclear activity, and in almost all cases the
disks are aligned with the major axes of the galaxies in which they
reside. The origin of the dust (the two possibilities being internal
from stellar mass loss, or external from merging or interactions with
other galaxies) remains undetermined (Ferrarese \& Ford 1999; Tran et
al. 2001).

By 1996, it had become apparent that the kinematics of dust disks was
a powerful tool to constrain the central potential in early type
galaxies, and many such studies began to be published. At present,
SBHs have been claimed in 11 galaxies thanks to dynamical studies of
gas/dust disks; for all, the SBH sphere of influence was resolved (see
Table\ref{tab:allmasses} for a complete list). These studies proved
even more valuable because they could target the very objects for
which stellar dynamical studies fail, i.e. bright, pressure supported
ellipticals like M87. In fact, stellar and gas dynamical studies are
complementary. Gas disks are not common in the elongated,  
fainter ellipticals with higher central surface brightness that are the preferred targets of stellar
dynamical studies.  Furthermore, the very presence of the disks
makes stellar dynamical studies observationally challenging, because
of the large amount of dust obscuration. The downside is that there
are few objects for which a dynamical measurement of $\mh$ based on
stellar and gas dynamics can be compared. Only one, IC1459 (Cappellari
\etal 2002) has been studied so far; we will return to IC1459
later in this section.  Studies to explore the possibility of using
the kinematics of cold disks to constrain the central potential in
spiral galaxies are also underway (Marconi \etal 2003; Sarzi et
al. 2001)

\begin{figure}
\centerline{\parbox[l][5.8cm]{9.2cm}{\vskip 3cm THIS FIGURE IS INCLUDED IN THE FULL VERSION OF THE
MANUSCRIPT AVAILABLE AT http://www.physics.rutgers.edu/~lff/publications.html.}}
%\begin{minipage}[t]{5.8cm}
%\psfig{figure=f22a.eps,height=5.8 cm}
%\end{minipage}
%\begin{minipage}[t]{5.8cm}
%\psfig{figure=f22b.eps,height=5.8 cm}
%\end{minipage}
\caption{Examples of  nuclear dust disks in early-type galaxies.  The
$R-$band images are taken with the Wide Field and Planetary Camera 2
on board \hstnsp. All galaxies are within 40 Mpc. From Tran \etal (2001).}\label{dustdisks}
\end{figure}

Constraining the central mass using gas dynamical data proceeds
through the following steps:

\begin{enumerate}

\item 
The stellar surface brightness profile is deprojected to give a
   stellar luminosity density (an inclination angle for the galaxy is
   assumed).  This is converted to a mass density by adoption of a
   (generally constant) mass to light ratio, which enters the models
   as a free parameter. Given the mass density, the stellar potential
   is calculated using Poisson's equation. The stellar contribution to
   the circular velocity at any given radius is then simply determined
   by the stellar mass enclosed within that radius.

\item
The contribution to the circular velocity due to the dust/gas disk
   itself is computed in a similar way. The mass of the disk can be
   measured directly if images are available in multiple passbands, so
   that the amount of extinction can be determined. This requires a
   knowledge of the visual mass absorption coefficient, reddening law,
   and gas-to-dust mass ratio, all of which are reasonably well
   constrained (see Ferrarese \& Ford 1999). The mass of the disk
   itself generally turns out to be too small (of the order $10^5$ -
   $10^6$ \msun) to be of any dynamical relevance (e.g. Ferrarese \&
   Ford 1999; Barth \etal 2001)

\item
An additional contribution to the rotational velocity comes from a
   central point mass $\mh$, which is obviously kept as a free
   parameter in the models.

\item
The rotational velocity (which is calculated in the plane of the disk)
   is projected along the line of sight. This requires knowledge of
   the inclination angle of the disk. Although this should be a free
   parameter, it can not always be constrained from the data
   (e.g. Harms \etal 1994; Sarzi \etal 2001). In these cases, the
   inclination angle is fixed to the value measured from the images,
   under the assumption that the disk is flat all the way into the
   center.

\item
The comparison between the line of sight velocities predicted by the
model and the observed velocities, depends on four additional
parameters: the $x$ and $y$ projected positions of the center of the
slit relative to the kinematical center of the disk, the angle between
the line of node of the disk and the slit's major axis, and the
systemic velocity of the disk. All are treated (if possible) as free
parameters.

\item
Finally, the model must be artificially ``degraded'' to simulate the
   observing conditions: in particular, the blurring introduced by the
   finite pixel size, as well as slit width must be taken into
   account. This step, which is very nicely described in Barth et
   al. (2001), depends on the emission line surface brightness within
   the disk, as well as on the intrinsic and instrumental velocity
   dispersion of the line. The reason for this is that the
   contribution to the velocity observed at any given position is
   given by the integrated contribution at each neighboring positions,
   entering  with a weight which depends on the broadening and
   strength of the line at each point.

\item
The free parameters in the model, namely the inclination angle of the
disk, the position of the center of the slit relative to the
kinematical center of the disk, the angle between the line of node of
the disk and the slit's major axis, the systemic velocity of the disk,
the mass-to-light ratio of the stellar population, and the mass of the
central SBH, are varied until the best fit to the data are obtained.

\end{enumerate}

For gas dynamical studies to be robust, a two dimensional velocity
field is essential (to be fair, the same is true of stellar dynamical
studies, for which such data are never available -- de Zeeuw 2003;
Verolme \etal 2002). In the case of gas kinematics, the 2-D velocity
field is necessary to verify the critical underlying assumption that
the gas motion is indeed gravitational and the gas is in
equilibrium. It must be pointed out that as long as the motion is
gravitational, the gas does not necessarily have to be in circular
rotation in a geometrically thin disk for the data to be useful.  In
fact, since the very first studies, it has been noted that the
emission lines are broadened beyond what is expected from simple
thermal broadening and instrumental effects (e.g. Ferrarese et
al. 1996; Ferrarese \& Ford 1999; van der Marel \& van den Bosch 1998;
Verdoes Kleijn \etal 2000; Barth \etal 2001). The origin of the
broadening is not well understood and has been interpreted either as
microturbolence within the disk (Ferrarese \& Ford 1999; van der Marel
\& van den Bosch 1998; Marconi \etal 2003) or as evidence of
non-circular motion, as could be expected if the gas is fragmented
into collisionless clouds that move ballistically, providing
hydrostatic support against gravity (Verdoes Kleijn \etal 2000; Barth
\etal 2001; Cappellari \etal 2002). In their analysis of  NGC 3245,
Barth \etal conclude that the mass of the central SBH should be
increased by 12\% to account for non-circular motions.  In either
case, turbulence and asymmetric drifts can and have been successfully
incorporated in the modeling (e.g. Verdoes Kleijn \etal 2000; Barth
\etal 2001), although doing so requires high S/N data with a wide
spatial coverage.

Cases in which the gas is affected by non-gravitational motions or is
not in equilibrium can also be recognized from high resolution
kinematic maps of the ionized gas. In NGC 4041, a quiescent Sbc
spiral, Marconi \etal (2003) remark that the systematic blueshift of
the disk relative to systemic velocity might be evidence that the disk
is kinematically decoupled. They conclude that only an upper limit, of
$2\times 10^7$ \msun, can be put on the central mass.  Cappellari et
al. (2002) conclude that non-gravitational motions might indeed be
present in the case of IC 1459, for which the ionized gas shows no
indication of rotation in the inner 1\sec. IC 1459 is the only galaxy
for which a SBH mass estimate exists based both on gas and stellar
kinematics (Cappellari \etal 2002). Three-integral models applied to
the stellar kinematics produce $\mh = (2.6 \pm 1.1) \times 10^9$
\msun, while the gas kinematics produces estimates between a few
$\times 10^8$ and $10^9$ \msun, depending on the assumptions made
regarding the nature of the gas velocity dispersion. Unfortunately,
the authors express strong reservations as to the reliability of
either mass estimate. As mentioned above, there is evidence that the
gas motion might not be completely gravitational, which would
invalidate the method entirely. On the other hand, they also note that
the \hst stellar spectra ``do not show any obvious evidence for the
presence of a BH'', and conclude that the ``BH mass determination [in
IC 1459] via the stellar kinematics should be treated with caution''.
 
Other systematics in gas dynamical study are of a more technical
nature.  Maciejewski \& Binney (2001) point out that  neglecting to
account for the finite size of a wide slit can introduce systematic
errors in the measured masses. The reason for this is that if the slit
is wider than the instrumental resolution, the position and velocity
information become entangled. As a consequence, intrinsically gaussian
emission lines profiles could become asymmetric or even display
multiple peaks. The degree to which this takes place is very sensitive
not only to the slit width (relative to the instrumental PSF), but
also to the gradient in the velocity field: spectra for galaxies with
large velocity gradients, and observed with wide slits, are affected
the most. From a practical point of view, if the features artificially
introduced by the slit in the line profiles are mistaken for distinct
kinematical components, the measurement of the rotational velocity,
and therefore the estimate of $\mh$, could be affected. Maciejewski \&
Binney indeed argue that this could be the case for Bower et al's 1998
\hstnsp/STIS observations of M84, observed with a 0\Sec2 slit, and
estimate that Bower \etal mass could have been overestimated by as
much as a factor 4 because of this effect. The earlier observations 
which used FOS aperture spectroscopy matched to the FOS resolution (Harms
\etal 1994; Ferrarese \etal 1996; Ferrarese \etal 1999) as well as
the FOC observations of M87 (Macchetto \etal 1997), which used a very
narrow slit, are not affected by this bias. The most recent
investigations (Cappellari \etal 2002; Verdoes Kleijn \etal 2000;
Barth \etal 2001) take the finite slit width into account.

\begin{figure}[t]
\centerline{\psfig{figure=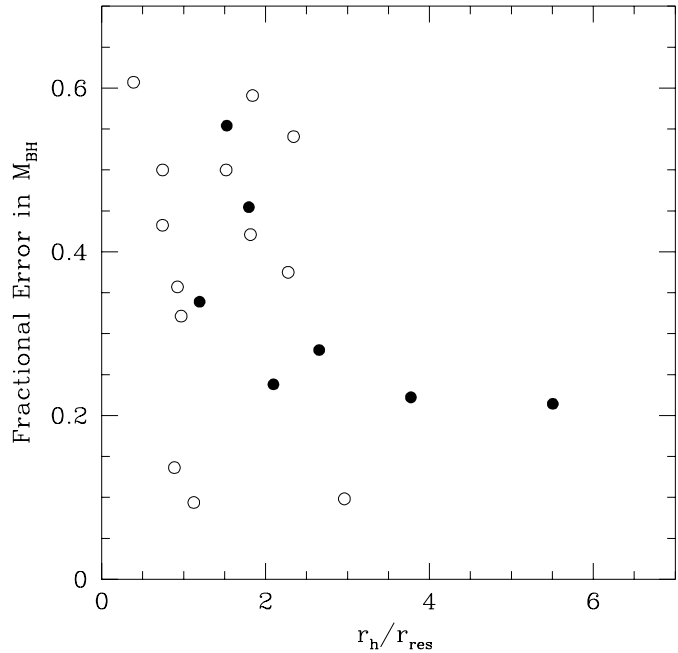,height=10.0 cm}}
\caption{The percent error quoted on all SBH detections made
using resolved stellar (open circles) or gas (filled circles) dynamical studies 
(Table~\ref{tab:allmasses}), plotted
against the degree by which the SBH sphere of influence is resolved
by the data.}
\label{reliable}
\end{figure}

In view of the above discussion, gas kinematics is a transparent and
therefore potentially powerful tracer of the central potential -- in
our view more so than stellar kinematics.  The criticism
that gas can be subject to non gravitational motions is a strength
rather than a weakness. Non gravitational motions can be  recognized
from the data, and detecting them leaves no doubt that the data are
unsuitable for dynamical studies.  The modeling is more transparent
than in the case of stellar dynamics.  Either the Keplerian motion is
detected or it isn't -- in the first case a SBH mass can be measured,
in the second it can't. Proof of this is that there have been no SBH
detections based on gas kinematical data which did not resolve the
sphere of influence. This is in sharp contrast with stellar dynamical
cases, where one third of the claimed detections are based on data for which the
sphere of influence is unresolved. Moreover, for many of the stellar
dynamical SBH detections the data shows no evidence of Keplerian motion,
even when the sphere of influence is resolved
(e.g. Gebhardt \etal 2003). Figure~\ref{reliable} shows
the percent error quoted on all SBH detections listed in
Table~\ref{tab:allmasses}, plotted against the degree with which the
SBH sphere of influence is resolved by the data. The errors and masses
are as quoted in the original papers; filled and open circles are
detections from gas and stellar kinematics respectively.  While, as
expected, detections based on gas kinematics became increasingly less
secure as the resolution (relative to the SBH sphere of influence)
worsens, this does not appear to be the case for detections claimed
based on stellar dynamical studies. In fact, rather surprisingly, the
most {\it secure} detections (as secure, at face value, as the
detections in the Milky Way and NGC 4258) based on stellar kinematics
appear to be in galaxies for which the SBH sphere of influence was
either not or only marginally resolved (e.g. NGC 3384, NGC 1023).

\section {BLACK HOLE MASSES IN LUMINOUS AGNs.}
\label{sec:agn}

Even for the heftiest SBHs, state of the art telescopes and
spectrographs lack the ability to spatially resolve the sphere of
influence in galaxies farther than $cz \sim 10000$ km
s$^{-1}$. Beyond, dynamical estimates of the central mass, using the
methods described in the preceding sections, become infeasible. A way
to measure black hole masses in more distant, quiescent galaxies or
low luminosity AGNs has yet to be devised: if M87 were at the distance
of one of the closest quasars, 3C273, its BH mass could not be
measured. For some AGNs and quasars, however, alternative techniques
are available.

Two of these techniques, fitting accretion disk models to
multiwavelength continua spectra (\S~\ref{sec:bbb}; Shields 1978;
Malkan 1983), and the study of the 6.4 KeV emission line due to iron
fluorescence (\S~\ref{sec:kalpha}; see Fabian \etal 2000; Reynolds \&
Nowak 2003 for recent reviews) rely on the existence of geometrically
thin, optically thick accretion disks. The study of the 6.4 keV
emission in particular gives us the most direct and powerful view of
the region within a few gravitational radii of the central engine. Its
use in deriving SBH masses is currently hampered by the limited energy
resolution and collective area of available X-ray satellites, but
might be within reach of future missions (e.g. XEUS and Constellation
X). Spectral fitting to the UV/optical continuum is a powerful
diagnostic of the physics of accretion in particular classes of
AGNs. However, because of their complexity, the models cannot yet
reliably measure SBH masses in individual objects.

The other two techniques we will discuss rely on spectral observations
of the broad optical emission lines which characterize the spectra of
Seyfert 1 galaxies and quasars. One of these, reverberation mapping,
has great potential. The method, which can be applied to
Seyfert 1 galaxies and quasars with a variable continuum source, is
extremely demanding  from an observational standpoint, but the
theoretical formalism is well described, and a better understanding
of the systematics is well underway. Photoionization models applied to
the BLR, on the other hand, require only simple observations, but rely
on a heavy set of assumptions. Their usefulness is mainly statistical.

Finally, we will discuss some ``secondary mass estimators''.  These
are based on empirical relations between SBH masses (or quantities
related to the masses) and other, easily measurable, AGN
properties. Such relations can be calibrated using galaxies for which
the SBH mass is measured independently, for instance through
reverberation mapping. Although they can lead to mass estimates in
error by a factor of several in individual objects, these relations
have the obvious practical advantage that they can be easily applied to large
samples of objects. Hopes for studying the redshift evolution of SBH
scaling relations (\S~\ref{sec:scale}) lie largely in
these relations.

\subsection{Spectral fitting to the Big Blue Bump}
\label{sec:bbb}

Twenty years have passed since the realization that the blue and
ultraviolet spectra of most quasars and Seyfert 1 galaxies show a
marked excess flux over the extrapolated infrared power law spectrum
(Shields 1978; Richstone \& Schmidt 1980; Malkan \& Sargent 1982). In
two ground breaking papers, Shields (1978) and Malkan (1983)
attributed the origin to what became affectionately known as the ``big
blue bump'' to thermal emission from a geometrically thin, optically
thick accretion disk surrounding the central SBH.

\begin{figure}[t]
\begin{minipage}[t]{5.5cm}
\psfig{figure=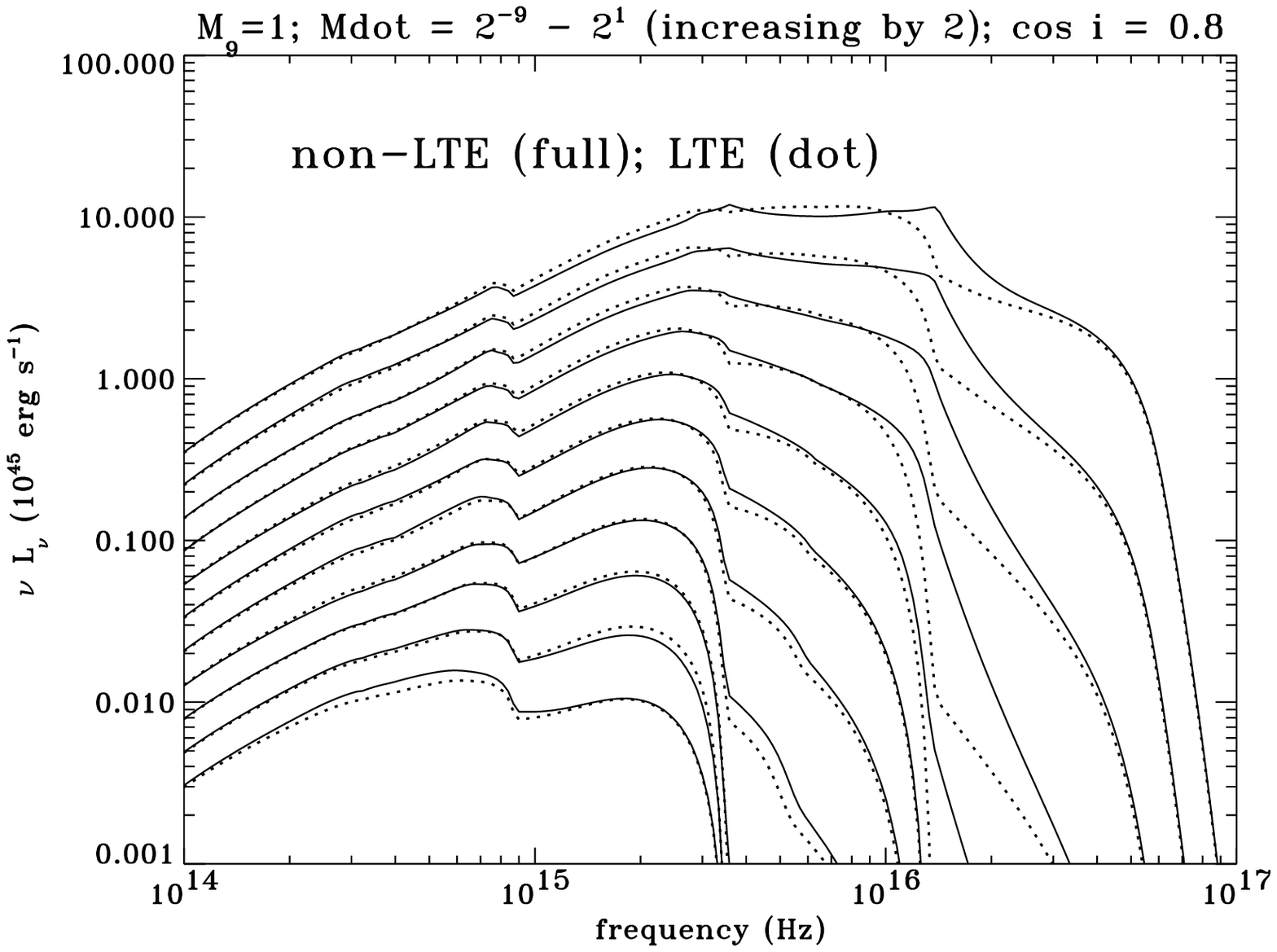,height=4.2 cm}
\end{minipage}
\begin{minipage}[t]{5.5cm}
\psfig{figure=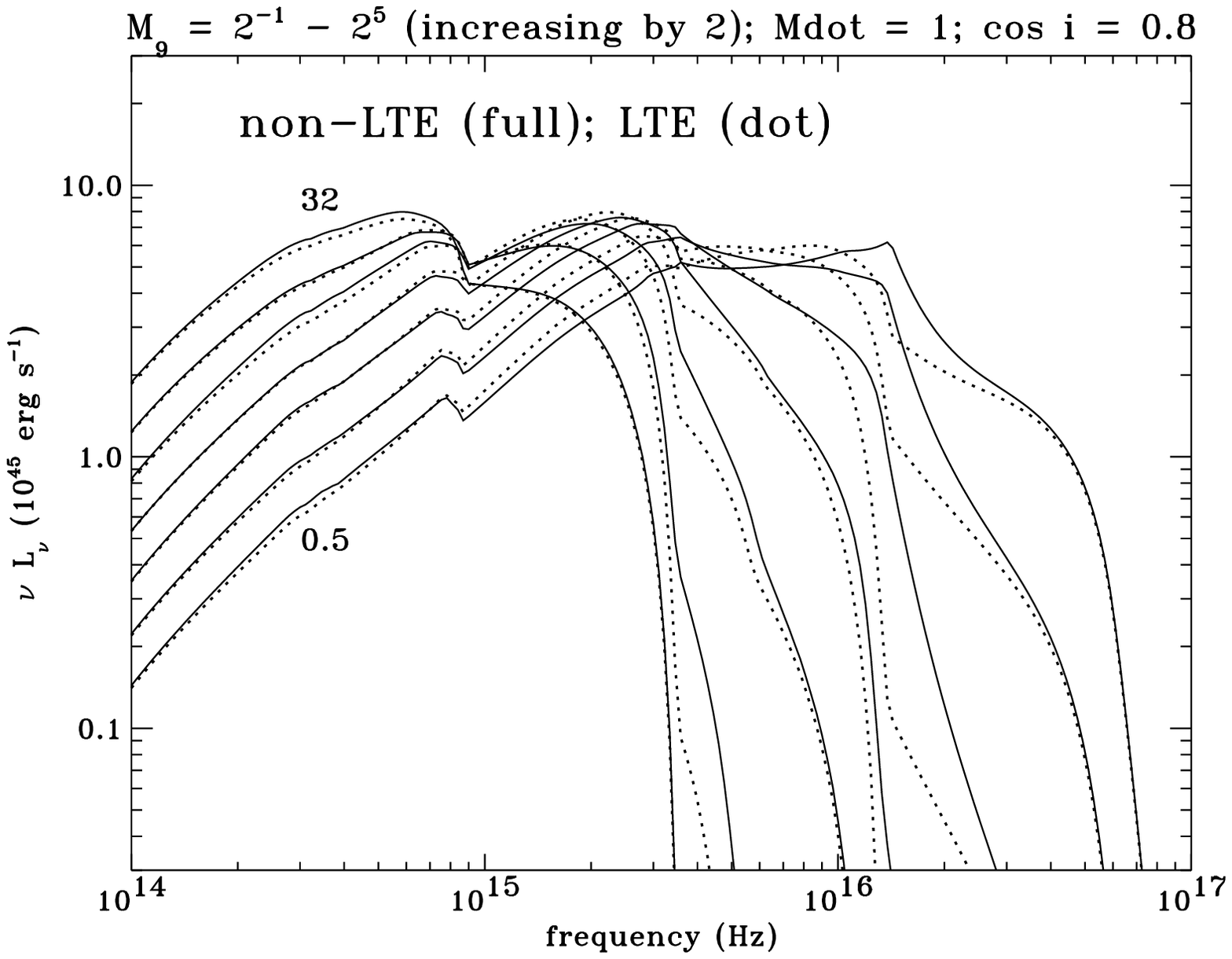,height=4.2 cm}
\end{minipage}
\caption{The face-on standard disk spectrum for: (left)  a fixed central mass 
($10^9$ \msun) and
varying accretion rate (i.e. total luminosity, from $2^{-9}$ to 2
\msun yr$^{-1}$); and (right) a fixed mass accretion
rate (1 \msun yr$^{-1}$) and varying central mass (from $0.5 \times
10^9$ to $32 \times 10^9$ \msun). In both cases, the disk
inclination is approximatively 37\deg (close to face on).  Full
lines represent non-LTE models, while dotted lines represent LTE models.
The models are from Hubeny et al. (2000).}
\label{bbbmodel}
\end{figure}

A comprehensive review of spectral models for accretion disks,
detailing the assumptions as well as shortcomings of the models, and
comparison with the observations, can be found in Krolik (1998).  In
its basic form, the model assumes local heat balance (i.e. the heat
generated by viscosity within the disk is radiated locally through the
disk surface), a Keplerian velocity field, and negligible pressure
gradient and velocity (i.e. small accretion rates) in the radial
direction. To zeroth-order, the spectrum can then be calculated by
dividing the disk in concentric annuli, each radiating as a blackbody
with effective temperature steadily decreasing outwards.  While
several problems can be identified, most notably the fact that the
accretion rates derived often exceed the limit above which the disk
ceases to be geometrically thin (Szuszkiewicz, Malkan \& Abramowicz
1996), this simple model can reproduce the blue/UV part of the
spectrum reasonably well, given only three parameters: the mass of the
central SBH, the mass accretion rate, and the inclination angle of the
disk. More complex and realistic models have been explored since then,
and include relativistic effects in inclined disks surrounding both
Kerr and Schwarzschild SBHs (Sun \& Malkan 1989), modification to the
black body spectrum due to electron scattering, comptonization and
irradiation of the surface of the disk by an external X-ray source
(Ross, Fabian \& Mineshige 1992; Siemiginowska \etal 1995), and the
presence of an inner shock front inside which the flow becomes
supersonic, and the disk gives way to a hot, ionized, pressure supported
torus. Geometrically thick disks able to support accretion rates
comparable to the Eddington rate have also been studied (Szuszkiewicz,
Malkan \& Abramowicz 1996; Sincell \& Krolik 1998; Wang \etal 1999;
Hubeny et al. 2000). Figure~\ref{bbbmodel} (from
Hubeny et al. 2000) represents recent simulations showing the
effects of varying SBH mass and accretion rate 
on the emerging integrated spectral energy distribution, for a disk
seen nearly face-on.

Spectral fitting to the optical/UV/soft X-ray continuum provides  a
powerful tool to study the physical mechanisms that regulate AGN
activity. Its value in measuring SBH masses is however limited. The
potential systematic errors in the derived masses are large, due to
the complex, and largely unknown, physical processes at
play. Furthermore, the models rarely do a good job at fitting the
available data, and often the SBH mass and other parameters enter the
models in a degenerate way. For instance, Szuszkiewicz, Malkan \& Abramowicz (1996) and Sincell
\& Krolik (1998) shows that the shape of the spectrum is
determined mainly by the ratio of the SBH mass to the accretion rate,
rather than by the two quantities independently.

\subsection{The Iron K$\alpha$ Emission Line}
\label{sec:kalpha}

X-ray production in Galactic black hole candidates as well as Seyfert
nuclei is generally credited to inverse Compton scattering of soft
optical/UV photons off a population of hot electrons (Thorne \& Price
1975; Sunyaev \& Truemper 1979; Zdziarski \etal 1994). While the
details of the model are not entirely agreed upon (see for instance
the excellent review in Reynolds \& Nowak 2003) the general picture is
that of a geometrically thin, optically thick accretion disk
generating UV photons, and surrounded by a Comptonizing population of
thermal electrons in the form of a hot corona, produced by evaporation
at the surface of the disk. The hard X-rays produced in the hot corona
irradiate and are in turn reprocessed by the cold surface of the
accretion disk, producing Compton reflection and fluorescence in the
low-ionization Compton thick material (e.g. Reynolds 1996). Indeed,
the detection of a hard reflection `hump' above 10 keV, and of the
iron K$\alpha$ fluorescence line emission at 6.4 keV by the Ginga and
ASCA X-ray satellites respectively, provide the most direct evidence
for the existence of thin accretion disks in these objects (Piro,
Yamauchi \& Matsuoka 1990; George \& Fabian 1991; Matt, Perola \& Piro
1991; Matt \etal 1992; Fabian \etal 1995).  Moreover, the study of
fluorescent iron emission strongly suggests that the central regions
in Seyfert 1 galaxies are under the influence of the extremely strong
gravitational field expected in the presence of a supermassive black
hole.

\begin{figure}[t]
\centerline{\psfig{figure=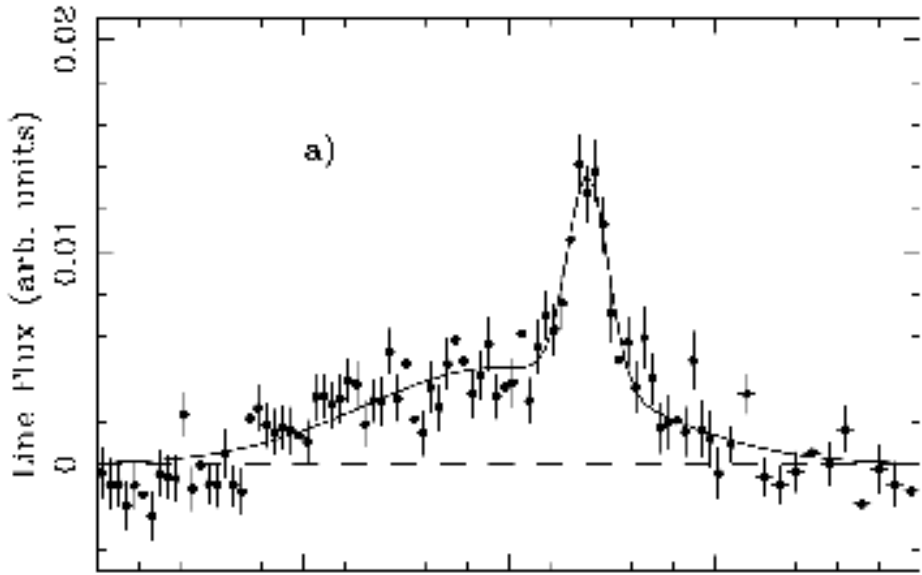,height=8.0 cm}}
\caption{Composite line profile produced by combining the data for 14
Seyfert 1 galaxies observed with ASCA (from Nandra \etal 1997).}
\label{kalpha}
\end{figure}

Broad K$\alpha$ line emission seems to be an almost ubiquitous
property of the X-ray spectra of Seyfert 1 galaxies, being detected in
$\sim 80\%$ of cases (Nandra \etal 1997; Reynolds 1997;  but see also
Done \etal 2000; Gondoin \etal 2001a\&b).  The first detection of
K$\alpha$ iron emission  was made by Exosat in the Seyfert 1 galaxy
MCG-6-30-15 (Nandra \etal 1989), but it was only thanks to the higher
energy resolution of ASCA that the line width and profile could be
studied in detail (Tanaka \etal 1995; Nandra \etal 1997). The line
mean energy (6.34 $\pm$ 0.04 Kev, Nandra \etal 1997) implies that the
line originates from fluorescence in the cold, low ionization
($<$FeXVI), thin (0.1\% to 1\% of the total disk's
thickness) outer layer of the accretion disk (Fabian \etal 2000). The line is expected
to be intrinsically narrow, but ASCA revealed it to be extremely broad
and significantly skewed. Figure~\ref{kalpha} is a composite line
profile produced by combining the data for 14 Seyfert 1 galaxies
observed with ASCA (from Nandra \etal 1997). The full width at zero
intensity exceeds $\sim 2$ keV, or $0.3c$. The narrow ($\sim 0.1$ eV)
blue peak of the line is centered around the rest wavelength of 6.4
keV, but the line itself extends dramatically to the red.  The line
profile can be reproduced remarkably well if it is assumed to
originate from a rapidly spinning accretion disk within a few
gravitational radii of the central SBH. The disk rotation splits the
line in two (in analogy with  HI line emission in the outer disks of
spiral galaxies).  Relativistic beaming enhances the blue
over the red component, and gravitational redshift strongly smears
the emission from the innermost regions of the disk into an extended red wing. While the blue peak moves
progressively to higher frequencies the higher the inclination (more
edge on) of the disk, the redward extent of the line is more strongly
influenced by relativistic effects, and is a sensitive function of the
innermost radius of the line emitting region.  Alternative
explanations to describe the line profile were suggested by Fabian
\etal (1995) but quickly rejected; these include Comptonization
producing a net redshift of the line, jet and outflow geometries
producing a marked asymmetric profile, or photoelectric absorption
cutting into the blue side of the line.

In the surviving disk interpretation, the iron K$\alpha$ line becomes
a powerful testbed for the immediate environment around the central
SBH, as well as a unique tool for testing our knowledge of the
properties of space and time in very strong gravitational fields.  The
line profile can not only be used to determine the inclination of the
accretion disk but also the {\it spin} of the central black hole
(albeit with some ambiguity: see discussion in Reynolds \& Nowak
2003); if the line emitting gas extends all the way to the innermost
stable orbit $r_{st}$, a much broader and more extended red wing is
expected for a maximally rotating SBH than for a static one, as a
consequence of the shrinkage of $r_{st}$ in a Kerr metric
(\S~\ref{sec:forma}; Fabian \etal 1989; Laor 1991; Martocchia \& Matt
1996; Reynolds \& Nowak 2003). Indeed, ASCA and XMM observations of
MCG-6-30-15 are best explained under the assumption of an almost
maximally spinning SBH (Iwasawa 1996; Dabrowski \etal 1997; Wilms et
al. 2001). We mention in passing that in the cases studied in detail,
the line profile points to an average inclination angle for the disk
of 30$\deg$ (Nandra \etal 1997), consistent with the postulates of
the unification models of AGNs (Antonucci 1993). Most of the emission
is produced at 20$r_g$, with an inner radius for the emitting region
extending to a few $r_g$.

The next step in the study of the Iron K$\alpha$ line will require the
increase in spectral resolution and collective area afforded by the
next generation of X-ray satellites. If the Fe K$\alpha$ line
responds to flares in the X-ray continuum produced in the corona above
the accretion disk, then the disk structure can be studied using
`reverberation mapping' techniques (Young \& Reynolds 2000; Fabian et
al. 2000), which can potentially lead to a robust measurement of the
SBH mass (see \S~\ref{sec:revmap}). The timescale expected for the
line to react is of order of the light crossing time at one
gravitational radius, $t \sim GM/c^3 \sim 500 \mh/(10^8 {\rm
M_{\odot}})$ s, which is unfortunately shorter than the integration
time required by current satellites to collect enough X-ray photons
for a meaningful measurement. It should be mentioned, however, that
preliminary results, based on limited variability studies conducted
with ASCA in NGC 3516 and NGC 4151, point to a picture far more
complex than expected (or hoped). The variability pattern of the line
seems to change with time and, even more damming, at least in one case
line and continuum variations seem to be uncorrelated (Nandra et
al. 1997; Nandra \etal 1999; Wang \etal 2001; Takahashi \etal 2002).

\subsection{Reverberation Mapping}
\label{sec:revmap}

According to the unification scheme of AGNs (Antonucci 1993; Urry \&
Padovani 1995; Figure~\ref{agn}), the central black hole and accretion
disk are completely surrounded by an obscuring, geometrically thick
molecular torus. The existence of the torus is strongly supported by
polarization maps (e.g. Watanabe et al. 2003), infrared observations 
(e.g. Reunanen et al. 2003), as well as by the \H2O maser observations discussed in
\S~\ref{sec:maser}. The geometry and extent of the torus are dictated
by the requirement that, to account for the well known polarization
properties of Seyfert 1 and Seyfert 2 galaxies, the Broad Line Region
(BLR) must be entirely enclosed within it, while the Narrow Line
Region (NLR) -- which extends from a few to several thousand parsecs
(Pogge 1989) -- must be largely outside. According to the standard
model, the BLR consists of many ($10^{7-8}$, Arav \etal 1997, 1998;
Dietrich \etal 1999), small, dense ($N_e \sim 10^{9-11}$ cm$^{-3}$),
cold ($T_e \sim 2\times10^4$ K) photoionized clouds (Ferland et
al. 1992; but see also, e.g., Smith \& Raine 1985, 1988; Pelletier \&
Pudritz 1992; Murray \etal 1995; Murray \& Chiang 1997;
Collin-Souffrin \etal 1988).  Even if direct observational
constraints on the BLR geometry are lacking, the emerging picture,
which we will discuss in more detail below, is that of a BLR which
extends from a few light days to several tens of light weeks from the
central engine. Although a factor 10 to 100 times larger than the
inner accretion disk, and three orders of magnitudes larger than the
Schwarzschild radius of the SBH which powers the AGN activity, the BLR
is, and will remain, spatially unresolved even using space based
instrumentation. In the early '80s it was realized that, while
preventing direct imaging, the closeness of the BLR to the ionizing
continuum source might open a new, unconventional way to map its
properties. Short-term continuum variability is a staple of AGNs (and
indeed is undisputed evidence of the compactness of the central
engine); if the delayed response of the line emitting region to
changes in the continuum source is regulated by light-travel time from
the latter to the former, then the time lag between variations in the
ionizing continuum and in the flux of each velocity component of the
line emission directly probes the spatial structure and emissivity
properties of the BLR. The technique is widely known as
``reverberation'' or ``echo'' mapping (Blandford \& McKee 1982;
Peterson 1993; Peterson 2002). It requires long-term, careful
monitoring of both continuum and broad emission lines, and it can of
course only be applied to those AGNs in which the BLR is directly
observable, namely Type 1 Seyferts and quasars.

\begin{figure}[t]
\centerline{\parbox[l][10cm]{9.2cm}{\vskip 3cm THIS FIGURE IS INCLUDED IN THE FULL VERSION OF THE
MANUSCRIPT AVAILABLE AT http://www.physics.rutgers.edu/~lff/publications.html.}}
%\centerline{\psfig{figure=f26.eps,height=10.0 cm}}
\caption{A schematic view of the central region of an AGN.}
\label{agn}
\end{figure}

In practice, reverberation mapping has so far failed to deliver a
picture of the central regions of AGNs, not because of fundamental
flaws in the theoretical reasoning, but rather because of inadequate
observational data, a fact that could, and hopefully will, be
rectified by future, dedicated spacecraft missions (Peterson 2003).
Past and current experiments have however led to accurate measurements
of the time delay between velocity-integrated line emission
variations, and continuum flux, which can (but see Krolik 2001) be
taken as a measure of the average BLR size. If assumptions are made
regarding the geometry of the BLR, and it is further argued that the
motion of the clouds is gravitational, then the binding mass can be
rather trivially derived from the virial theorem: $M = fr\sigma^2 /
G$. Here, the BLR radius $r$ is known from the reverberation
measurements, the mean velocity $\sigma$ is given by the emission line
width, and the factor $f$  depends on the geometry and  kinematics of
the BLR.

As a mass estimator, reverberation mapping has several important
advantages. First, the region probed is only a factor 1000 beyond the
Schwarzschild radius of the central SBH. This might seem large, but it
is still at least a thousand times  closer to the central engine than
the innermost radius reached by `traditional' methods based on
resolved kinematics. The implied mass densities
(Table~\ref{tab:methods}) are in excess of $10^{10}$ \msun~pc$^{-3}$,
leaving little doubt that if  the measurements can indeed establish a
mass reliabily, this must belong to a supermassive black hole, since
no cluster configuration could be stable under these conditions
(e.g. Rees 1984). Second, bright Type 1 AGNs, the ideal targets of
reverberation mapping, cannot be easily probed using standard
techniques, precisily because the bright AGN continuum overwhelms the
dynamical signature of the gas and dust needed for resolved dynamical
studies. Although it can be argued that Type 1 AGNs comprise only 1\%
of the total galaxy population, exploring the SBH mass function in as
many and as diverse as possible galactic environments is essential to
understand the evolutionary histories of SBHs. Third, unlike SBH
masses derived from resolved kinematics, those inferred from
reverberation mapping do not depend on the cosmology adopted. 
Because the BLR size is measured directly in physical units from the time
delay, the distance to the host galaxy does not enter the
analysis. This means that the redshift to which reverberation mapping
measurements can be pushed is limited only by the requirement that the
AGN must be bright enough to be easily detected.  Forth, reverberation
mapping is intrinsically unbiased with respect to the mass of the SBH,
since spatially resolving the sphere of influence (whose size depends
on the mass) is irrelevant. Very small SBHs can be probed as long as
the monitoring can be carried out on short enough timescales, and
viceversa for very massive SBHs.

Time lags, in some cases in multiple emission lines, have now been
measured for 37 AGNs (\S~\ref{sec:agnmon}); the inferred virial
masses, listed with the appropriate references in
Table~\ref{tab:revmap},  have formal random uncertainties which rival
those of resolved gas and stellar dynamical studies. However,
systematic uncertainties can in principle be a factor of several or
more (Krolik 2001).   To date, the most compelling evidence supporting
the reliability of reverberation mapping is observational; AGN
reverberation masses obey the same scaling relations observed in
quiescent galaxies, for which SBH masses are measured using ``proven''
traditional techniques (\S~\ref{sec:revmapobs}).

\subsubsection{Results of AGN Monitoring Programs}
\label{sec:agnmon}

Reverberation mapping was not originally devised with the intent of
measuring SBH masses -- but rather as a way to probe the kinematical
and morphological structure of the BLR. However, for the reasons
outlined in the previous section, it holds high promise of becoming
the method of choice for mass determinations, replacing resolved
kinematical studies at least in galaxies beyond a few hundred
megaparsecs. It is therefore instructive to review the observational
requirements of the method, the  monitoring programs that have led to
the masses listed in Table~\ref{tab:revmap}, and what we have learned
from them about the structure of the BLR.

Although variability in quasars has been known since the early '60s,
and in Seyfert galaxies since the late '60s, methodical UV/optical
monitoring did not start until the early '80s. The most extensive
efforts have been carried out by the International AGN Watch (Peterson
1999). The project started in the late 1980s as an IUE and
ground-based optical program to monitor NGC 5548, to-date the most
extensively monitored AGN, and was soon expanded to include ground
based and eventually \hst monitoring.  Since then, the project has
monitored eight additional Seyfert 1 galaxies, including five  in more
than one emission line (see Wandel, Peterson \& Malkan 1999 for a
compilation). Additional programs were led by Bradly M. Peterson at
Ohio State University (nine previously unobserved Seyfert 1s); Shai
Kaspi at Wise/Steward Observatory (17 quasars, Kaspi \etal 2000), and
the LAG (Lovers of AGNs) consortium,  to which goes our prize for the
most clever acronym. LAG, led by the late M.V. Penston,  monitored
eight Seyfert galaxies and quasars, three of which had not been
previously observed (Stirpe \etal 1994 and references therein). With
the exception of the International AGN Watch, which included UV,
optical and X-ray monitoring (although not for all objects), all
studies focused on optical monitoring of the continuum and H$\beta$
emission, although H$\alpha$ emission was also monitored for some
objects.

The fact that more progress has not been made on the observational
front might at first seem surprising. In fact, monitoring programs are
exceedingly demanding, to the point that the efforts mentioned above
should be considered nothing short of heroic. On the upside, Type 1
AGNs are normally very bright, so large telescopes are generally not
necessary for variations in the continuum to be easily detected --
indeed, all ground based monitoring has been conducted with one to
two-meter class telescopes. Photometric conditions are desirable but
not strictly necessary, since the continuum and line flux can be
calibrated against a non-variable spectral component (e.g. a narrow
emission line) or by simultaneously observing a nearby (non-variable!)
star (Peterson 2002). On the downside, observing conditions must be
controlled and stable, or spurious variations can easily be
introduced.  For instance, the amount of light from the non-variable
galactic component (NLR or underlying galaxy) entering the
spectrograph depends on seeing conditions, pointing accuracy and
guiding stability. If these conditions are not controlled, a change in
any of them could easily be mistaken for a {\it bona-fide} variation
in the AGN continuum flux.  The time sampling of the data should
ideally be 10 to 30 times shorter, and the total duration of the
experiment 10 to 30 times longer, than the expected time lag (Krolik
2001). This implies that spectra should be obtained a few hours to
several weeks apart, depending on the size of the BLR under study, and
that monitoring should extend over a several year period for the most
massive systems. This mandates the use of a dedicated telescope.

\begin{figure}[t]
\centerline{\psfig{figure=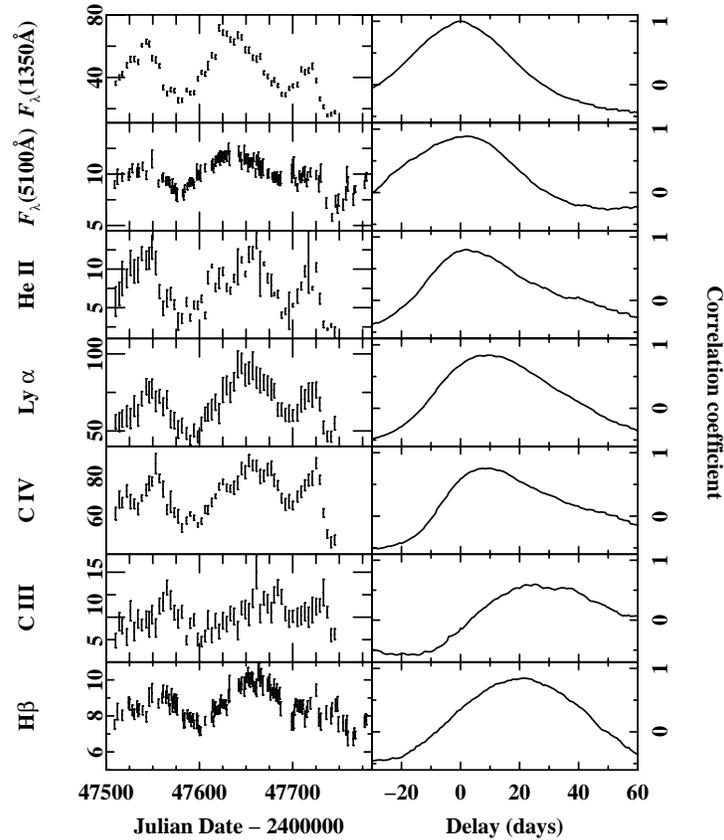,height=13.0 cm}}
\caption{Light curves and cross correlation functions (relative to the
UV continuum) for the UV and optical continuum (top two panels) and
several emission lines in NGC 5548. From Peterson (2002)}
\label{N5548}
\end{figure}

The results of these AGN monitoring programs are beautifully described
in Peterson (2002). They conclusively demonstrated that
although variations in the AGNs' continua are aperiodic (periodicities
have been searched for but never found), they correlate tightly with
and precede variations in the broad emission lines flux
(Figure~\ref{N5548}; Peterson 2002; Onken \& Peterson 2002; Onken et
al. 2003)\footnotemark.  This is, of course, a necessary condition for
reverberation mapping studies to be viable. It is strong evidence
that, from the perspective of the BLR, the continuum source must be
point-like; if the BLR and the continuum emission were spatially
coexisting, the time delays would be chaotic, since it should be quite
possible to have emission line variations from one part of the BLR
precede variations in the continuum from a different part (indeed a
similar situation might affect the 6.4 keV Fe K$\alpha$ line emission,
as discussed in \S~\ref{sec:kalpha}).

\footnotetext{Curiously, the physical drive for the continuum
variability itself is still unclear. Some clues are given by the fact
that in all of the objects studied so far, there is no apparent time
delay between variations in the UV and in the optical, arguing against
mechanical instabilities in the accretion disk or random fluctuations
in the accretion rate as responsible for the fluctuations (these
mechanism would propagate from the inner and hotter to the outer and
cooler part of the disk, stimulating variations in the UV continuum
first and in the optical later). The leading theory is that continuum
variations might be caused as the disk reprocesses the hard X-rays
produced in the surrounding corona.  Further clues as to the
fundamental physics that regulates the continuum variability
(including perhaps the mass of the central SBH, Edelson \& Nandra
1999) are held in the power density spectrum of the X-ray fluctuations
(e.g. Belloni \& Hasinger 1990; Haardt \& Maraschi 1991, 1993;
Zdziarski \etal 1994; Stern \etal 1995; Bao \& Abramowicz 1996).}

The time delay between continuum and optical emission line variations,
measured by cross-correlating the continuum and emission light curves,
can be of days to weeks. This time delay is a direct measure of the
``responsivity weighted'' size of the BLR\footnotemark, which is
approximated by the radius where physical conditions (e.g. the
particle density) produce the strongest line response for a given
continuum change. The responsivity weighted size of the BLR depends on
the ionization potential of the line and on the continuum flux, and
therefore can change with time. In NGC 5548, the only AGN for which
there is sufficient information on the long term behavior of a single
line (H$\beta$) to measure both the line widths and lags as a function
of time, the lag is observed to become longer as the continuum
brightens (Peterson \etal 2002). This is exactly what is expected. As the
continuum brightens, the radius of the Str\"{o}mgren sphere, i.e. the
depth to which the BLR is ionized, increases, and therefore the
responsivity weighted size of the emitting region becomes
larger. Furthermore, when more than one line is monitored, the highest
ionization lines respond most rapidly, and are hence produced closer,
on average, to the central continuum source. Although the data are
quite sparse, radial (out or inflow) motions do not seem to be
dominant. A radial flow would be reflected in a difference between
the time lag measured for the blue and red wings of the line; such
difference is generally not detected (e.g. Koratkar \& Gaskell 1991;
Wanders \& Horne 1994; Wanders \& Peterson 1996; O'Brien \etal 1998;
Sergeev \etal 2002), although a mild radial flow has been claimed by
Kollatschny (2003) in the Seyfert 1 Mrk 110.

\footnotetext{This is true only if the BLR structure does not change
over the duration of the monitoring program. This condition is
satisfied for programs shorter than the dynamical cloud crossing time,
typically a few to several years. A longer monitoring program has been
conducted only for NGC 5548.}

When using reverberation mapping to calculate SBH masses, the BLR
velocity which enters the virial equation, $M = fr\sigma^2 / G$, must
be measured.  This is done (relatively) easily from the FWHM of the
emission lines. To avoid contamination from the non-varying
components of the AGN, the line width is measured from the ``rms''
spectrum, where the rms is calculated relative to the mean of all
spectra obtained during the monitoring experiment. By definition,
non-variable components (for instance narrow absorption lines, and the
underlying galaxy's spectrum) cancel out in the rms spectrum, leaving
only the clean contribution from the BLR. Krolik (2001) points out that 
the line flux and velocity might have different
responsivity functions, i.e. produce the strongest response at
different locations within the BLR. If the velocity measured from the
rms spectrum is not representative of the velocity at the radius
derived from the lag measurements, systematic errors in the derived
masses can easily ensue.

\begin{table}[t] 
\scriptsize
\caption{Reverberation Mapping Radii and Masses$^{\rm a}$}
\begin{tabular}{llcccc} 
\hline 	     Object & z & $R_{BLR}$  & $\lambda
L_{\lambda}$(5100\AA)$^{\rm b}$ & v$_{FWHM}$(rms) & $M$(rms)  \\ & &
(lt-days)  & $10^{44}$ergs s$^{-1}$ & km s$^{-1}$ & $10^7$\msun \\
\hline    3C 120      &0.033  & $42^{+27}_{-20}$     & $0.73\pm0.13$ &
$2210\pm120$ & $3.0^{+1.9}_{-1.4}$\\ 3C 390.3    & 0.056  &
$22.9^{+6.3}_{-8.0}$ &$0.64\pm0.11$       & $10500\pm800$&
$37^{+12}_{-14}$\\ Akn 120     &0.032  & $37.4^{+5.1}_{-6.3}$ &
$1.39\pm0.26$       & $5850\pm480$ &$18.7^{+4.0}_{-4.4}$\\ F 9 & 0.047
& $16.3^{+3.3}_{-7.6}$ &$1.37\pm0.15$       & $5900\pm650$ &
$8.3^{+2.5}_{-4.3}$\\ IC 4329A &0.016  & $1.4^{+3.3}_{-2.9}$  &
$0.164\pm0.021$     & $5960\pm2070$&$0.7^{+1.8}_{-1.6}$\\ Mrk 79 &
0.022  & $17.7^{+4.8}_{-8.4}$ & $0.423\pm0.056$     & $6280\pm850$ &
$10.2^{+3.9}_{-5.6}$\\ Mrk 110 &0.035  & $18.8^{+6.3}_{-6.6}$ &
$0.38\pm0.13$       & $1670\pm120$ &$0.77^{+0.28}_{-0.29}$\\ Mrk 335 &
0.026  & $16.4^{+5.1}_{-3.2}$ & $0.622\pm0.057$     & $1260\pm120$ &
$0.38^{+0.14}_{-0.10}$\\ Mrk 509& 0.034  & $76.7^{+6.3}_{-6.0}$ &
$1.47\pm0.25$       & $2860\pm120$ &$9.2^{+1.1}_{-1.1}$\\ Mrk 590 &
0.026  & $20.0^{+4.4}_{-2.9}$ & $0.510\pm0.096$     & $2170\pm120$ &
$1.38^{+0.34}_{-0.25}$\\ Mrk 817& 0.031  & $15.0^{+4.2}_{-3.4}$ &
$0.526\pm0.077$     & $4010\pm180$ &$3.54^{+1.03}_{-0.86}$\\ NGC 3227
& 0.0038 &$12.0^{+14.9}_{-10.0}$& $0.0202\pm0.0011$  &
$4360\pm1320$&$3.6\pm1.4$\\ NGC 3516    & 0.0088 & $7.4^{+
5.4}_{-2.6}$ & & $3140\pm150$ & $1.68\pm0.33$\\ NGC 3783    & 0.0097 &
$10.4^{+4.1}_{-2.3}$ & $0.177\pm0.015$     & $2910\pm190$ &
$0.87\pm0.11$\\ NGC 4051    & 0.0023 & $ 5.9^{+3.1}_{-2.0}$ &
$0.00525\pm0.00030$ & $1110\pm190$  & $0.11^{+0.08}_{-0.05}$\\ NGC
4151 & 0.0033 & $ 3.0^{+1.8}_{-1.4}$ & $0.0720\pm0.0042$   &
$5230\pm920$ & $1.20^{+0.83}_{-0.70}$\\ NGC 4593    & 0.0090 &
$3.1^{+7.6}_{-5.1}$  & & $4420\pm950$ & $0.66\pm0.52$\\ NGC 5548    &
0.017  & $21.2^{+2.4}_{-0.7}$ & $0.270\pm0.053$     & $5500\pm400$ &
$9.4^{+1.7}_{-1.4}$\\ NGC 7469    & 0.016  & $ 4.9^{+0.6}_{-1.1}$ &
$0.553\pm0.016$     & $3220\pm1580$& $0.75^{+0.74}_{-0.75}$\\ PG 0026
& 0.142  & $113^{+18}_{-21}$    & $7.0\pm1.0$         & $1358\pm91$  &
$2.66^{+0.49}_{-0.55}$\\ PG 0052     & 0.155  & $134^{+31}_{-23}$    &
$6.5\pm1.1$         & $4550\pm270$ & $30.2^{+8.8}_{-7.4}$\\ PG 0804 &
0.100  & $156^{+15}_{-13}$    & $6.6\pm1.2$         & $2430\pm42$  &
$16.3^{+1.6}_{-1.5}$\\ PG 0844     & 0.064  & $24.2^{+10.0}_{-9.1}$&
$1.72\pm0.17$       & $2830\pm120$ & $2.7^{+1.1}_{-1.0}$\\ PG 0953 &
0.239  & $151^{+22}_{-27}$    & $11.9\pm1.6$        & $2723\pm62$  &
$16.4^{+2.5}_{-3.0}$\\ PG 1211     & 0.085  & $101^{+23}_{-29}$    &
$4.93\pm0.80$       & $1479\pm66$  & $2.36^{+0.56}_{-0.70}$\\ PG 1226
& 0.158  & $387^{+58}_{-50}$    & $64.4\pm7.7$        & $2742\pm58$  &
$23.5^{+3.7}_{-3.3}$\\ PG 1229     & 0.064  & $50^{+24}_{-23}$     &
$0.94\pm0.10$       & $3490\pm120$ & $8.6^{+4.1}_{-4.0}$\\ PG 1307 &
0.155  & $124^{+45}_{-80}$    & $5.27\pm0.52$       & $5260\pm270$ &
$33^{+12}_{-22}$\\ PG 1351     & 0.087  & $227^{+149}_{-72}$   &
$4.38\pm0.43$       & $950\pm130^{c}$ & $3.0^{+2.1}_{-1.3}$\\ PG 1411
& 0.089  & $102^{+38}_{-37}$    & $3.25\pm0.28$       & $2740\pm110$ &
$8.8^{+3.3}_{-3.2}$\\ PG 1426     & 0.086  & $95^{+31}_{-39}$     &
$4.09\pm0.63$       & $5520\pm340$ & $37^{+13}_{-16}$\\ PG 1613     &
0.129  & $39^{+20}_{-14}$     & $6.96\pm0.87$       & $2500\pm140$ &
$2.37^{+1.23}_{-0.88}$\\ PG 1617 & 0.114  & $85^{+19}_{-25}$     &
$2.37\pm0.41$       & $3880\pm650$ & $15.4^{+4.7}_{-5.5}$\\ PG 1700 &
0.292  & $88^{+190}_{-182}$   & $27.1\pm1.9$        & $1970\pm150$ &
$5.0^{+11}_{-10}$\\ PG 1704     & 0.371  & $319^{+184}_{-285}$  &
$35.6\pm5.2$        & $400 \pm120$ & $0.75^{+0.63}_{-0.81}$\\ PG 2130
& 0.061  & $200^{+67}_{-18}$    & $2.16\pm0.20$       & $3010\pm180$ &
$20.2^{+7.1}_{-2.4}$\\ \hline
\multicolumn{6}{p{1\columnwidth}}{\hangindent20pt\hangafter1 $^{\rm
a}$ Data for the PG quasars are from Kaspi \etal (2000), while data
for all  other objects (with the exceptions noted below) are
originally from Wandel, Peterson  \& Malkan (1999), as listed in Kaspi
\etal (2000). Data for NGC 3783 are from Onken  \& Peterson
(2002). Data for NGC 3227, NGC 3516, \& NGC 4593 are from Onken et
al. 2003. Data for NGC 4051 are from Peterson \etal (2000).}\\
\multicolumn{6}{p{1\columnwidth}}{\hangindent20pt\hangafter1 $^{\rm
b}$Assuming $H_0=75$ km s$^{-1}$ Mpc$^{-1}$.} \\
\multicolumn{6}{p{1\columnwidth}}{\hangindent20pt\hangafter1 $^{\rm
c}$H$\beta$ was not observed, therefore the rms velocity refers to
H$\alpha$.}\\
\end{tabular}
\label{tab:revmap}
\end{table}

\subsubsection{Observational Support for the Reliability of 
Reverberation Mapping Masses}
\label{sec:revmapobs}

Reverberation mapping has always enjoyed high esteem within the AGN
community; however, it has been viewed with considerable skepticism
within the circles of stellar dynamicists, at least as a way to
estimate SBH masses.  Although there is no galaxy for which a SBH mass
estimate exists based on both reverberation mapping and resolved
stellar or gas kinematics, early studies indicated that for galaxies
of comparable magnitude, masses derived from reverberation mapping,
and those obtained using traditional resolved stellar dynamics,
differed by as much as a factor 50 (e.g. Wandel 1999; McLure \& Dunlop
2000; Ho 1999; Richstone \etal 1998). The blame for the discrepancy
fell, only too easily, on the reverberation mapping results, which
were vindicated only when it was realized that the stellar dynamical
masses used in the comparison (from Magorrian \etal 1998) were
seriously flawed. A discussion of this issue and its resolution can be
found in Merritt \& Ferrarese (2001c, see also Wandel 2002).

There still are, however, serious  theoretical considerations which
could potentially affect mass estimates from reverberation mapping
(Krolik 2001).  Of these, perhaps the most important are the
assumptions made on the velocity and geometric structure of the BLR.

\begin{figure}[t]
\centerline{\psfig{figure=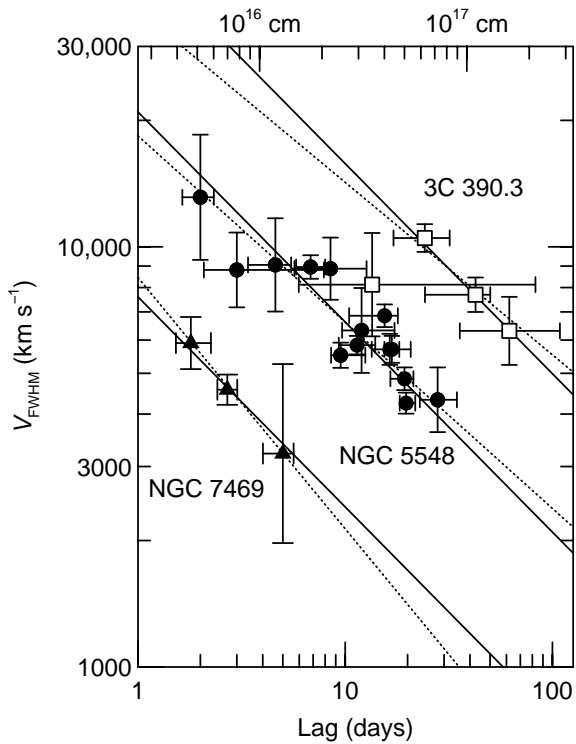,height=10.0 cm}}
\caption{Line widths as a function of time lag for several emission
lines in NGC 7469, NGC 5548 and 3C390.3. The velocities closely follow
a virial relation, shown by the solid lines. The dashed lines show the
actual fits to the data. From Peterson \& Wandel (2000).}
\label{virial}
\end{figure}

The virial hypothesis is of course at the heart of reverberation
mapping as  a mass estimator. If the hypothesis is invalidated, for
instance if the BLR is in a radial flow, the whole method is
undermined, at least as far as mass estimates are
concerned. Reassuringly, the virial hypothesis has recently received
strong observational support from the work of  Peterson \& Wandel
(2000)  and Onken \& Peterson (2002). If the motion of the gas is
gravitational, using the lags derived from different emission lines in
the same AGN must lead to the same mass measurement. NGC 5548 was the
first galaxy for which this was indeed verified. The highest
ionization lines are observed to have the shortest time lag, so that
the virial product $rv^2$ remains constant (Figure~\ref{virial}). The
implied central mass is $\mh = 6 \times 10^7$ \msun. The same has
now been observed in three additional galaxies, NGC 7469, NGC 3783,
and 3C390.3; in all cases, the time lag (i.e. the BLR radius) and
measured line width obey a virial relationship. Although there exist
non-equilibrium kinematical configurations which can mimic a virial
relationship (e.g.  cloud outflows or disk winds could also explain
the observations, Blumenthal \& Mathews 1975; Murray \etal 1995;
Emmering, Blandford \& Shlosman 1992; Chiang \& Murray 1996; Bottorff
\etal 1997), these alternative explanations need to be very finely
tuned to account for the mounting quantity of observational evidence,
and are therefore becoming more and more untenable.  The H$\beta$ line
in NGC 5548 also maintains a virial relationship as a function of
time: as the continuum brightens and the lags become larger, the line
width is observed to become narrower, in the exact fashion necessary
to maintain a virial relation (Peterson \etal 2002).

The geometry of the BLR is less securely constrained. For lack of
evidence which favors one model over another, most studies (Wandel,
Peterson \& Malkan 1999; Kaspi \etal 2000; Onken \etal 2003; Sergeev
\etal 2002; Vestergaard \etal 2000) assume that the BLR is spherical
and characterized by an isotropic velocity distribution, although
different assumptions are also made. For instance, McLure \& Dunlop
(2000) assume a thin disk geometry, leading to velocities $1.7$ times
and black holes masses three times greater than in the spherical,
isotropic case.

\begin{figure}[t]
\centerline{\psfig{figure=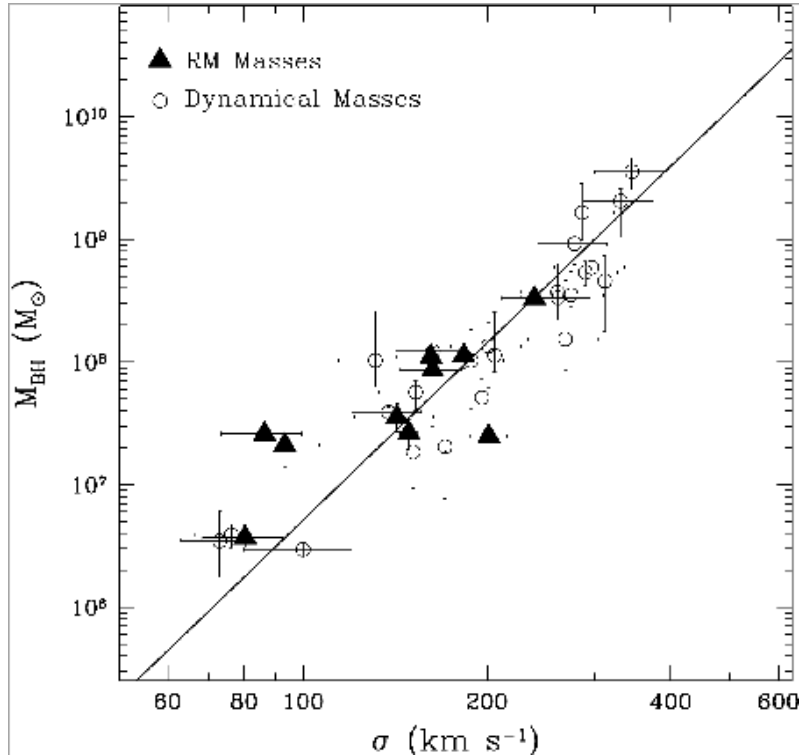,height=10.0 cm}}
\caption{The location of reverberation mapped AGNs (shown as
triangles) in the $\ms$ plane.  The quiescent galaxies which define
the relation are shown by the circles.  Adapted from Ferrarese et
al. (2001).}
\label{AGNms}
\end{figure}

Perhaps the strongest support for reverberation mapping masses comes
by comparing them with those measured, using traditional methods
(\S~\ref{sec:sdyn},~\ref{sec:maser},~\ref{sec:gasdyn}), for the SBH
hosted in ``similar'' (kinematically or morphologically) but quiescent
galaxies. SBH masses measured through resolved stellar or gas dynamics
are tightly connected to the velocity dispersion of the host bulge
(Ferrarese \& Merritt 2000; Gebhardt \etal 2000b, see
\S~\ref{sec:ms}), so tightly that the $\ms$ relation can be used to
estimate SBH masses with 30\% accuracy from a single measurement of
$\sigma$. Gebhardt et al.  (2000c) presented tantalizing evidence that
AGNs might follow the same  $\ms$ relation as quiescent galaxies.
Ferrarese \etal (2001) started a systematic program to map all
   reverberation mapped AGNs onto the $\ms$ plane. The results are
encouraging (Figure~\ref{AGNms}); the remarkable agreement between SBH
masses measured in AGNs and quiescent galaxies with similar velocity
dispersion is strong observational support for the reliability of the
AGN masses. By the time this review appears in press, bulge velocity
dispersion measurements will have been obtained for all of the
reverberation mapped galaxies (Onken et al. 2004). These studies
promise to shed light on the value of the geometric factor $f$, at
least in a statistical sense.

\subsubsection{The Future of Reverberation Mapping}
\label{sec:revmapfut}

Observationally, the main limitation of current reverberation mapping
studies lies in the  less than optimal temporal sampling and duration
of the experiments (Krolik 2001; Peterson 2002). This is the main
obstacle which has so far prevented a direct determination of the
geometry and kinematics of the BLR -- the original aim of the method.
Information about both is contained in the ``transfer function'',
i.e. the time-smeared emission line response to a $\delta$-function
outburst in the continuum:

\begin{equation}L(v_z,t) = \int_0^\infty \Psi(v_z,\tau)C(t-\tau)d\tau\end{equation}

Here, $L(v_z,t)$ is the emission line flux observed at line of sight
velocity $v_z$, $\Psi(v_z,t)$ is the transfer function, and
$C(t-\tau)$ is the continuum light curve. The integral of the transfer
function over time gives the line profile (flux as a function of line
of sight velocity), while the integral of the transfer function over
velocity gives the delay map (normalized flux as a function of time
delay). Solving for the transfer function is a classical inversion
problem; in practice, it requires extremely high S/N data. In the best
cases so far, it has been possible to solve only for the 1D, or
velocity independent transfer function, where both $\Psi(\tau)$ and
$L(t)$ represent integrals over the emission line width.

\begin{equation}L(t) = \int_0^\infty \Psi(\tau)C(t-\tau)d\tau\end{equation}

\begin{figure}[t]
\centerline{\psfig{figure=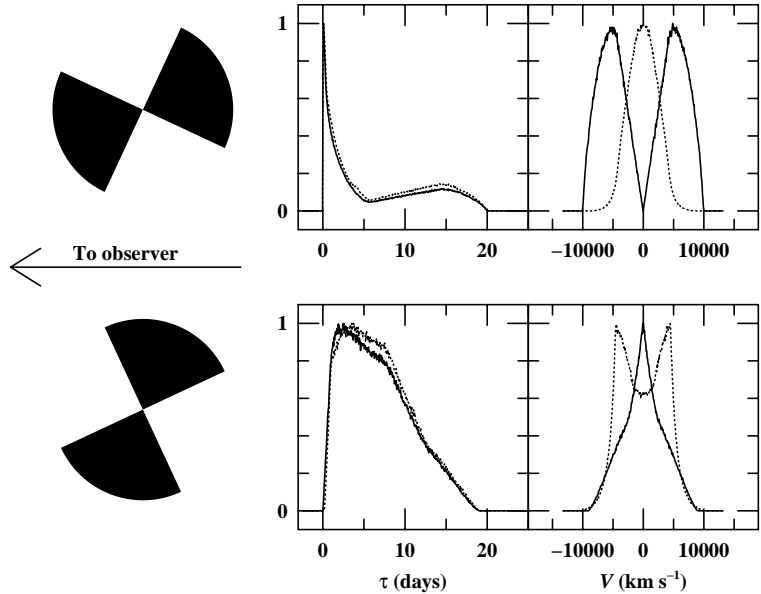,height=8.0 cm}}
\caption{The one-dimentional transfer functions (middle panel) and line
profiles (right panel) for two different biconical outflow models
(solid lines), compared to the case in which the clouds are in circular, Keplerian, isotropic
orbits (dotted lines). The upper and lower panel differ only in the spatial
orientation of the outflow, which includes the observer's line of
sight in the lower panel. The Keplerian and outflow cases can be
easily distinguished only if both the one-dimensional transfer
function and the line profiles are available. From Peterson (2002).}
\label{transfunc}
\end{figure}

Examples of transfer functions are given in Peterson (2002). They can
be calculated for different geometries (e.g. planar or
spherical, as well as more complicated cases), anisotropy in the line
emission (which could arise, for instance, if the BLR is optically
thick to both continuum and emission line radiation, so that most of
the emission  would emerge from the side of the the BLR facing the
continuum source), anisotropy in the continuum emission (originating,
for instance, if the continuum is emitted in a biconical beam with
given opening angle),  distance-dependent responsivity functions (due,
for instance, to geometrical dilution or a varying covering factor for
the BLR). While very different scenarios can correspond to very
similar 1D transfer functions, they are easily distinguishable using
the 2D transfer function, i.e. when both time delay maps and line
profiles are available (Figure~\ref{transfunc}).

Attempts at recovering transfer functions have been made in the case
of the CIV and HeII lines in NGC 4151 (Ulrich \& Horne 1996), the CIV
and H$\beta$ lines in NGC 5548 (S. Collier, as cited by Peterson
2002), and the  H$\alpha$,  H$\beta$,  H$\gamma$, HeI$\lambda$5876 and
HeII$\lambda$4686 lines in Mrk 110 (Kollatschny 2003). In the latter
case, the transfer function seems to resemble the one expected for a
disk in Keplerian rotation, inclined approximatively 30\deg~ relative
to the line of sight (although, as mentioned earlier, the red wing of
the lines seems to respond slightly faster than the blue wing,
suggesting the presence of gas infall). Generally, however, the
data are not of high enough quality for a
satisfactory analysis. For instance, in the case of NGC 5548, the same
transfer function has been argued to be evidence of no outflows
(Wanders \etal 1995), radial outflows (Chiang \& Murray 1996;
Buttorff \etal 1997), and radial inflows (Done \& Krolik 1996).

An unambiguous understanding of the morphology and kinematics of the
BLR is not just a must for quantifying the  uncertainties and
systematics affecting reverberation masses, it also addresses one of
the long standing problems in extragalactic astrophysics -- the nature of the
innermost structure of AGNs. Achieving such understanding requires
long term, well sampled, high time resolution observations over a
broad energy range, from the X-rays to the optical. {\it Kronos}, a
concept for an optical/UV/X-ray space telescope, is designed to meet
this goal in a selected sample  of nearby AGNs (Peterson \& Horne
2003).

\subsection{Secondary Mass Estimators Based on Reverberation Mapping}
\label{sec:sec}

\begin{figure}[t]
\centerline{\psfig{figure=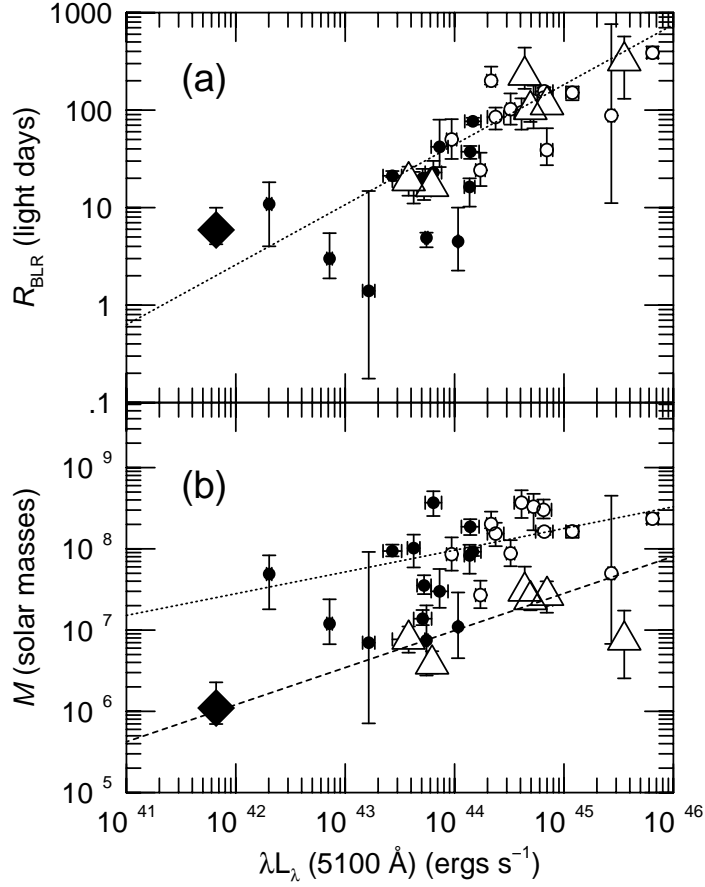,height=12.0 cm}}
\caption{Top panel: relationship between BLR size and optical
continuum luminosity for reverberation mapped AGNs.  Filled circles
represent Seyfert galaxies,  open circles are QSOs, and large
triangles are Narrow Line Seyfert 1 galaxies. The filled diamond is
NGC 4051, one of the best studied galaxies. The dotted line is
the best-fit regression line. Bottom panel:
Relationship between the reverberation-based virial mass and continuum
optical luminosity for AGNs. The dotted line is the best-fit to all objects 
excluding the Narrow Line Seyfert 1 galaxies; while the dashed line is the 
best fit to the Narrow Line Seyfert 1 galaxies only.
From Peterson \etal (2000).}
\label{masslum}
\end{figure}

Since the early days of reverberation mapping, observers noted a correlation of both
the BLR radius and the central mass with continuum luminosity 
(Figure~\ref{masslum};  Koratkar \& Gaskell 1991; Wandel et
al. 1999; Kaspi \etal 2000; Peterson \etal 2000).   In the most
recent characterization of the $\rl$ relation, given by  Vestergaard
(2002), the BLR size (measured from the H$\beta$ broad lines)
correlates with the monochromatic continuum luminosity, measured at
5100 \AA, as:

\begin{equation}R_{BLR} = (30.2 \pm 5.6)\left[{\lambda L_{\lambda}(5100 {\rm \AA})} \over
{10^{44} {\rm ~erg s^{-1}}} \right] ^{0.66 \pm 0.09} {\rm ~lt~
days}\end{equation}

\noindent albeit with large scatter (Peterson \etal calculated a
reduced $\chi^2$ for the fit of 15.7).  A smaller scatter has been
claimed if the AGN luminosity is measured  at 3000 \AA~ instead
(McLure \& Jarvis 2002), although it should be noted that the 
analysis is complicated by the fact that at 3000
\AA~ the AGN continuum is severely contaminated by emission lines.

The slope of the relation holds interesting information about the
physical nature of the ionization process. As pointed out by Koratkar
\& Gaskell (1991) one would expect  the BLR size to correlate with the
0.5 power of the continuum luminosity if the shape of the ionizing
continuum in AGNs is independent of luminosity, and all AGNs are
characterized by the same ionization parameter and BLR density.  In
this case, a stronger optical continuum corresponds to a higher
ionizing flux, which will be able to ionize the BLR to a larger
depth. The fact that the slope of equation 15 seems marginally
inconsistent with this simple expectation, suggests that the physical
parameters within the central region (e.g. continuum shape, BLR
density, column density, and ionization parameter) are likely not
universal.

The $\rl$ relation has tremendous appeal because it allows us to estimate the BLR size from a quick, simple measure
of the continuum luminosity, bypassing the need for long monitoring
programs. Once the BLR size is known, the SBH mass is easily derived
from the virial relation

\begin{equation}\mh = 1.5 \times 10^5 \left({R_{BLR}} \over {\rm lt~days} \right)
\left({v} \over {\rm km~s^{-1}} \right)^2 {\rm
~M_{\odot}}\end{equation}
\noindent (under the assumption of isotropy for the BLR).

The $\rl$ and virial relation, both of which are based on optical
measurements, prompted a recent study by Vestergaard (2002) which
investigates the viability of SBH mass estimators based on UV data, in
particular the CIV $\lambda$1549 emission (to replace H$\beta$) and
continuum luminosity at 1350 \AA~ (rather than 5100 \AA). This is an
extremely important step, since the spectral region chosen by
Vestergaard is redshifted into the optical (and therefore accessible
from the ground) for AGNs with redshifts $1 < z < 5$. Vestergaard's
calibration gives:

\begin{equation}\log{{\mh}\over{M_{\odot}}} = (6.2 \pm 0.03) + \log \left\{ \left[
{\rm FWHM(CIV)} \over {1000 {\rm~km ~s^{-1}}} \right]^2 \left[{\lambda
L_{\lambda}(1350 {\rm \AA})} \over {10^{44} {\rm ~erg~ s^{-1}}}
\right]^{0.7}\right\} \end{equation}

Vestergaard estimates 1$\sigma$ uncertainties of a factor 2.5 and 3
respectively in the mass determined for any single object from optical
data (equations 15 and 16) and UV data (eq. 17, the larger scatter is
a direct consequence of the fact that the UV relation is calibrated
using the optical data). In spite of the large scatter, the undeniable
value of these relations stems from the fact that they are easily
applicable to  large samples of objects, for which direct
reverberation mapping measurements would be unrealistic. While the UV
relations have not yet received direct applications (although many
must be in the works!), several studies have employed the optical
relations (e.g. Bian \& Zhao 2003; Shields \etal 2003; Netzer 2003;
Woo \& Urry 2002a\&b; Oshlack \etal 2002; Wandel 2002; Boroson 2002;
McLure \& Dunlop 2001).

Figure~\ref{masslum} shows the relation between reverberation-based
masses and continuum luminosity (measured at 5100 \AA). The relation
follows from, but has larger scatter than, the correlations between
BLR size and broad line widths with luminosity. A similar relation
exists between SBH masses and the 10$\mu$m and $2-10$ keV nuclear
luminosities (the latter for Compton-thin sources only, Alonso-Herrero
\etal 2002). The larger scatter in these relations, relative to the
$\rl$ relation, indicates that luminosity, rather than mass, is the
main factor determining BLR sizes. Nevertheless, these relations
provide important clues as to the nature of the nuclear activity and
accretion processes  in different classes of AGNs. For instance,
Peterson \etal (2000) note that narrow line Seyfert 1  galaxies
(NLS1s) seem to have smaller SBH masses (by a factor 10) than regular
Seyfert 1 galaxies with comparable continuum luminosity, indicating
that they are accreting at higher rates and/or efficiencies. Based on
these findings, Mathur (2000) further speculate that NLS1s might
represent an early evolutionary  stage of Seyfert 1 galaxies.

\subsection{The Photoionization Method}
\label{sec:photo}

In the assumption that the BLR emission line gas is  photoionized and
not stratified, line ratios can be used to determine the average
physical properties of the gas, in particular the density $n_e$ and
the ionization parameter $U$, i.e. ratio of the ionizing photon flux
to the electron density (this ratio is normally divided by the speed
of light, $c$, in order to make $U$ dimensionless):

\begin{equation} U = {{\int_{\nu_0}^{\inf} L_{\nu}d\nu/h\nu} \over {4 \pi r^2 c n_e}}\end{equation}

Here, $L_{\nu}$ is the monochromatic luminosity of the central source.
In practice, these measurements suffer large uncertainties, and an
average value of $Un_e$ ends up being assumed for all objects.  Once a
value of $Un_e$ is agreed upon, the radius $r$ of the BLR is simply
inversely proportional (via the ionizing flux, knowing which requires
UV spectra) to the square root of $Un_e$, and the central mass can
then be derived under the virial assumption, just as in the case of
reverberation mapping. BLR sizes derived using the photoionization
method are in good agreement with reverberation mapping estimates
(Wandel, Peterson \& Malkan 1999). Indeed, using
this method, Padovani, Burg \& Edelson (1990) conducted the first
study of SBH demographics in Seyfert 1 galaxies, and concluded that
local AGNs cannot be the only repository of the SBHs residing in high
redshift quasars, further justifying the search for SBHs in
quiescent galaxies.

In contrast to reverberation mapping techniques, the photoionization
method is completely independent of the assumed geometry of the BLR
(for example it is insensitive to whether the gas is confined in many
separate clouds or has a continuous distribution). The downside is, of
course, that $Un_e$ is not accurately known, although reasonable
estimates can be obtained given high quality data. More importantly,
the photoionization method relies on a single-zone approximation for
the gas, which is certainly an oversimplification. Nevertheless, 
the method has appeal since it can
have statistical validity when applied to  large samples of objects.

\clearpage

\section{SCALING RELATIONS FOR SUPERMASSIVE BLACK HOLES.}
\label{sec:scale}

{\it If man were restricted to collecting facts, the sciences would
only be a sterile nomenclature and he would never have known the
great laws of nature. It is in comparing the phenomena with each
other, in seeking to grasp their relationships, that he is led to
discover these laws.}

\hskip 0.5in Pierre-Simon Laplace, in ``Exposition du systeme du
monde''

\subsection {The $\ml$ Relation}
\label{sec:ml}

Using the eight SBH detections available at the time, Kormendy \&
Richstone (1995) noticed that supermassive black hole masses correlate
with the blue luminosity of the surrounding hot stellar component,
this being the bulge of spiral galaxies, or the entire galaxy in the
case of ellipticals. Whether the observed correlation was simply the
upper envelope of a distribution which extends to smaller masses was
unclear: while there should be no observational bias against detecting
large SBHs in small bulges, a failed detection of a small SBH in a
large system could have gone unreported at the time (but none has been
reported since). Kormendy \& Richstone point out that the existence of
the correlation indicates that SBH and bulge formation are tightly
connected or even, based on the claimed absence of a SBH in the
bulge-less spiral M33, that the presence of a bulge might be
essential for SBH formation.

\begin{figure}[t]
\centerline{\psfig{figure=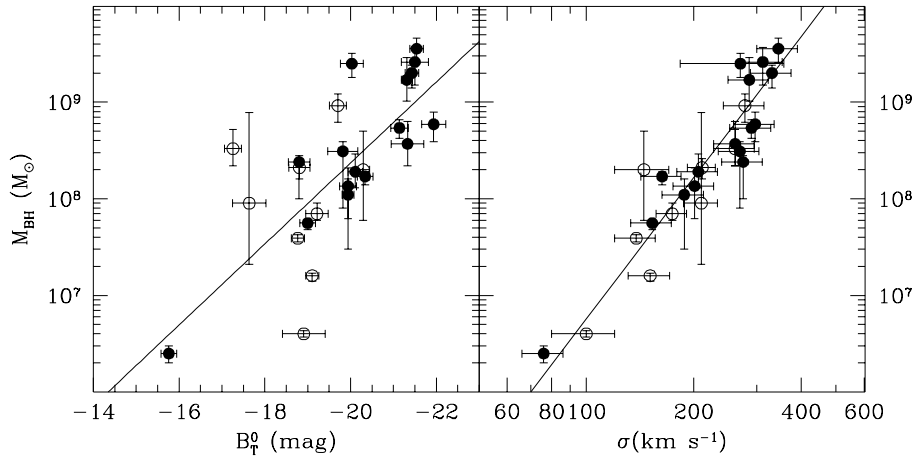,height=6.0 cm}}
\caption{The $\mlb$ (left) and $\ms$ relations for all SBH detections
for which $r_h/r_{res} > 1.0$. Filled symbols show elliptical
galaxies, while open symbols show spiral galaxies and lenticulars. The
solid lines are the best fits to the data, accounting for errors in
both coordinates as well as intrinsic scatter.}
\label{mlms}
\end{figure}

Further SBH detections have confirmed the existence of a
correlation. The most up-to-date $\mlb$ relation, using all SBH
detections for which the sphere of influence is resolved by the
observations, is shown in Figure~\ref{mlms}. A best fit, accounting
for errors in both coordinates as well as intrinsic scatter gives:

\begin{equation}\log(\mh) = (8.37\pm0.11) - (0.419\pm 0.085)(B_T^0+20.0)\end{equation}

The scatter in the $\mlb$ relation is the subject of some debate. Using
the 12  reliable SBH masses known at the time, Ferrarese \& Merritt
(2000) reported a reduced $\chi^2_r = 23$ when the magnitudes are
measured in the $B-$band,  and a scatter of 0.6 dex in $\mh$ at any
given luminosity. This result has not changed significantly in the
past three years. The expanded sample shown in Figure~\ref{mlms} has a
scatter around the best fit line of 0.79 dex in $\mh$.  McLure \&
Dunlop (2001) were the first to point out that if the sample is
restricted to  elliptical galaxies, the scatter is reduced
significantly, down to 0.33 dex. This is comparable to the
scatter in the $\ms$ relation to be discussed in the next
section. Elliptical galaxies are distinguished from lenticulars and
spirals in Figure~\ref{mlms}, they indeed define a relation with
significantly reduced scatter, 0.40 dex in $\mh$ in the updated sample
presented in this review. This is only
slightly larger than calculated by McLure \& Dunlop, due to the
addition of  Cyg A (Tadhunter \etal 2003) and NGC 5845 (Gebhardt et
al. 2003), both of which deviate (significantly in the case of Cyg A)
from the best fit relation.

A natural question is whether the Hubble dependence of the
scatter (and possibly characterization) of the $\mlb$ relation is
indicative of a SBH formation and/or evolution history which differs
in elliptical and spiral galaxies. This seems unlikely for several
reasons. First and foremost is the fact that the characterization of the
$\ms$ relation, to be discussed in the next section, does not seem to
depend on Hubble type. McLure \& Dunlop note that disentangling the
bulge from the disk light, especially in late type spirals, and
especially in the $B-$band, is a difficult and uncertain
process.   For instance, Ferrarese \& Merritt (2000) used the
Simien \& De Vaucouleurs relation (that is know to have large scatter) to estimate the
fraction of total light contained in the bulge as a function of Hubble
type. It is quite possible, therefore, that the overall larger scatter
exhibited by spiral galaxies is simply due to an inaccurate
determination of $B_T^0$ for the bulge component.

This conclusion is supported by the recent study of Marconi \& Hunt
(2003). The authors used $K-$band images from the recently released
2MASS database, and performed an accurate bulge/disk decomposition for
the nearby quiescent spiral galaxies with dynamically measured
masses. When only SBH masses based on data which resolved the sphere
of influence are used, Marconi \& Hunt find that the scatter in the
$K-$band $\ml$ relation is very small (0.31 dex), comparable to the
scatter in the $\ms$ relation, no matter whether spiral galaxies are
included or not. This is not altogether surprising. Both the $\ms$ and
$\ml$ relations betray the existence of a correlation between the mass
of supermassive black holes, and the mass of the host bulge. Near IR
magnitudes are a better tracer of mass than $B-$band magnitudes (and
are also less sensitive to the disk component in spiral galaxies). If
mass is the underlying fundamental parameter as Marconi \& Hunt
suggest, the scatter in the $\ml$ relation should depend on the
photometric band in which the magnitudes are
measured\footnotemark. Marconi \& Hunt reach a second noteable
conclusion, namely that the scatter in the $\mlk$ relation depends on
the sample of galaxies used, increasing significantly (from 0.31 dex
to 0.51 dex) when including galaxies for which the data do not resolve
the SBH sphere of influence. This is a strong endorsement of the
argument, first proposed by Ferrarese \& Merritt (2000), that
resolving the sphere of influence is a necessary condition for a
reliable mass determination to be made.

\footnotetext{It must however be pointed out that 
Marconi \& Hunt measure a smaller scatter for the
$B-$band $\ml$ relation than previously reported.}

\subsection {The $\ms$ Relation}
\label{sec:ms}

\begin{figure}[t]
\centerline{\psfig{figure=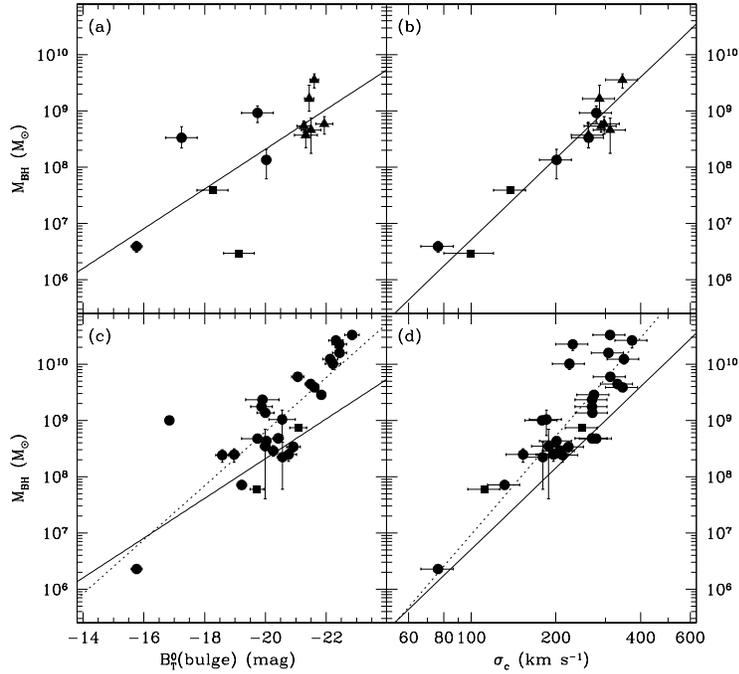,height=10.0 cm}}
\caption{The $\ml$ (left panels) and $\ms$ (right panels) relations
for all SBH masses (as available in 2000) for which the sphere of
influence was (top panels) or was not (bottom panel) resolved
by the data. The solid and dotted lines are the best-fit regression lines
to the galaxies for which the SBH sphere of influence was and was not resolved
respectively. From Ferrarese \& Merritt (2000).}
\label{fm01}
\end{figure}

Bulge magnitudes and velocity dispersions are correlated through the
Faber-Jackson relation. The $\ml$ relation, therefore, immediately
entails a correlation between $\mh$ and $\sigma$.  In spite of the
ample attention devoted to the $\ml$ relation since 1995, five years
went by before the first $\ms$ relation was published (Ferrarese \&
Merritt 2000; Gebhardt \etal 2000b). The reason for the long delay is
easily understood. Figure~\ref{fm01} (bottom panels) shows the $\ml$
and $\ms$ correlations using all SBH masses, regardless of their
accuracy, available in 2000. Both relations have large intrinsic
scatter, and there doesn't appear to be any obvious advantage in
preferring one to the other. The breakthrough came with the
realization that the scatter in the $\ms$ relation is significantly
dependent on sample selection. The upper panels of Figure~\ref{fm01}
only show  SBH masses (as were available in 2000) derived from data
which resolved the sphere of influence. While no significant changes
(besides the obvious decrease in sample size) are noticeable in the $\ml$
relation, the scatter in the $\ms$ relation decreases significantly
when the restricted sample is used. Based on these findings, Ferrarese
\& Merritt (2000) concluded that the reliability of the SBH mass
depends critically on the spatial resolution of the data, and that the
$\ms$ relation is tighter, and therefore more fundamental, than either
the Faber-Jackson or the $\ml$ relation.

The latter conclusion was also reached by Gebhardt \etal (2000b), who
more than doubled the sample used by Ferrarese \& Merritt (2000) by
including SBH masses for 13 additional galaxies (subsequently
published in Gebhardt \etal 2003). There are some noticeable differences
between the Ferrarese \& Merritt and Gebhardt \etal study, besides the
size of the sample. Gebhardt \etal use ``luminosity-weighted
line-of-sight [velocity] dispersions inside a radius R'', while
Ferrarese \& Merritt used central velocity dispersions
normalized to an aperture of radius equal to 1/8 of the galaxy
effective radius (Jorgensen et al. 1995, the same definition is used in studies of the fundamental
plane of elliptical galaxies). Based on the fact that the $\sigma$
measured by Gebhardt \etal (2000b) and those used by Ferrarese \&
Merritt (2000) differ slightly but systematically from each other,
Tremaine et al. (2002) argue that the latter are flawed.

The original slope measured by Gebhardt \etal for the $\ms$ relation
($3.75\pm0.3$) was considerably flatter than the one measured by
Ferrarese \& Merritt ($4.8\pm0.5$). The reasons were identified by
Merritt \& Ferrarese (2001b) in the different algorithm used to fit the
data, the $\sigma$ value adopted for the Milky Way, and the inclusion
in the Gebhardt \etal sample of SBH masses for which the data did not
resolve the sphere of influence. The first two issues were recognized
and corrected in Tremaine \etal (2002), who unfortunately never
addressed the third issue. Tremaine \etal (2002) propose a slope for
the $\ms$ relation of $4.02 \pm 0.32$, and trace the reason for the
difference between their value and that of Ferrarese \& Merritt to the
aforementioned discrepancies in the adopted definition of the velocity
dispersion. To date, there has been no reconciliation of the issue.
Independent investigation of the $\ml$ (\S~\ref{sec:ml}, Marconi \& Hunt 2003)
and $\mc$ relations (S~\ref{sec:mc}, Graham et
al. 2001) have led to the conclusion that resolving the SBH sphere of
influence significantly affects the characterization and scatter of
those relations, arguing that masses based on data with poor spatial
sampling (several of which are included in the Tremaine \etal study)
should not be used. On the other hand, Tremaine et al.'s claim needs
to be pursued, if nothing else because central velocity dispersions
are commonly  used in studies of the fundamental plane of elliptical
galaxies. Unfortunately, the data and procedure used by  Gebhardt et
al. (2000b) to measure what Tremaine \etal claim to be more reliable
values of $\sigma$ are to this date unpublished, and therefore not
reproducible. It is unclear, for instance, if and how inclination and
aperture corrections were accounted for.

Figure~\ref{mlms} shows the most current version of the $\ms$
relation, including only SBH for which the data resolved the sphere of
influence (from Table~\ref{tab:allmasses}, which now includes more
galaxies than used in any previously published study of the $\ms$
relation), using published central values of the velocity dispersions,
and the fitting routine from Akritas \& Bershady (1996):

\begin{equation}{{M_{\bullet}} \over {10^8 \rm M_{\odot}}} = (1.66 \pm 0.24) \left({{\sigma} \over {200~ {\rm km~s^{-1}}}}\right)^{4.86 \pm 0.43}\end{equation}

\noindent (using the fitting algorithm used in Tremaine et al. (2002) actually
produces a slightly steeper slope, $5.1 \pm 0.4$).
The reduced $\chi^2$ is 0.880, indicating that the intrinsic scatter
of the relation is negligible. The scatter around the mean is only
0.34 dex in $\mh$. Neither the scatter, slope or zero point of the
relation depend on the Hubble type of the galaxies considered, as can
be judged qualitatively from Figure~\ref{mlms}.

It is, of course, a trivial exercise for anyone to produce their own
fit to their preferred sample.  Whatever the slope of the $\ms$
relation might turn out to be, there is one point on which there seems
to be universal agreement: because of its negligible scatter, the
$\ms$ relation is of fundamental relevance for many issues related to
the studies of SBHs. In particular:

\begin{itemize}

\item 
The $\ms$ relation allows one to infer SBH masses with 30\% accuracy
from a single measurement of the large scale bulge velocity
dispersion. Through $\sigma$, it has therefore become possible to
explore the role played by the SBH mass in driving the character of
the nuclear activity, not only in individual galaxies (Barth et
al. 2002), but also in different classes of AGNs (Marchesini,
Ferrarese \& Celotti 2004; Barth \etal 2003; Falomo \etal 2003).

\item
Studies of SBH demographics (\S~\ref{sec:demo}), both in quiescent
(Merritt \& Ferrarese 2001a; Ferrarese 2002a; Yu \& Tremaine 2003; Aller
\& Richstone 2003; Whythe \& Loeb 2002, 2003) and active galaxies
(Ferrarese \etal 2001) have relied heavily on the relation.

\item The $\ms$ relation has become the litmus test of models of  SBH
formation and evolution. Reproducing its slope, normalization and,
above all, scatter, and maintaining it in spite of the merger events
which inevitably take place during galaxy evolution, is currently the
biggest challenge faced by the models (Adams, Graff, \& Richstone
2000; Monaco, Salucci \& Danese 2000; Haehnelt, Natarajan, \& Rees
1998; Silk \& Rees 1998; Haehnelt \& Kauffmann 2000; Cattaneo,
Haehnelt \& Rees 1999; Loeb \& Rasio 1994).

\end{itemize}

\begin{figure}[t]
\centerline{\psfig{figure=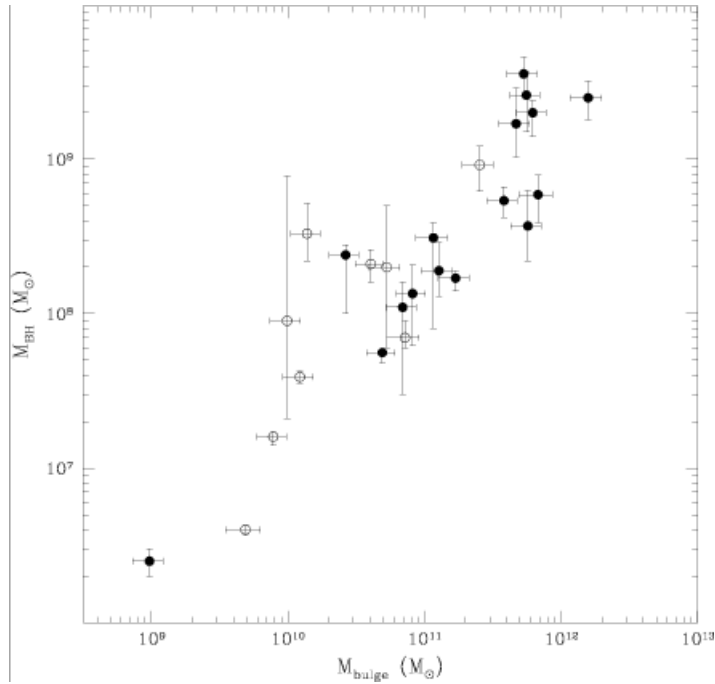,height=9.0 cm}}
\caption{The relation between SBH and virial bulge masses
(following Marconi \& Hunt 2003).}
\label{mmrel}
\end{figure}

Figure~\ref{mmrel} shows the relation between black hole and bulge mass,
where the latter has been derived $M_{bulge} \sim 3 R_e \sigma /G$, 
following Marconi \& Hunt (2003), from which we also adopted 
the  values of the effective radii, $R_e$.

\subsection{A Visual Comparison of the $\ml$ and $\ms$ Relation}
\label{sec:msvsml}

\begin{figure}[t]
\centerline{\psfig{figure=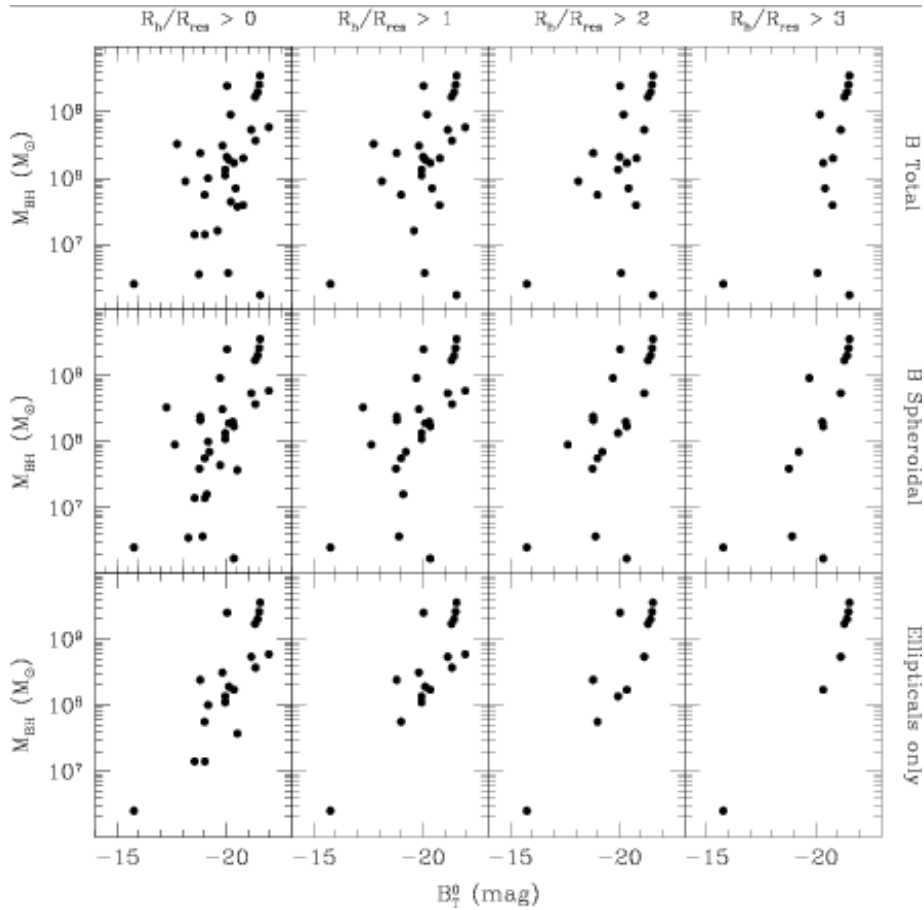,height=12.0 cm}}
\caption{Top panels: The $\ml$ relation obtained when total $B-$band
magnitude is used, regardless of the Hubble Type of the host galaxy.
Middle panels: as for the upper panel, except bulge, rather than total
magnitudes are used. Bottom panels: The $\ml$ relation for elliptical
galaxies only.  From left to right, the sample is increasingly
restricted in terms of how well the data resolves the sphere of
influence of the measured SBH, as indicated by the labels at the top.}
\label{mlcomp}
\end{figure}

\begin{figure}[t]
\centerline{\psfig{figure=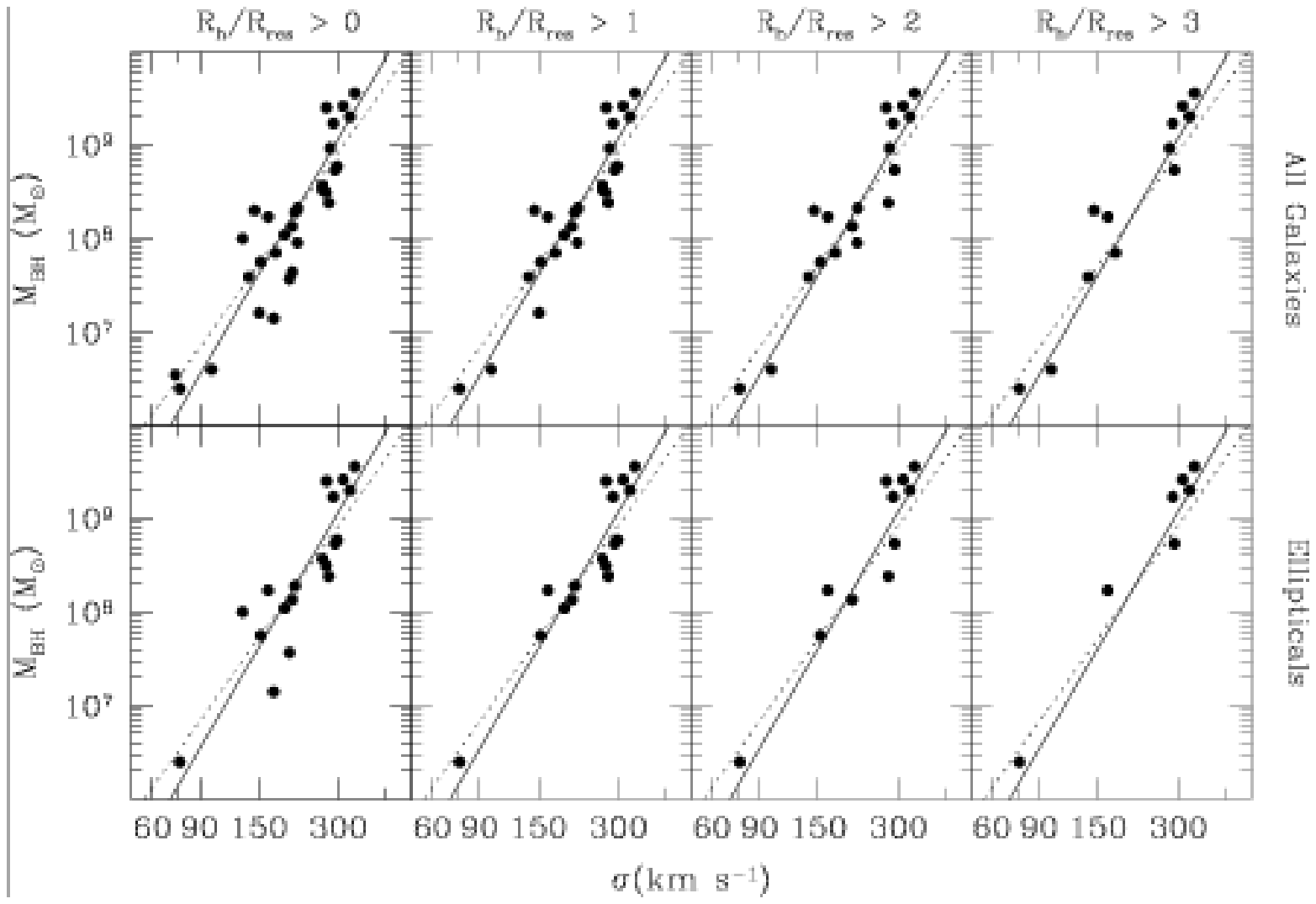,height=9.0 cm}}
\caption{The $\ms$ relation obtained regardless of the Hubble Type of
the host galaxy (top panels) and for elliptical galaxies only (bottom
panel). From left to right, the sample is increasingly restricted in
terms of how well the data resolves the sphere of influence of the
measured SBH, as indicated by the labels at the top. The two lines represent
the fit from Tremaine et al. (2002, dashed line) and this paper (equation 20,
solid line)}
\label{mscomp}
\end{figure}

Figures~\ref{mlcomp} and~\ref{mscomp} are included as a  way to assess, qualitatively,
the scatter and character of the $\ml$ and $\ms$
relations as a function of Hubble Type and the ability of the data to
spatially resolve the sphere of influence. The figures  support the
following conclusions, which can be rigorously proven by fitting the
data given in Table~\ref{tab:allmasses}: 1) the scatter in the
$B-$band $\ml$ relation decreases when bulge, instead of total,
magnitudes are used, as pointed out by  Kormendy \& Gebhardt (2001). 2)
The scatter in the $\ml$ relation decreases when the sample is
restricted to elliptical galaxies only, as first pointed out by McLure
\& Dunlop (2000). 3) The scatter further decreases if the sample is
restricted to galaxies for which the sphere of influence is well
resolved, independently on the Hubble Type of the galaxy considered.
4) For all samples, the scatter in the $B-$band $\ml$ relation is
larger than in the $\ms$ relation (c.f. Marconi \& Hunt 2003).

\subsection{Black Hole Masses and Light Concentration}
\label{sec:mc}

\begin{figure}[t]
\centerline{\psfig{figure=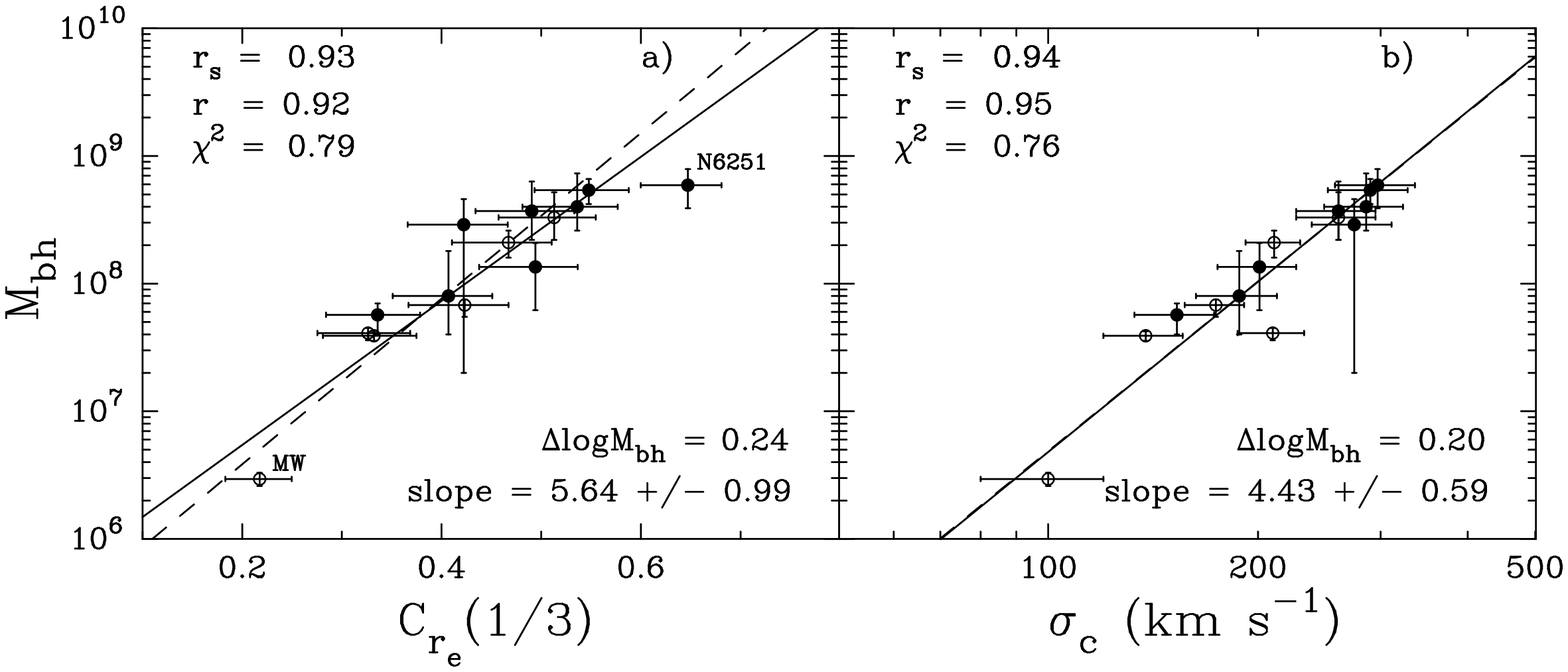,height=5.0 cm}}
\caption{Correlations between supermassive black hole mass and bulge
concentration. Only data for which the SBH sphere of influence was
resolved are shown. From Graham \etal (2001).}
\label{mbhc}
\end{figure}

Graham \etal (2001) found a remarkably tight correlation between SBH
masses and the concentration of bulge light, defined as the ratio  of
flux inside one-third of the half-light radius to the one within the
entire half-light radius (Figure~\ref{mbhc}). The existence of the
correlation is not entirely surprising. At least in ellipticals (which
comprise most of the sample) the shape of the brightness profile
correlates with galaxy luminosity (Ferrarese \etal 1994; Lauer et
al. 1995; Rest \etal 2001). Larger and more luminous galaxies have
shallower brightness profiles (larger radii enclosing $1/3$ of the light
and thus less light concentration) and host
more massive black holes. As is the case for the $\ms$ relation,
however, what is surprising is the scatter, which appears to be
negligible. As noticed by Ferrarese \& Merritt (2000) for the $\ms$
relation and, later, by Marconi \& Hunt for the $\ml$ relation, Graham
\etal find that the scatter in the $\mc$ relation decreases
significantly when only SBH masses derived from data which resolves
the sphere of influence are used.

The $\mc$ and $\ml$ relations have the practical advantage  of needing only
imaging data, generally more readily available than  the
spectroscopic data necessary to measure $\sigma$. The $\mc$ relation is,
however, dependent on a parametric characterization of the light
profile (Graham \etal use a modified Sersic law) which might not
prove to be a good fit for some galaxies. Graham \etal point out that
cD and merging/interacting galaxies might be particularly problematic,
and exclude NGC 6251 and (judging from their Figure 2) M87 and NGC
4374 from their analysis. It is interesting to notice that while
outliers in the $\mc$ relation, these galaxies fit the $\ms$ relation.

Regrettably, there have not yet been theoretical studies targeting the
$\mc$ relation specifically, although a number of authors have studied
the effect of a SBH (or a SBH binary) on the density profile of the
surrounding stellar population (Young 1980; Quinlan, Hernquist \&
Sigurdsson 1995; Lee \& Goodman 1989; Cipollina \& Bertin 1994;
Sigurdsson, Hernquist \& Quinlan 1995; Quinlan \& Hernquist 1997;
Nakano \& Makino 1999, van der Marel 1999; Milosavljevic \& Merritt
2001).

\subsection{Black Hole Masses and Dark Matter Haloes}
\label{sec:mm}

Kormendy \& Richstone (1995) first pointed out that the existence of
the $\ml$ relation indicates that SBH and bulge formation are tightly
connected or even that the presence of a bulge might be a necessary
condition for SBH formation. Six years later,  based on the
observation that the scatter in the $\ml$ relation increases mildly
when total magnitude is substituted for bulge magnitude, Kormendy \&
Gebhardt (2001) argued that SBH masses do not correlate with total
mass. These conclusions, however, can be questioned on several grounds.  
$B-$band observations are a poor tracer of mass, AGNs
in bulge-less galaxies do exist (e.g. Filippenko \& Ho 2003), and most
self regulating theoretical models of SBH formation predict the
fundamental connection to be between $\mh$ and the total gravitational
mass of the host galaxy, rather than the bulge mass   (Adams, Graff,
\& Richstone 2000; Monaco, Salucci, \& Danese 2000; Haehnelt,
Natarajan \& Rees 1998; Silk \& Rees 1998; Haehnelt \& Kauffmann 2000;
Cattaneo, Haehnelt \& Rees 1999; Loeb \& Rasio 1994).  Unfortunately,
finding observational support for the existence of a relation between
SBHs and the total gravitational mass of the host galaxy is
very difficult: measuring dark matter halo masses makes the
measurement of $\mh$ look almost trivial!

\begin{figure}
\begin{minipage}[t]{6.0cm}
\psfig{figure=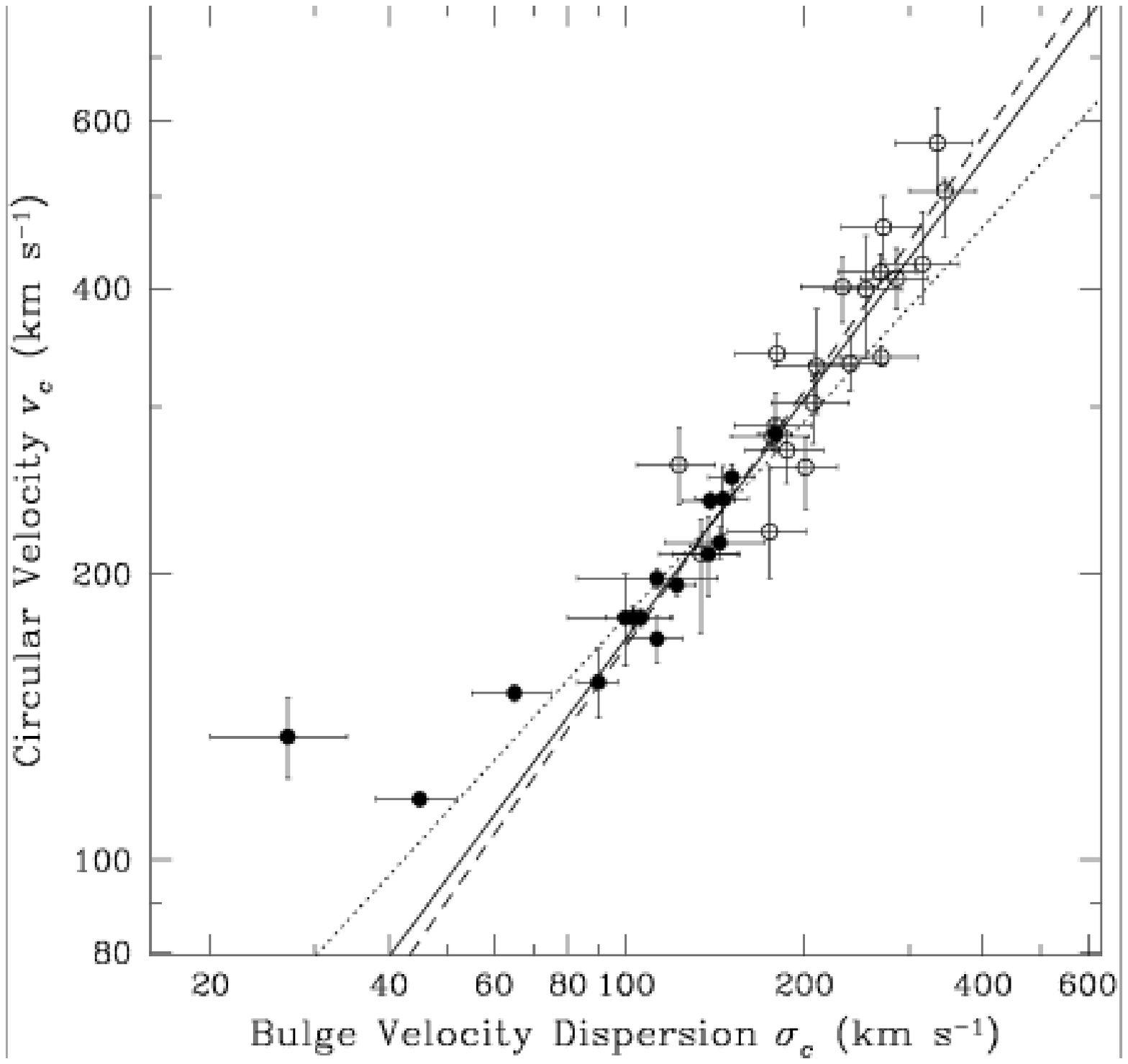,height=5.2 cm}
\end{minipage}
\begin{minipage}[t]{6.0cm}
\psfig{figure=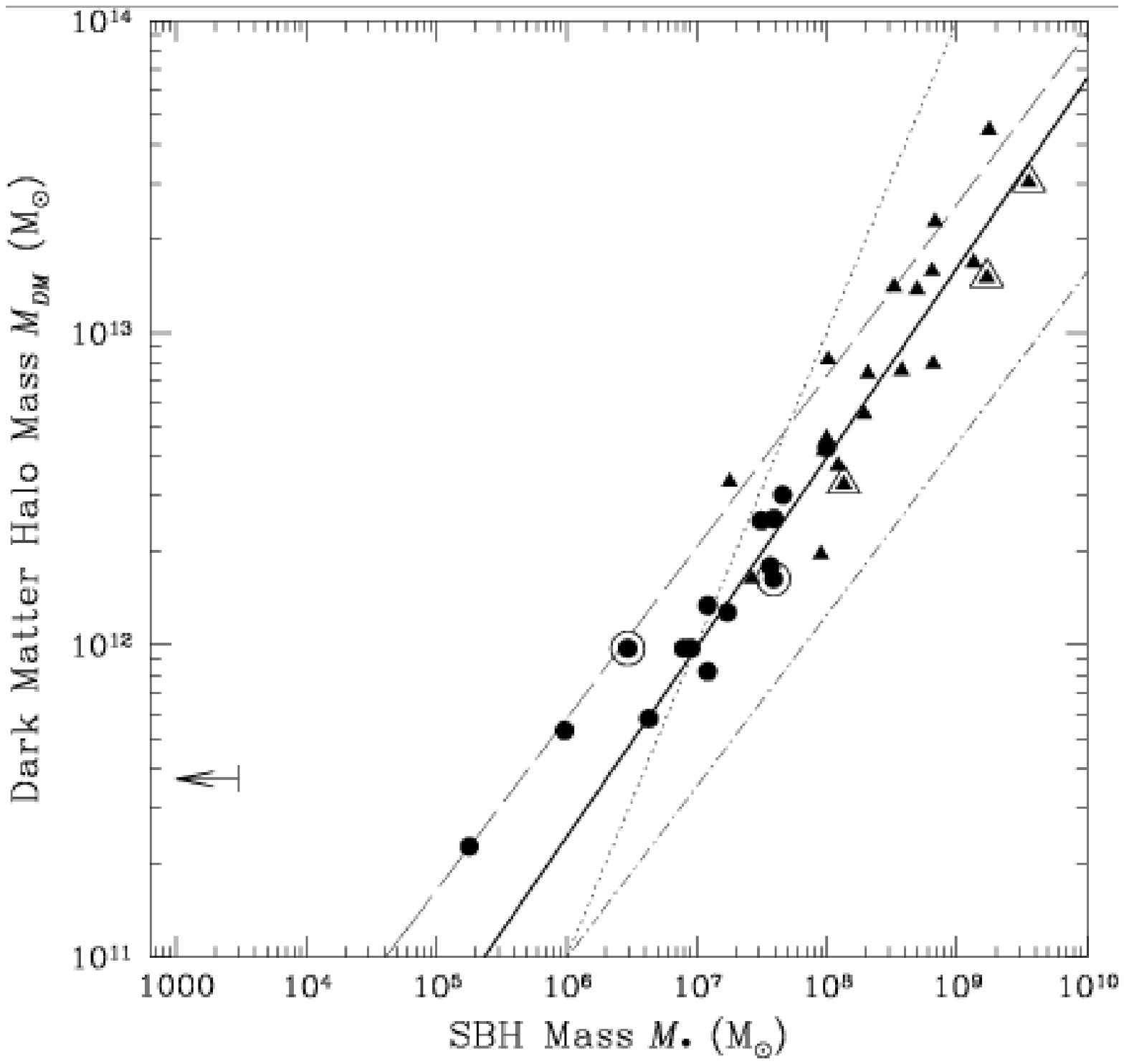,height=5.2 cm}
\end{minipage}
\caption{Left: Correlation between bulge velocity dispersion
$\sigma_c$ and disk circular velocity $v_c$ for a sample of 20
elliptical galaxies (from Kronawitter \etal 2000, open symbols) and
16 spiral galaxies with H I rotation curves extending beyond the
galaxies' optical radius (filled symbols). The galaxy to the far left,
with the smallest value of $\sigma_c$, is NGC 598 (M33). The dotted
line is a fit to all spiral galaxies with the exception of NGC 598,
while the solid line is the fit to all spirals with $\sigma_c > 70$ km
s$^{-1}$. The dashed line is a fit to the entire sample of ellipticals
plus spirals with $\sigma_c > 70$ km s$^{-1}$. Right: $\sigma_c$ has
been transformed into SBH mass using the $\ms$ relation, and $v_c$ has
been transformed into dark matter halo mass following the
prescriptions of Bullock \etal (2001). Filled circles are spirals;
filled triangles are ellipticals. Symbols accentuated by a larger open
contour identify galaxies having a dynamical estimate of $\mh$, which
was used in the plot. The upper limit on the SBH mass in NGC 598 is
marked by the arrow. The dotted line represents a constant ratio
$M_{DM}/\mh = 10^5$. The solid line corresponds to the best fit to the
data. The dashed and dot-dashed line show the best fits which would be
obtained if different prescriptions to relate $v_c$ and $M_{halo}$ were
used.}
\label{mbhmhalo}
\end{figure}

Ferrarese (2002b) provided the first -- albeit indirect --
observational evidence that SBH and dark matter halos might be tightly
connected.  Figure~\ref{mbhmhalo} shows that in spiral galaxies as
well as in ellipticals, the bulge velocity dispersion, which is
typically measured within a radius of 0.5 kpc, and the large scale
circular velocity, measured at radii which range from 10 to 50 kpc,
are tightly connected. For the spirals, the disk circular velocity is
measured directly from HI observations which extend beyond the optical
radius of the galaxy, well within the region where the rotation
velocity stabilizes into a flat rotation curve. For the elliptical
galaxies, the circular velocity is derived from dynamical models
applied to the stellar kinematics (Gerhard \etal 2001; Kronawitter et
al. 2000).  The best fit $\vs$ relation gives:

\begin{equation}\log{\vc} = (0.84 \pm 0.09) \log{\sigma_c} + 
(0.55\pm0.29)\end{equation}

The existence of the $\vs$ relation, with virtually unchanged
characterization,  has since been confirmed using a larger sample of
galaxies by Baes \etal (2003).

The interest of the $\vs$ relation reaches beyond the realm of SBHs. For
instance, characterizing the slope and normalization of the relation
might help in constraining theoretical models and numerical
simulations following the formation and evolution of galaxies
(e.g. Steinmetz \& Muller 1995). For the purpose of this review,
however, the $\vs$ relation clearly betrays a correlation between the
masses of SBHs and those  of the surrounding dark matter halos
(Figure~\ref{mbhmhalo}), although the exact characterization of such a
relation remains to be explored. The velocity dispersion is easily
transformed to $\mh$ using the $\ms$ relation but, unfortunately, the
relation between circular velocity and dark matter halo mass is more
uncertain.  In a CDM-dominated universe, the dark matter halo mass is
uniquely determined by the halo velocity measured at the virial
radius. The latter is defined as the radius at which the mean density
exceeds the mean universal density by a constant factor, generally
referred to as the ``virial overdensity''. Based on this definition,
and by virtue of the virial theorem, it immediately follows that the
halo mass must be proportional to the third power of the virial
velocity, but unfortunately the constant of proportionality depends on the
adopted cosmology and may vary with time (e.g., Bryan \& Norman 1998;
Navarro \& Steinmetz 2000).  To convert the $\vc$ relation to a $\mm$
relation,  the virial velocity must also be related to the measured
circular velocity, a step which entails further assumptions on the
value and mass dependence of the halo concentration parameter.  Using
the $\ms$ relation, and the $\Lambda$CDM cosmological simulations of
Bullock \etal (2002), Ferrarese (2002b) derives

\begin{equation}{{\mh} \over {10^8~{\rm M_{\odot}}}} \sim 0.10 {\left({M_{DM}}  
\over {10^{12}~ {\rm M_{\odot}}}\right)}^{1.65}.\end{equation}

The dependence of $\mh$ on $M_{DM}$ is not linear, a result that seems
quite robust no matter what the details of the adopted cosmological
model are. The ratio between SBH and halo mass decreases from $\sim 2
\times 10^{-4}$ at $M_{DM} = 10^{14}$ \msun~to $\sim 10^{-5}$ at
$M_{DM} = 10^{12}$ \msun, meaning that less massive halos seem less
efficient in forming SBHs. Ferrarese (2002b) notes that this tendency
becomes more pronounced for halos with $M_{DM} < 5 \times 10^{11}$
\msun, and further note that, since there is no direct evidence that
SBH of masses smaller than $10^6$ \msun~ exist, SBH formation might
only proceed in halos with  a virial velocity larger than $\sim 200$
km s$^{-1}$, in qualitative agreement with the formation scenario
envisioned by Haehnelt \etal (1998) and Silk \& Rees (1998).

Although there is no doubt that the $\vs$ relation implies that the
formation of SBHs is controlled, perhaps indirectly, by the properties
of the dark matter halos in which they reside, it is unclear at this
time whether the connection between SBHs and dark matter halos is of a
more fundamental nature than the one between SBHs and bulges,
reflected in the $\ms$ relation. Ferrarese (2002b) notes that the
scatter in the $\mm$ relation could be as small or perhaps even
smaller than the one in the $\ms$ relation for a given choice of the
dark matter density profile and cosmological parameters, although
``secure conclusions will have to await an empirical characterization
of the $\mm$ relation, with both $\mh$ and $M_{DM}$ determined
directly from observations''.

\subsection{Black Hole Masses and Core Radio Power}
\label{sec:mr}

\begin{figure}[t]
\centerline{\psfig{figure=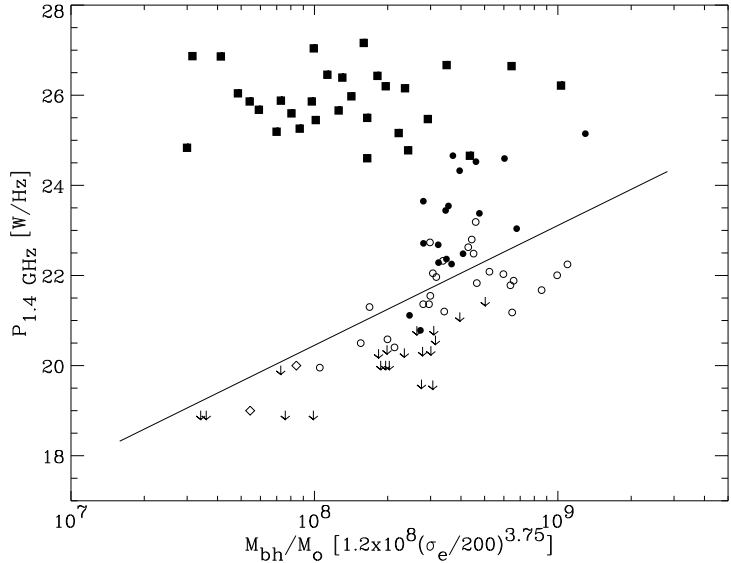,height=8.0 cm}}
\caption{The radio luminosity at 1.4 GHz vs. SBH mass for radio
selected AGNs (filled squares), and optically selected radio quiet
galaxies (open circles) and radio loud AGNs (filled circles). $\mh$
was derived from the bulge central velocity dispersion. The radio
quiet sample seem to follow the relation (shown by the solid line)
first proposed by Franceschini \etal (1998) for a sample of galaxies
with dynamically measured SBH mass. From Snellen \etal (2003).}
\label{mbhradio}
\end{figure} 

One final correlation is worth mentioning. An early study by
Franceschini, Vercellone \& Fabian (1998)  found a correlation
between $\mh$ and both the core and total radio power (measured at
6cm) for the dozen nearby weakly active galaxies with measured $\mh$
available at the time.  Subsequent studies have extended the
correlation to high luminosity radio quasars, for which SBH masses are
estimated  from the H$\beta$ line width and the $r_{BLR}-L_{5100}$
correlation discussed in \S~\ref{sec:sec}. These studies find either a
looser correlation (Laor 2000; Lacy \etal 2001; Wu \& Han 2001) or no
correlation at all (Ho 2002; Oshlack \etal 2002; Woo \& Urry
2002). This possibly indicates that other parameters (for instance the
accretion rate) play a more fundamental role than $\mh$ in determining
the radio properties. The most recent (and perhaps cleanest)
investigation is that of Snellen \etal (2003), who estimated $\mh$
for a sample of powerful radio galaxies using the $\ms$
relation. Because the $\ms$ relation is tighter than the
$r_{BLR}-L_{5100}$ relation used in all other studies, the results are
more easily interpretable. Snellen \etal find that a correlation
between $\mh$ and core radio power exists for optically selected
inactive galaxies only, but breaks down in powerful AGNs
(Figure~\ref{mbhradio}).  The correlation between $\mh$ and radio
power is clearly worth investigating further, not only because of its
predictive power, at least in radio quiet galaxies, but also because
it holds important clues as to the role played by the SBH mass and
accretion rate in shaping the character of the nuclear activity.

\section{BLACK HOLE DEMOGRAPHICS}
\label{sec:demo}

Tracing the mass function of SBHs from the quasar era to the present
day can yield important clues about the formation and growth of
SBHs. It is, however, a road that can prove treacherous, unless we are
fully aware of the uncertainties associated with this process.  For
instance, in 1998, a widely cited study by Magorrian \etal led to  a
cumulative mass density recovered in local SBHs in excess, by a factor
five, relative to the one required to power quasars during the
optically bright phase ($z \sim 2-3$). Based on these findings,
Richstone \etal (1998) suggested that ``a large fraction of black hole
growth may occur at radiative efficiencies significantly less than
0.1''. This possibility was also put forth by Haehnelt, Natarayan \&
Rees (1998), although the authors do mention alternative explanations,
including the possibility that the Magorrian \etal (1998) local SBH mass density could have been ``strongly
overestimated''. The latter was shown to be the case by Ferrarese \& Merritt (2000).
Merritt \& Ferrarese's (2001a) reassesment of the local SBH mass density 
gave values in acceptable agreement with the SBH mass density at
high redshifts. Today, the emerging
picture is one in which the total cumulative SBH mass density does not
seem to depend on look-back time, although the SBH mass function (the mass
density per unit SBH mass interval) {\it might}. Again, much has been
read into this (e.g. Yu \& Tremaine 2002) although, as we will see,
firm conclusions are still premature. In the following  we will
briefly discuss how SBH mass functions are derived, paying particular
attention to potential systematic errors which might affect the
results.

Lacking a dynamical way of measuring SBH masses in high redshift
quasars, these have been inferred directly from quasar counts in the
optical  (Soltan 1982; Small \& Blandford 1992; Chokshi \& Turner
1992) and, most  recently, X-rays. Andrzej Soltan (1982) first pointed
out that under the assumption  that the QSO luminosity is produced by
gas accretion onto the central SBH, i.e. $L_{bol} = \epsilon \dot{M_{acc}}
c^2$, the total cumulative SBH mass density  can be derived directly
from the mean comoving energy density in QSO light. Once a  bolometric
correction and a conversion efficiency $\epsilon$ of mass into energy
are assumed,  the latter is  simply the integral, over luminosity, of
the optical quasar   luminosity function multiplied by luminosity.
Based on these arguments, Soltan concluded that the SBHs powering high
redshift ($z > 0.3$) quasars comprise a total mass density of $\sim 5
\times 10^4$ M$_{\odot}$ Mpc$^{-3}$, each SBH having a mass in the
$10^8 - 10^9$ M$_{\odot}$ range. It was Soltan's argument which first
lead to the inescapable conclusion that most, if not all, nearby
galaxies must host dormant black holes in their nuclei. This
realization has been the main driver for SBH searches in nearby
quiescent galaxies and has kindled interest in the accretion
crisis in nearby galactic nuclei (Fabian \& Canizares 1988),
ultimately leading to the revival of accretion mechanisms with low
radiative efficiencies (Rees \etal 1982; Narayan \& Yi 1995).

Small \& Blandford (1992) proposed a different formalism to describe
the time evolution of the SBH mass function, based on a continuity
equation of the type:

\begin{equation}%2 
\frac{\partial N}{\partial t}  + \frac{\partial}{\partial M} (N<\Mdot>) = S(M,t)
\end{equation}

\noindent where $N(M,t)$ is the number density of SBHs per unit
comoving volume and unit mass,  $<\Mdot(M,t)>$  is the mean growth
rate of a SBH of mass $M$, and $S(M,t)$ is a source function.  Under
the assumption that during the optically bright phase quasars and
bright AGNs accrete at an essentially constant $\Mdot/M$, 
the mass accretion rate $\Mdot$ is related to the luminosity  as
$L = \lambda M c^2 / t_S = \epsilon \Mdot/(1-\epsilon) c^2$, where $\lambda$ is the
fraction of the Eddington rate at which the SBH is accreting, and
$t_S$ is the Salpeter time (equation 2). $N(M,t)$ is related to the
optical luminosity function $\Phi(L,t)$ as $\Phi(L,t) =
N(M,t)\delta(M,t)M$. $\delta(M,t)$ is the QSO ``duty cycle'', i.e. the
fraction of the time which the QSO spends accreting from the
surrounding medium, so that $<\Mdot(M,t)>=\delta(M,t)\Mdot(M,t)$.
Under the assumption that SBHs are originally in place with a minimum
mass, and that subsequently no SBHs are formed or destroyed (for
instance through merging processes), equation (23) can then be written
as

\begin{equation}%2 
\frac{\partial N}{\partial t} = - {\left(\frac{\lambda c}{t_E}\right)}^2 \frac{1}{\epsilon} \frac{\partial \Phi}{\partial L},
\end{equation}

\noindent which can be integrated for a given set of initial conditions 
(Small \& Blandford 1992; Yu \& Tremaine 2002; Ferrarese 2002a; 
Marconi \& Salvati 2002; Marconi et al. 2004; Shankar et al. 2004). 

\begin{figure}[t]
\centerline{\psfig{figure=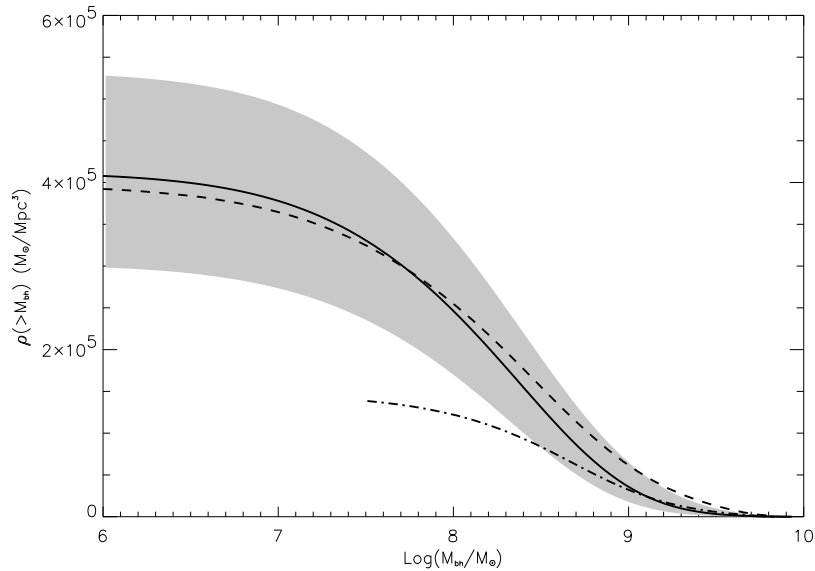,height=8.0 cm}}
\caption{Comparison between the black hole mass function in high
redshift QSOs (dot-dashed line); X-ray selected Type-2
AGNs (dashed line)) and local
quiescent galaxies (solid line, with 1$\sigma$ uncertainty shown by the gray area).
From Shankar et al. (2004)}
\label{democomp}
\end{figure}

The most recent application of these arguments is by Shankar et al. (2004).
Starting from the 2dF QSO Survey from Croom et al. (2003), the authors
find a total cumulative mass density in SBH accreted during the bright AGN phase of 
$\sim 1.4 \times 10^5$ \msun~Mpc$^{-3}$ for a radiative
efficiency $\epsilon = 0.1$ (Figure~\ref{democomp}). This value is consistent 
with several other estimates in the literature (see Table~\ref{tab:sbhdens}), although
the exact value is sensitive to the bolometric correction that needs to be applied 
to the optical fluxes (uncertain by at least 30\% and perhaps as
much as a factor two, Elvis et al. 1994, Vestergaard 2003) and, to a lesser
extent, the luminosity function used in the analysis. In this regard, it is worth pointing out that 
the magnitude limits of the 2dF ($0.3 < z < 2.3$, Boyle \etal 2000, Croom et al. 2003)
and Sloan QSO surveys ($3.0 < z < 5.0$, Fan \etal 2001) correspond to
Eddington limits on the SBHs masses of $4.5\times 10^7$ \msun~and $7.3
\times 10^8$ \msun~respectively. This implies that, at $z > 3$ in
particular, QSO surveys probe only the most massive supermassive black
holes, in sharp contrast with the local SBH sample
(e.g. Figure~\ref{mlms}), requiring a not necessarily trivial extrapolation
to smaller masses. Furthermore, the QSO
luminosity function is not sampled in the $2.3 < z < 3.0$ range. Extrapolating 
within this range is reasonable but not
altogether satisfactory, and could produce an overestimate of the
cumulative SBH mass density below $10^8$ of a factor of a
few (Ferrarese 2002a)\footnotemark.

\footnotetext{While the  extrapolation to $z= 2.3$ of the mass density
of the SDSS QSOs joins smoothly with those measured from the 2dF survey
at the same redshift for $\mh \lae 10^8$ \msun, this is not true for
smaller masses, where the SDSS mass density extrapolated to
$z=2.3$ overpredicts the QSO mass density (per unit redshift) derived
from the 2dF data by an order of magnitude.}

SBH mass densities derived from optical surveys fail to account for the
contribution from obscured populations of AGNs that are known to exist
from X-ray observations.  Early  studies relied on spectral synthesis models of the X-ray background, 
the general consensus being that luminous X-ray absorbed  AGNs,
possibly showing a fast redshift evolution, are necessary to reproduce the
$2-10$ keV integrated source spectrum (Fabian \& Iwasawa 1999; 
Salucci \etal 1999; Gilli, Salvati \& Hasinger 2001; Elvis \etal 2002). 
Without redshift measurements, estimates of the contribution of these obscured 
sources to the SBH mass density were
necessarily based on the assumption that they share the same redshift evolution 
as their better studied, unobscured counterparts. Such estimates generally yielded values in 
excess, by a factor of several, of the mass density in SBH recovered locally
in quiescent or weakly active galaxies (see below). For example, Barger \etal (2001), 
integrated the accretion rate density inferred from the bolometric luminosity of a
sample of 69 hard X-ray sources to obtain a total cumulative mass density of 
SBHs of $\sim 2 \times 10^6$  \msun Mpc$^{-3}$ (for $\epsilon = 0.1$), 
a factor $5-10$ larger than recovered locally (Yu \& Tremaine 2002; Ferrarese 2002a; Marconi \& Salvati 2001).

The launch of the Chandra and the XMM-Newton satellites has made it possible to resolve most of the 
X-ray background into individual sources; optical followups revealed that the bulk of the 
sources are at a redshift $z \sim 0.7$, which is significantly lower than the one which 
characterizes the optically bright phase of the quasars (Alexander \etal 2001; Barger \etal 2002; Hasinger
2002; Rosati \etal 2002; Cowie \etal 2003).  The redshift distribution of the obscured
sources affects both their intrinsic luminosity and the bolometric
correction to the observed X-ray fluxes (which differs for high
luminosity and low luminosity AGNs).  Although accounting for this has led to a downward
revision in the cumulative SBH mass density in obscured AGNs
($\sim 4.1 \times 10^5$ \msun Mpc$^{-3}$ according to Shankar et al. 2004, see also Table~\ref{tab:sbhdens}),  
the mass accreted could be enough to account  
for all SBH found locally if X-ray selected AGNs accrete with $\sim$ 10\% radiative efficiency 
and radiate at $\sim$ 30\% of the Eddington luminosity (Shankar et al. 2004).
The current scenario is therefore of a 
population of unobscured, high luminosity quasars which dominate the accretion at 
high redshift, and a lower redshift  population of obscured AGNs accreting at much 
lower rates\footnotemark, but possibly accounting for the bulk of the 
accreted mass. 

\footnotetext{For instance, Cowie \etal estimate that obscured AGNs accrete at no more 
than a few $10^{-5}$ \msun yr$^{-1}$ Mpc$^{-3}$.} 

\begin{figure}[t]
\centerline{\psfig{figure=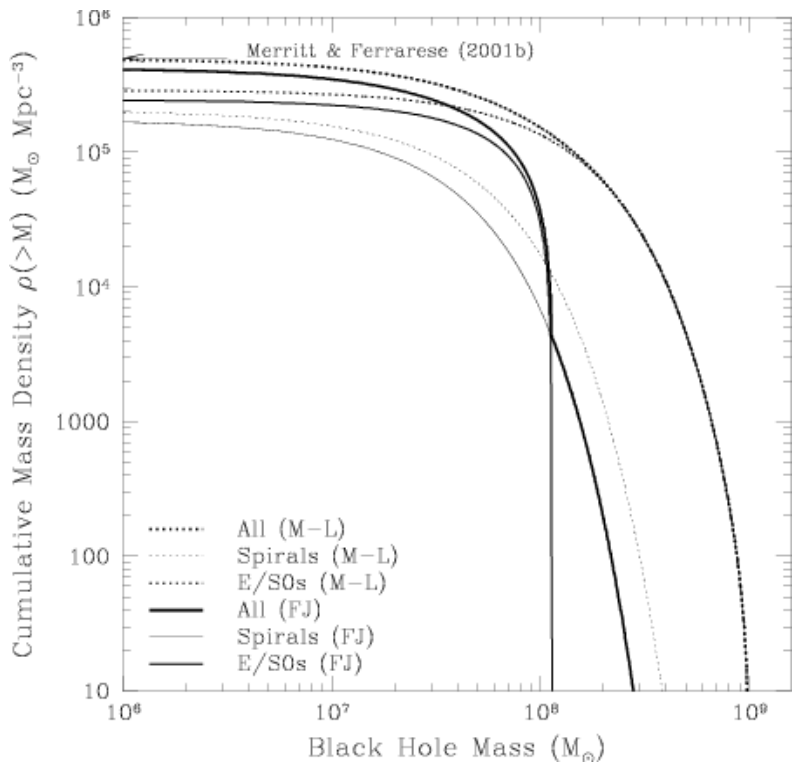,height=10.0 cm}}
\caption{The mass function in local black holes for spirals (thin line),
E/S0 (thicker line) and a complete  sample of galaxies (thickest line). Dotted lines
are derived from the $\ml$ relation, solid lines  from the $\ms$
combined with the Faber-Jackson relation (as described in the text).}
\label{demolocal}
\end{figure}

\scriptsize
\begin{table}[t]
\begin{tabular}{ll}
\multicolumn{2}{c}{\bf \normalsize Summary of SBH Mass Densities}\\
\hline\hline 
\multicolumn{1}{c}{$\rho_{\bullet}$} &
\multicolumn{1}{c}{Reference}\\
\multicolumn{1}{c}{(\msun Mpc$^{-3}$)}  &
\multicolumn{1}{c}{} \\
\hline   
\multicolumn{2}{c}{Local Quiescent Galaxies ($z < 0.025$)} \\
$2.3^{+4.0}_{-1.5} \times 10^5$  & Wyithe \& Loeb (2003)\\
$2.4 \pm 0.8 \times 10^5$  & Aller \& Richstone (2002)\\
$\sim 2.5 \times 10^5$  & Yu \& Tremaine (2002)\\
$2.8 \pm 0.4 \times 10^5$   & McLure \& Dunlop (2004)\\
$4.2 \pm 1.0  \times 10^5$   & Shankar et al. (2004)\\
$\sim 4.5 \times 10^5$   & Ferrarese (2002a)\\
$4.6^{+1.9}_{-1.4} \times 10^5$   & Marconi et al. (2004)\\
$\sim 5 \times 10^5$  & Merritt \& Ferrarese (2001a)\\
$\sim 5.8 \times 10^5$  & Yu \& Tremaine (2002)\\
\hline
\multicolumn{2}{c}{QSOs Optical Counts ($0.3 < z < 5.0$)}\\
$\sim 1.4 \times 10^5$  & Shankar et al. (2004)\\ 
$\sim 2 \times 10^5$   & Fabian (2003)\\
$\sim 2.1 \times 10^5$   & Yu \& Tremaine (2004)\\ 
$\sim 2.2 \times 10^5$  & Marconi et al. (2004)\\
$2 - 4 \times 10^5$   & Ferrarese (2002a)\\
\hline
\multicolumn{2}{c}{AGN X-ray Counts ($z(peak) \sim 0.7$)}\\
$\sim 2 \times 10^5$  & Fabian (2003)\\
$\sim 4.1 \times 10^5$  & Shankar et al. (2004)\\
$4.7 - 10.6 \times 10^5$  & Marconi et al. (2004)\\ 
\hline\hline
\end{tabular}
\label{tab:sbhdens}
\end{table}
\normalsize

The SBH mass function in local ($z < 0.1$), optically-selected AGNs
has received considerably less attention for several
reasons. Compared to QSOs, lower-luminosity AGNs have a small ratio of
nuclear to stellar luminosity, making it difficult to assess what
fraction of the total luminosity is due to accretion onto the central
black hole. Furthermore, the past accretion history is not known, and
it is quite likely that a significant fraction of the SBH mass
predates the onset of the present nuclear activity.  Finally, it is
unlikely that lower luminosity AGNs are accreting at a constant
fraction of the Eddington rate. Padovani, Burg \& Edelson (1990)
estimated a total cumulative mass density in local Seyfert 1 galaxies
of $\sim 600$ \msun Mpc$^{-3}$, based on the
photoionization method (\S~\ref{sec:photo}) applied to  the CfA
magnitude limited sample.  A recent revision of this result
(Ferrarese 2002a), led to an estimate a factor $\sim 8$ larger
(Figure~\ref{democomp}). This notwithstanding, the main conclusion
reached by Padovani \etal still holds: ``the bulk of the mass related
to the accretion processes connected with past QSO activity does not
reside in Seyfert 1 nuclei. Instead, the remnants of past activity
must be present in a much larger number of galaxies''.  This is true
even after correcting for the contribution of AGNs other than Seyfert
1 galaxies (in the local universe, the ratio of Seyfert 2 to Seyfert 1
galaxies is $\sim$ four, while  LINERs are a factor of a few more
numerous than Seyferts, e.g. Maiolino \& Rieke 1995;  Vila-Vilaro
2000).

Finally, in local quiescent galaxies - the necessary repository of the SBHs
which powered quasar activity at high redshifts -  the SBH mass
function can be calculated from the $\ml$ or, preferably due to its
smaller scatter, the $\ms$ relation. If the mean ratio between the
mass of the SBH and that of the host bulge is known,  the observed
mass density of spheroids (e.g. Fukugita \etal 1998) can be easily
transformed into a SBH mass density. This approach was adopted by
Merritt \& Ferrarese (2001a). The $\ms$ relation was used to estimate
$\mh$ for a sample of 32 early type galaxies with a dynamical
measurement of the total mass (from Magorrian \etal 1998). For this
sample, the frequency function $N[\log(\mh/M_{\rm bulge})]$ is well
approximated by a Gaussian with $\langle\log(\mh/M_{\rm bulge})\rangle
\sim -2.90$ and standard deviation $\sim 0.45$. This implies
$\mh/M_{\rm bulge} \sim 1.3 \times 10^{-3}$ or, when combined with the
mass density in local spheroids from Fukugita \etal (1998),
$\rho_{\bullet} \sim 5 \times 10^5$ \msun Mpc$^{-3}$.

Alternatively, $\rho_{\bullet}$ can be derived by combining the $\ms$
relation with the velocity function of galaxies.  Compared to the
previous method, this has the added bonus of producing an analytical
representation of the cumulative SBH mass density as a function of
$\mh$. The process, however, can be rather involved.  The velocity
function is generally constructed starting from a galaxy luminosity
function (e.g. Marzke \etal 1998; Bernardi et al. 2003). This needs to be converted to a
bulge luminosity function, which entails the adoption of a ratio
between total and bulge luminosity (e.g. from Fukugita \etal
1998). Both steps need to be carried out separately for galaxies of
different Hubble types. The derived bulge luminosity function can be
combined directly with the $\ml$ relation to give a SBH mass function,
or it can first be transformed to a velocity dispersion function
(through the Faber-Jackson relation, which again depends on the Hubble
type), and then combined with the $\ms$ relation. Both approaches have
drawbacks. The $\ml$ relation has significant (and possibly
Hubble-type dependent) scatter, while the Faber-Jackson relation has
large scatter and is ill defined, especially for bulges.  The approach
above was followed by Salucci \etal (1999), Marconi \& Salvati (2001),
Ferrarese (2002a), Yu \& Tremaine (2002),  Aller \& Richstone
(2002), Marconi et al. (2004) and Shankar et al. (2004). 
Figure~\ref{demolocal} shows the cumulative SBH mass functions
separately for the E/S0 and spiral populations, derived by Ferrarese
(2002a) from the $\ml$ relation and the $\ms$ relation combined with
the Faber-Jackson relations for ellipticals and spirals. While the two
distributions differ in the details, there is little difference in the
total mass density,  which falls in the range $(4-5) \times 10^5$
\msun Mpc$^{-3}$.

\begin{figure}[t]
\centerline{\psfig{figure=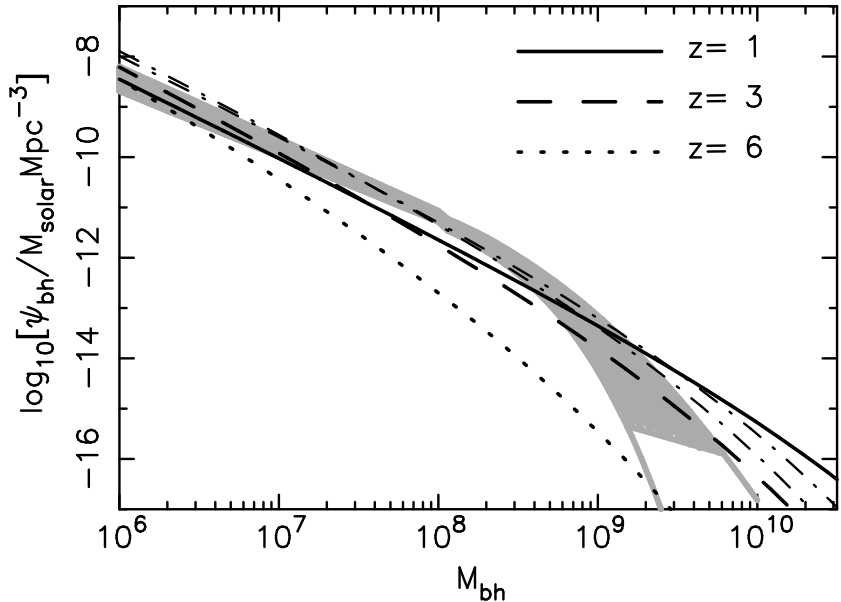,height=8.0 cm}}
\caption{The mass function of SBHs derived by Wyithe \& Loeb (2003)
from the velocity function of Sheth \etal~(2003) and the $M_{\rm
bh}-\sigma$ relation of Merritt \& Ferrarese~(2001b) (gray area). The
lower bound corresponds to the lower limit in density for the observed
velocity function, while the grey lines show the extrapolation to
lower densities. The black lines show the mass function computed at
$z=1$, 3 and 6 from the Press-Schechter~(1974) halo mass function.}
\label{sbhmf_wl}
\end{figure}

Wyithe \& Loeb  (2003) and the Shankar et al. (2004)
used yet another, more direct approach, and employed the
velocity dispersion function directly measured by Sheth \etal (2003)
from SDSS data. The only disadvantage of this approach is that the SDSS velocity
dispersion function is defined only for early-type galaxies, and
therefore incomplete below 200 \kms, i.e. in the regime
populated by spiral galaxies  ($\mh \lae 1.5 \times10^8$ \msun). The
mass function derived by Wyithe \& Loeb is shown in Figure~\ref{sbhmf_wl}. As pointed
out by the authors, the mass function differs from the one
derived using the methods described in the
preceding paragraphs (e.g. Yu \& Tremaine 2002), declining more gradually at the  high
mass end (where the velocity dispersion function is measured
directly).  By comparing the local mass function with that derived by
combining the Press-Schechter (1974) halo mass function with the $\mm$
relation, Wyithe \& Loeb find that SBHs with $\mh \lae 10^8$
\msun~formed during the peak of quasar activity ($z \sim 1 - 3$),
while more massive black holes were already in place at higher
redshifts, as high as $z \sim 6$ for SBHs of a few $\times 10^9$
\msun.   Integrating over $\mh$, the total cumulative mass density is
estimated to be $\rho_{\bullet} = (2.3^{+4.0}_{-1.5}) \times 10^5$
\msun~Mpc$^{-3}$ (see also Shakar et al. 2004).

These results indicate, that, given the current uncertainties, the
total cumulative mass function in SBHs hosted by local 
galaxies is of the correct order of magnitude to account for
the AGNs energetics, although the relative contribution
of high redshift optical QSOs and local obscured AGNs
remains somewhat uncertain (Marconi et al. 2004; Shankar et al. 2004;
Table~\ref{tab:sbhdens}).
The fine details remain to be worked out. For instance,
Yu \& Tremaine (2002) claimed a larger fraction of very massive black
holes, $M > 10^9$, in high redshift QSOs than have been recovered
locally, although this result is not supported by more recent studies
(Marconi et al. 2004, Shankar et al. 2004, see Figure~\ref{democomp}). 
The case is complicated by the fact that, as noted by Wyithe \& Loeb 2003
and Ferrarese 2002a,  the local SBH mass function is {\it not} defined
in the high ($\geq 10^{10}$ \msun) mass range .  In this context,
targeting SBHs in brightest cluster galaxies (\S~\ref{sec:future}),
will be particularly illuminating.

\section {THE FUTURE OF SUPERMASSIVE BLACK HOLE STUDIES.}
\label{sec:future}

The past sections dealt with the measurements of  SBH masses in a
(relatively) large number of nearby galaxies, and the scaling
relations which have ensued between SBH masses and the properties of their host galaxies or
nuclei.   A graphic summary is given in Figure~\ref{methods}. It is a
remarkable fact that most of the information contained in the Figure
has surfaced only in the past few years; even more remarkable to think
that, in spite of the tremendous progress made, we are still only at
the beginning of a long, fascinating journey. We have yet to find out,
for instance, if SBHs spin and how fast, whether binary SBHs exist and
how quickly they coalesce following galaxy mergers, what are the
detailed modalities of accretion, and what controls the character and
level of nuclear activity. We have not yet probed the morphology and
kinematics of the gas within a few thousands of Schwarzschild radii
from the central SBH. On a grander scale, we do not know how
supermassive black holes form. We are, however, for the first time, in
a position in which it does not appear hopelessly optimistic to think
that answers will be found in the near future.

\begin{figure}[t]
\centerline{\parbox[l][9cm]{9.2cm}{\vskip 3cm THIS FIGURE IS INCLUDED IN THE FULL VERSION OF THE
MANUSCRIPT AVAILABLE AT http://www.physics.rutgers.edu/~lff/publications.html.}}
%\centerline{\psfig{figure=f43.eps,height=9.0 cm}}
\caption{The upper two panels show the primary methods for
determining SBH masses, and the classes of galaxies to which they are
most easily applied. ``2-D RM'' and ``1-D RM'' refer to two- and
one-dimensional reverberation mapping respectively; only the latter
has been addressed by current monitoring programs (\S 5), while the
former (\S 5), which would lead to a complete morphological and
kinematical picture of the BLR, will have to await a future dedicated
space mission. The third panel illustrates the fundamental relations
discussed in \S 6. Reverberation mapping can be used to build the
$\mh-L$ and $\ms$ relations for AGNs; by comparing them with the same
relations defined by quiescent galaxies, constraints can be set on the
reliability of reverberation mapping studies. Dashed lines represent
``future" connections; if 2-D reverberation mapping becomes a reality,
 the AGN $\mh-L$ and $\ms$ relations can be
compared directly to the relations observed in quiescent galaxies, to
evaluate the role played by $\mh$ in driving the character of the
nuclear activity. The lower panel shows secondary mass estimators. In
low redshift galaxies, both quiescent and active, SBH masses can be
estimated through the $\mh-L$, $\ms$ or $\mc$ (where $C$ is the
concentration parameter for the bulge, Graham \etal 2001) relations
provided that $L_{bulge}$, $\sigma$ or $C$ can be measured. For large
samples of early type galaxies, if $\sigma$ is not directly observed,
it can be derived using the fundamental plane if the surface
brightness profile and effective radius are known. In high redshift
AGNs, the bulge luminosity and velocity dispersion are difficult to
measure because of contamination from the active nucleus. A method
which has been applied in these cases is to use the width of the
[OIII]$\lambda$5007 emission as an estimate of stellar velocity
dispersion (Nelson \& Whittle 1995; but see also Boroson 2003); $\mh$
can then be derived from the $\ms$ relation. In Type 1 AGNs, the black
hole mass can be derived from the virial approximation if the AGN
luminosity, which correlates with BLR size (\S 5.2) is known. Figure
adapted from Peterson (2004).}
\label{methods}
\end{figure}

We will conclude this review with a brief discussion of what
observational  breakthroughs are required to catalyse further
progress. To this day, the field has been propelled forward largely by
a single event, the launch of the Hubble Space Telescope in 1990.
With (at the time)  a 10-fold boost in spatial resolution compared to
ground based optical telescopes, \hst could resolve the sphere of
influence of putative SBHs in a significant number of galaxies. \hst
also enabled the discovery of the nuclear, regular gas and dust disks
which opened a new, unexpected, and ultimately very successful
way to constrain the central potential (\S~\ref{sec:gasdyn}).
It remains true that the case for a singularity is airtight only in the Galactic center and
NGC 4258, both of which were studied using ground based
facilities. However, proper motion studies are not feasible in
extragalactic nuclei (although Galactic globular clusters might be
promising targets), and though water masers hold great promise
(Greenhill \etal 2003a), they have not yet and are unlikely to produce
more than a few (albeit superlative, if NGC 4258 is any indication)
more detections.  The natural question to ask is therefore, what will
be the role of \hst in the years to come? The answer is, unfortunately,
not as optimistic as one might  hope, making it necessary to look
elsewhere for new ways to boost our knowledge in this field.

To farther our understanding of SBH formation and evolution, it is
imperative to target SBH environments as diverse and extreme as
possible. For instance, if the evolution of SBHs follows a ``merger
tree'' scenario, the very small and very large SBHs, the dwarf and
brightest galaxies, occupy a precious niche in the parameter space,
the  former  as the building blocks and the latter as the end results of
the evolutionary chain. Dark matter dominated low surface brightness
galaxies can tell us whether it is the bulge, or the dark matter halo
that catalysizes the formation of SBHs (\S~\ref{sec:ms},~\ref{sec:mm}).
Unfortunately, not only have dwarf galaxies, brightest cluster
galaxies and low surface brightness galaxies not been probed yet,  the
sample of galaxies listed in Table~\ref{tab:allmasses} is
uncomfortably homogeneous. Most are early type galaxies in small
groups or clusters (M87 is the only cD) within 30 Mpc. Most detected
SBHs are in the $10^8 \lae \mh \lae 10^9$ \msun~ range, there are no
detections below $10^6$ \msun~(the ``building block" range) or above
$10^{10}$ \msun~(the brightest quasar range), and even the $10^6 \lae \mh
\lae 10^7$ \msun~ range is very poorly sampled.

How much can \hst broaden these horizons?  Figure~\ref{hstlimit1}
shows $\mh$ (calculated using the $\ml$ relation, after transforming
the  total luminosity to bulge luminosity following Fukugita \etal
1998) against distance (for $H_0 = 75$ \kms~Mpc$^{-1}$) for all
galaxies in the CfA redshift sample (Huchra \etal  1990).  Exposure
time requirements (\hstnsp's mirror is rather small and the throughput
is relatively low) are included in Figure~\ref{hstlimit2}, which is
restricted, for simplicity, to early type galaxies only. Based on
these two figures, it is clear \hst can only (and pretty much has
already) scrape the top of the iceberg. For instance, ironically, most
of the very largest SBHs ($\mh \gae 10^9$ \msun) are beyond the reach
of \hstnsp, since the low surface brightness which characterizes giant
ellipticals makes stellar dynamical studies prohibitively time
consuming\footnotemark.  \footnotetext{For instance, measuring $\mh$
in M87 ($d \sim 15$ Mpc) using stellar dynamics would require over 100
orbits of STIS time.}  Similarly, the less massive SBHs, typically
expected to be hosted in dwarf elliptical galaxies, are beyond  \hst
capabilities: the stellar population in the nearest dwarfs -- where
\hst can resolve the sphere of influence of a small, hundred thousand
solar masses SBH, is resolved in individual stars, each too faint  to
be handled by \hstnsp's small mirror\footnotemark.

\footnotetext{For instance, a constraint on  the central mass in NGC
147 ($d \sim 660$ kpc) would require several  hundred orbits with
STIS.}

It seems unavoidable that a 10-meter class, diffraction limited
(at 8500 \AA) telescope\footnotemark~ will be needed to fully explore the incidence of SBHs as
small as $10^6$ and as large as $10^{10}$ \msun, or to target galaxies
fully sampling the Hubble sequence. If one wishes to reach galaxies in the most
nearby rich Abell clusters, and study the influence of environment on
the formation and evolution of SBHs, the requirements are even more
stringent, demanding a 30-meter class, diffraction limited
telescope. While there are currently no plans to replace \hst with a
new optical space telescope (\hstnsp's replacement, the James Webb Space
Telescope, will operate at near and mid-infrared wavelengths), several
ground based 10m class telescopes are already in operation (Keck,
Gemini, VLT, Subaru, etc.). Designs for 30m (CELT - the California
Extremely Large Telescope; GSMT - the Giant Segmented Mirror
Telescope; VLOT - the Very Large Optical Telescope) and 100m (OWL -
the Overwhelmingly Large Telescope) are at various stages of
development. For all of these telescopes, diffraction limited
performance in the near infrared using Adaptive Optics techniques is a
priority, and will soon become routine operation. In the near
infrared, stellar kinematics can be obtained through observations of
the CO bandhead  at 2.3 $\mu$m; at these wavelengths, a ground based,
AO-equipped,  8m-class telescope will have resolution comparable to
the one \hst achieves at the CaII triplet near 8500 \AA. The
availability of Integral Field Spectrographs (IFUs), several of which
are either commissioned or in operation, is another important
advantage of ground-based telescopes over \hstnsp, since a full
kinematical map is critical to fully understand the dynamical
structure of galactic cores (e.g. de Zeeuw 2003)

\footnotetext{This is a somewhat loose requirement, which is made by
simply scaling the performance of HST/STIS to a telescope with larger
collective area/resolution. It is possible that a ground based telescope with
adaptive optics working at 8500 \AA~ might prove sufficient - although
more detailed modeling, accounting for the PSF shape, should be carried
out to verify this point.}

Multiple-conjugate adaptive optics (MCAO), producing diffraction
limited images over wide field of views, are  being developed at several
existing observatories (e.g. Gemini and the VLT) in the near infrared. In
the more distant future, the hope is that MCAO will become a reality in
the optical as well. The increased sensitivity, combined with the
critical increase in resolving power, will provide the perfect
conditions for systematic and complete surveys of SBHs in the local
universe.

\begin{figure}[t]
\centerline{\parbox[l][10cm]{9.2cm}{\vskip 3cm THIS FIGURE IS INCLUDED IN THE FULL VERSION OF THE
MANUSCRIPT AVAILABLE AT http://www.physics.rutgers.edu/~lff/publications.html.}}
%\centerline{\psfig{figure=f44.eps,height=10.0 cm}}
\caption{SBH mass vs. distance for all galaxies in the CfA Redshift
Sample (Huchra \etal 1990). Only for the galaxies which lie above the
solid lines the sphere of influence of the putative nuclear SBH can be
resolved by \hstnsp/STIS, an 8m and a 30m diffraction limited telescope.
A few nearby groups and clusters are marked. It should be noted that
because of the large scatter in the $\ml$ relation (a factor of
several in $\mh$), this figure has only statistical value. From
Ferrarese (2003).}
\label{hstlimit1}
\end{figure}

\begin{figure}[t]
\centerline{\psfig{figure=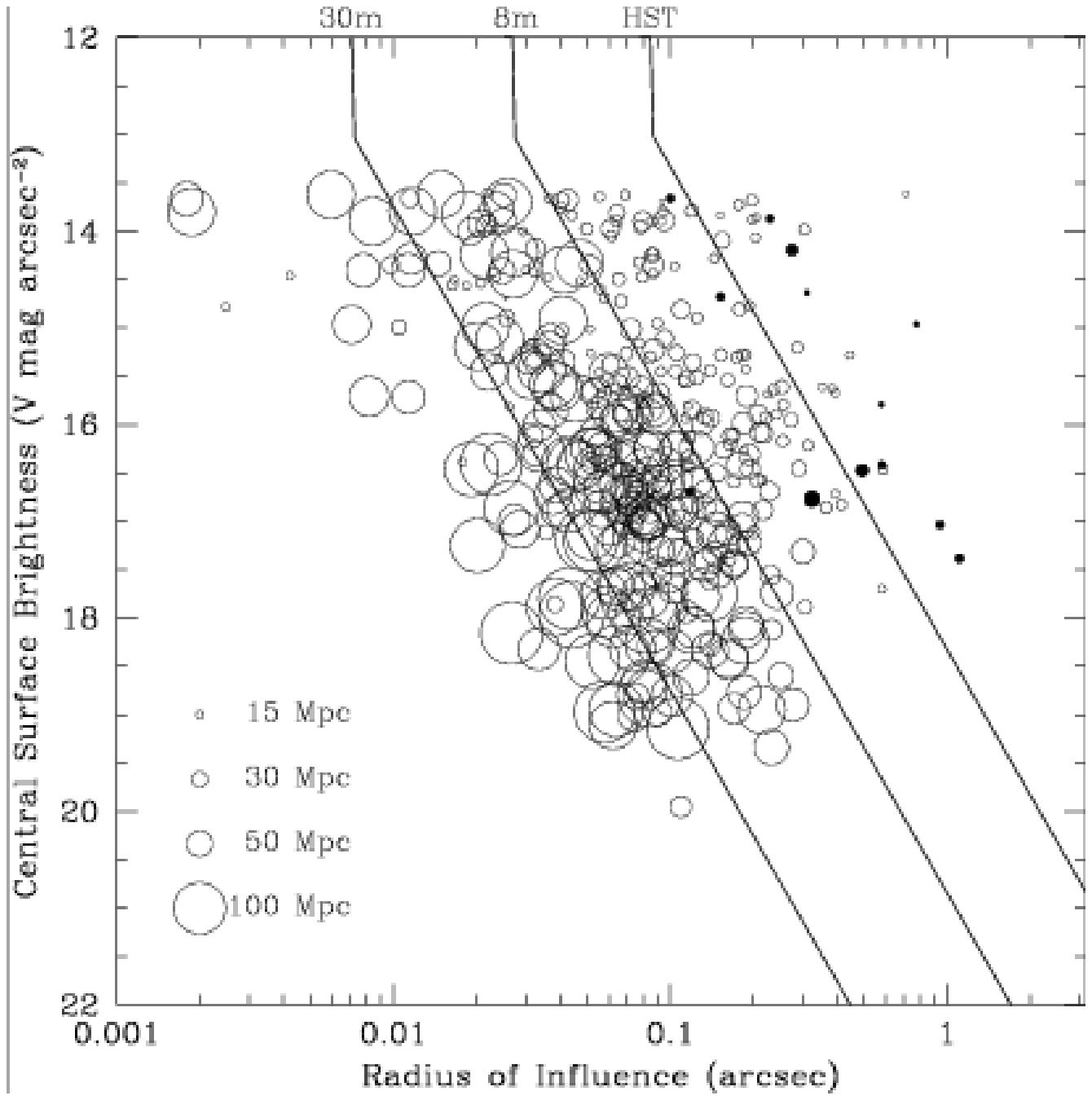,height=10.0 cm}}
\caption{SBH radius of influence vs. central surface brightness for
the Early Type galaxy sample of Faber \etal (1989). The size of the
symbols is proportional to the  galaxy distances, as shown in the
legend. Spectra with ideal signal-to-noise and spatial resolution for
dynamical studies can be collected in the equivalent of 3 \hst orbits
only for the galaxies to the right of the solid lines (shown for
\hstnsp/STIS, an 8m and a 30m diffraction limited telescopes). The solid
circles identify galaxies observed with \hstnsp. From Ferrarese (2003).}
\label{hstlimit2}
\end{figure}

Until then, reverberation mapping (\S~\ref{sec:revmap}) is likely to
replace traditional resolved dynamics as the method of choice to probe
the very smallest, largest and most distant SBHs.  The implementation
of queue scheduling at some large ground based optical facilities
(Gemini, SALT, and the VLT) will greatly increase the efficiency of the
monitoring programs.  The masses inferred from the time-delays will
receive a final stamp of approval once the morphology and kinematics
of the BLR are firmly established. The latter task, which requires
multi-wavelengths monitoring with very stringent conditions for the
time resolution and length of the program, will remain the prerogative
of dedicated space missions (Peterson \& Horne 2003; Peterson, Polidan
\& Robinson 2003).  There are three dozen galaxies for which time
delays have been measured to date, it doesn't seem unreasonable to
think that this sample could be doubled in the near future. Even then,
studying the redshift evolution of SBH scaling relations will more
realistically be done using secondary mass estimators tied to
reverberation mapping (\S~\ref{sec:sec}).

\begin{figure}[t]
\centerline{\psfig{figure=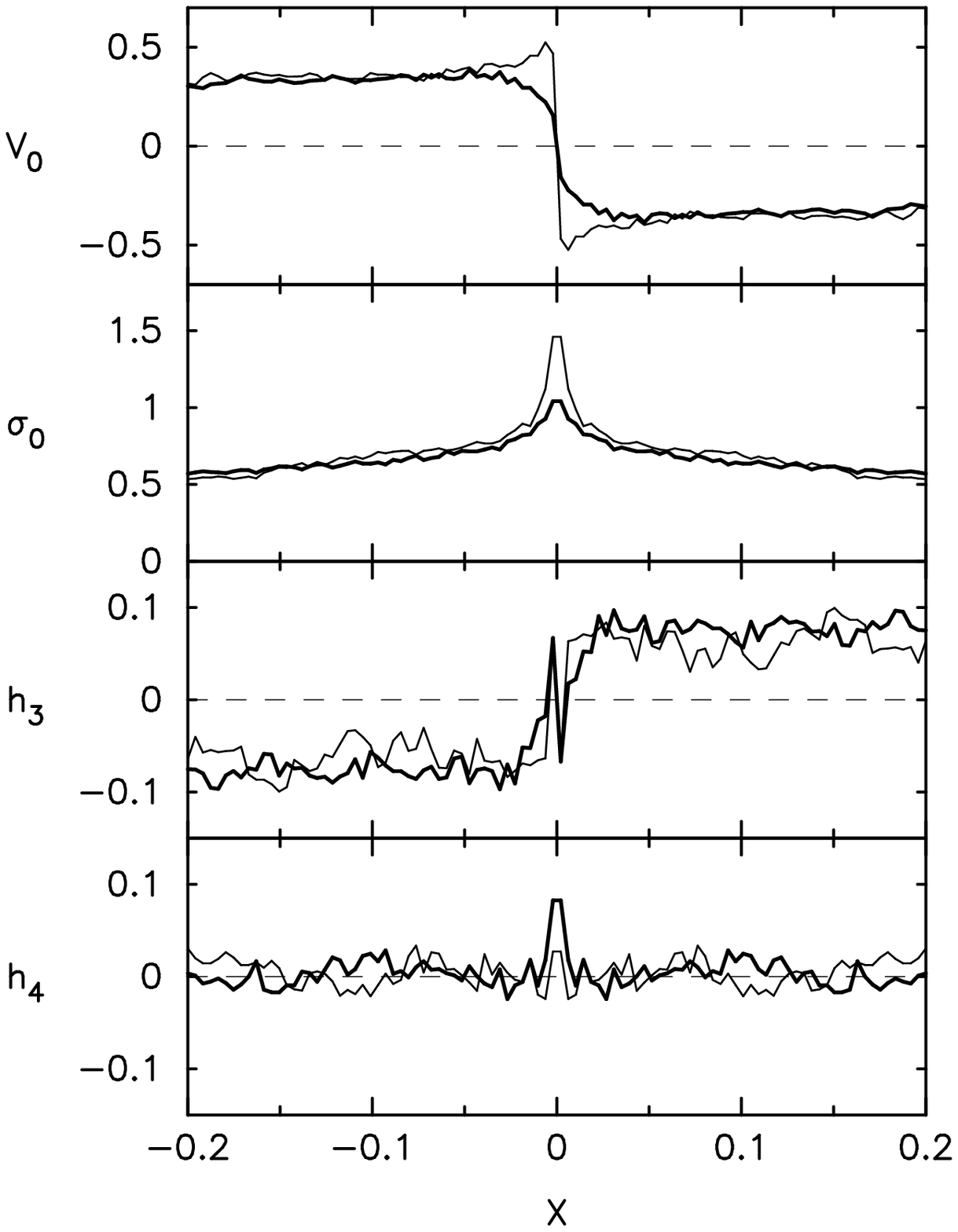,height=10.0 cm}}
\caption{Rotational Velocity, velocity dispersion and Gauss-Hermite expansion
of the LOSVD in the case of a BH binary (thick line) and a single BH
(thin line). From Milosavljevic \& Merritt (2002).}
\label{sbhbinary}
\end{figure}

Detecting SBH binaries is another important goal for future
research. The formation of SBH binaries following galaxy merging seems
unavoidable (Begelman, Blandford \& Rees 1980), and indeed the
coalescence of binary SBHs has been called into action to account for
the peculiar properties of the jets/lobes in some radio galaxies
(Merritt \& Ekers 2002; Liu, Wu \& Cao 2003). The existence of a SBH
binary has also recently been claimed in the radio galaxy 3C 66, based
on the positional change of the radio core detected in VLBI
observations (Sudou \etal 2003). The elliptical orbit traveled by the
core in a $1.05 \pm 0.03$ year period (apparently not a result of the
orbital motion of the Earth) is interpreted as due to a SBH binary
with a separation of $\sim 10^{17}$ cm, and total mass in the
neighborhood of $6 \times 10^{10}$ \msun (assuming that the radio core
is associated with the least massive of the SBHs). The presence of two
SBHs (about 1 kpc apart and therefore not strictly yet forming a
binary system) in the Ultraluminous Infrared Galaxy NGC 6240 is
betrayed by the Chandra X-ray detection of two optical/IR nuclei
(Komossa \etal 2003). From a dynamical prespective, the evolution
of a SBH binary is very uncertain (Ebisuzaki, Makino  \& Okumura 1991;
Milosavljevic \& Merritt 2002; Yu 2002), but current state-of-the-art
simulations (Milosavljevic \& Merritt 2002) predict the first stages of the
evolution, leading to the formation of a hard binary with separation
between a few hundredths to a few parsecs depending on the masses
involved, to proceed very rapidly, within one Myr after galaxy
merging. The subsequent evolution is dominated by the exchange of
energy between the binary and nearby stars, reducing the binary
separation to $\lae 0.2$ pc ($\propto \mh^{0.57}$) before the
hardening is stalled following the depletion of stars with which the
binary can interact.  Although on this scale the predicted stellar
rotational velocity and velocity dispersion differ significantly from
the case of a single SBH (Figure~\ref{sbhbinary}), resolving this
region in the optical requires a 100m-class diffraction limited
telescope even in the most nearby systems (Ferrarese 2003).

The ultimate proof of the existence of binary SBHs will likely not be made
using traditional methods. Indeed, progress in this field will
be tied to the most challenging, anticipated, and - in our view - exciting
of all future developments: the launch of 
gravitational wave missions, such as the proposed Laser
Interferometer Space Antenna (LISA, Phinney 2003). Such
missions will prove (or disprove) the existence of black
hole binaries by recording (or failing to record) their coalescence. They
will do so by  detecting the purely relativistic signature imprinted by the event 
on the fabric of space time, i.e. by observing the very 
signature of a strong field regime which has so far eluded all investigations. 
We started this review by identifying the ``birth" of the modern era of BH studies with the
1915 publication of  Albert Einstein's theory of general
relativity. It seems only fitting that we should return to Einstein's contribution
in our concluding remarks.

In 1916 Einstein predicted that
accelerating masses will emit gravitational waves that are manifested
as ripples (changes in the curvature) propagating through the fabric
of spacetime. 
Because the gravitational wave draws its energy from the 
energy of the binary, the objects in
the binary will spiral in with an ever-decreasing orbital
period, and will eventually coalesce. Because spacetime is very  ``stiff'', 
the amplitude of the
ripples, and the corresponding stretching and compression of objects
imbedded in spacetime, is exceedingly small and difficult to detect. A
neutron star (NS) binary system at 300 Mpc produces a spacetime strain
$h \sim 10^{-21}$ (Thorne 1994; Barish 2000), where the change in length by
compression and stretching is given by  $\Delta L = hL$. The amplitude of the
wave that produces the strain is directly proportional to the mass of
the accelerating object, and, in contrast to electromagnetic
radiation, inversely proportional to the distance to the source
(e.g. de Araujo, Miranda, \& Aguiar 2001). A gravitational wave with a
strain of $10^{-21}$ moving through the solar system will change the mean
Earth-Moon separation by $\sim4 \times 10^{-8}$ microns! The endeavors to measure
gravitation waves are testimonies to the ingenuity, sophistication,
and boldness of physicists. The payoff from the formidable task of
detecting gravitational waves will be an unprecedented opportunity to
observe and study general relativistic phenomena {\it in extremis}.

A worldwide network
of terrestrial gravitational wave observatories is already at work; the American LIGO
(Barish 2000; Abbott, B. 2004), the British-German GEO600 detector
(Danzmann, K. 1995; Willke et al. 2002; Abbott et al. 2004), the
French-Italian VIRGO detector (Acernese et al. 2002), the Japanese
TAMA300 detector (Ando et al. 2002), and five resonant-bar detectors
(Allen et al. 2000). These experiments are designed to detect
the characteristic ``chirp''  (rapid increase in frequency) produced by
the merger of neutron stars (NS), stellar 
black holes, and by the formation of black holes
in asymmetrical supernova explosions within tens of Mpc (Abbott, B. et
al. 2003). For instance, LIGO, which consists of two independent optical laser interferometers 
each with two equal 4-km long (1-m diameter) evacuated arms in an L-configuration 
(Barish 2000; Abbott, B. 2004), has a design
sensitivity $h_{rms}(\nu) \leq 10^{-21}$ (equivalent RMS strain amplitude spectral
density) for the strain (differential fractional change in the length
of the two arms) induced by gravitational waves with frequencies
between $\sim$30 Hz and 7000 Hz (Abbott et al. 2004).
Depending on the uncertain rate of NS/NS and BH/BH mergers in
galaxies (Lipunov, Postnov, \& Prokhorov 1997; Kalogera et al. 2001),
LIGO will be able to detect NS/NS mergers out to ~20 Mpc (Abbott et
al. 2004) when it achieves its design sensitivity. 
LIGO's first data run in 2002, which lasted for 17 days,  
was sensitive to binary
inspiral events at maximum distances between 30 and 180 Kpc, depending
on the instrument configurations during the run. 
The run established an upper limit $R < 170$ events per year per
equivalent Milky Way Galaxy (Abbott et al. 2003). 

Below $\sim$10 Hz, the critical range where gravitational waves from the coalescence of SBH binaries are emitted,  ground-based gravitational wave
observatories are crippled by the seismic noise
floor.  Space, once again, represents the answer. The next
extraordinarily ambitious step in gravitational wave astronomy is the
proposed joint NASA/ESA mission to launch a Laser Interferometer Space
Antenna (LISA) in ~2015. LISA will consist of three spacecraft at the
corners of an equilateral triangle with $5 \times 10^6$ km arms in an orbit
that trails the earth by 20 degrees (Danzmann 2000). The angle between
the plane of the triangle and the ecliptic will be 60\deg. Each
spacecraft has two interferometers in a Y-shaped arm that send/receive
a laser beam to/from each of the other spacecraft. The remarkable
return mirror in each of the interferometer arms is a highly polished,
free-flying 4-cm platinum-gold cube. The cubes within their vacuum
enclosures are maintained in a drag-free inertial orbit by sensing the
cubes  positions relative to the spacecraft and moving the spacecraft
with micro-newton field emission  thrusters  to follow the cubes. The
spacecraft must follow the orbits of the reference cubes with an
accuracy of 10 nanometers, and measure the distance between the
spacecraft with an accuracy of $10^{-5}$ microns. When these challenges are
met, LISA will be able to measure strains of $10^{-23}$ (SNR = 5) at
frequencies from $10^{-4}$ Hz to 1 Hz with a year of observations (Danzmann
2000).

LISA then will be able to observe the formation and merger of SBHs
throughout the Universe. With persistence and sustained funding,
gravitational wave observatories will usher in a new era in
relativistic astrophysics that will allow us to observe the most
extreme and most energetic events in the Universe.


\begin{thebibliography}{}


\bibitem[]{} Abbott, B., et al. 2003, arXiv:grqc/ 0308069 v1 21 Aug 2003 

\bibitem[]{} Abbott, B., et al. 2004, NIMPA, 517, 154 

\bibitem[]{} Acernese, F., et al. 2002, CQGra, 19, 1421

\bibitem[]{} Adams, F. C., Graff, D. S., \& Richstone, D. 2000, ApJ, 551, L31

\bibitem[]{} Akritas, M. G., \& Bershady, M. A. 1996, ApJ, 470, 706

\bibitem[]{} Alexander, D.M., et al. 2001, ApJ, 122, 2156

\bibitem[]{} Allen, C.R., etal. 1962, MNRAS, 124, 477

\bibitem[]{} Allen, Z.A., et al. 2000, PhRvL, 85, 5046 

\bibitem[]{} Alonso-Herrero, A., Ivanov, V.D., Jayawardhana, R., \& Hosokawa, T. 2002, ApJ, 571, L1

\bibitem[]{} Ando, M., et al. 2002, CQGra, 19, 1409 

\bibitem[]{} Antonucci, R.R.J. 1993, ARA\&A, 31, 473

\bibitem[]{} Antonucci, R.R.J., \& Miller, J.S. 1985, ApJ, 297, 621.

\bibitem[]{} Arav, N., Barlow, T.A., Laor, A., \& Blandford, R.D. 1997, MNRAS, 288, 1015

\bibitem[]{} Arav, N., Barlow, T.A., Laor, A., Sargent, W.L.W., \& Blandford, R.D. 1998, MNRAS, 297, 990

\bibitem[]{} Baade, W., \& Minkowski, R. 1954, ApJ, 119, 206

\bibitem[]{} Baade, W. 1955 ApJ, 123, 550

\bibitem[]{} Bacon, R., Emsellem, E., Combes, F., Copin, Y., Monnet, G., \& Martin, P.  2001, A\&A, 371, 409

\bibitem[]{} Baes, M., Buyle, P., Hau, G.K.T., \& Dejonghe, H. 2003, MNRAS, 341, L44

\bibitem[]{} Balick, B., \& Brown, R.L.\ 1974, ApJ, 194, 264

\bibitem[]{} Bao, G., \& Abramowicz, M.A. 1996, ApJ, 465, 646

\bibitem[]{} Barger, A. J. \etal 2001, AJ, 122, 2177

\bibitem[]{} Barger, A. J. \etal 2002, AJ, 124, 1839

\bibitem[]{} Barish, B.C. 2000, AdSpR, 25, 1165 

\bibitem[]{} Barth, A. \etal 2001, ApJ, 555, 685

\bibitem[]{} Barth, A., Ho, L.C., \& Sargent,  W.L.W. 2002, ApJ, 566, L13

\bibitem[]{} Barth, A., Ho, L.C., \& Sargent,  W.L.W. 2003, ApJ, 583, 134

\bibitem[]{} Begelman, M. C., Blandford, R. D., \& Rees, M. J. 1980, Nature, 287, 307

\bibitem[]{} Begelman, M.C. 2001, ApJ, 551, 897

\bibitem[]{} Belloni, T., \& Hasinger, G. 1990, A\&A, 227, L33

\bibitem[]{} Bernardi M., et al., 2003, AJ, 125, 1817

\bibitem[]{} Bian, W.-H., \& Zhao, Y.-H. 2003, PASJ, 55, 143

\bibitem[]{} Binney, J., \& Mamon, G.A. 1982, MNRAS, 200, 361

\bibitem[]{} Binney, J. \& Tremaine, S. 1987 {\it Galactic Dynamics}, Princeton University Press (Princeton)

\bibitem[]{} Blandford, R.D., \& McKee, C.F. 1982, ApJ, 255, 419 R.D., \& Begelman, M.C. 1999, MNRAS, 303, L1

\bibitem[]{} Blandford, R. D., \& Begelman, M. C.\ 1999, MNRAS, 303, L1

\bibitem[]{} Blumenthal, G.R., \& Mathews, W.G. 1975, ApJ, 198, 517

\bibitem[]{} Bolton, C.T. 1972, Nature, 235, 271 

\bibitem[]{} Boroson, T.A. 2002, ApJ, 565, 78

\bibitem[]{} Bottorff, M., Korista, K.T., Shlosman, I., \& Blandfrod, R.D. 1997, ApJ, 479, 200

\bibitem[]{} Boyle, B.J., \etal, 2000, MNRAS, 317, 1014

\bibitem[]{} Bower, G., \etal 1998, ApJ, 492, L111

\bibitem[]{} Braatz, J.A., Wilson, A.S., \& Henkel, C. 1994, ApJ, 437, L99

\bibitem[]{} Braatz, J.A., Wilson, A.S., \& Henkel, C. 1996, ApJS, 106, 51

\bibitem[]{} Braatz, J.A., Wilson, A.S., \& Henkel, C. 1997, ApJS, 110, 321

\bibitem[]{} Brucato, R.J., \& Kristian, J. 1972, ApJ, 173, 105

\bibitem[]{} Bryan, G. L., \& Norman, M. L. 1998, ApJ, 495, 80

\bibitem[]{} Bullock, J.S., \etal 2001, MNRAS, 321, 559

\bibitem[]{} Burbidge, G.R. 1962, in `Problems of Extra-Galactic Research'
Proceedings from IAU Symposium no. 15. Eds  George Cunliffe
McVittie. Macmillan Press, New York, p.258

\bibitem[]{} Cappellari, M. \etal 2002, 578, 787

\bibitem[]{} Cattaneo, A., Haehnelt, M. G., \& Rees, M. J. 1999, MNRAS, 308, 77

\bibitem[]{} Cecil, G., Wilson, A.S., \& Tully, R.B. 1992, ApJ, 390, 365 

\bibitem[]{} Cecil, G., et al. 2000, ApJ, 536, 675

\bibitem[]{} Chandrasekhar, S. 1931, ApJ, 74, 81

\bibitem[]{} Chandrasekhar, S. 1935, MNRAS, 95, 226

\bibitem[]{} Cheung, \etal 1969, Nature, 221, 626

\bibitem[]{} Chiang, J., \& Murray, N. 1996, ApJ, 466, 704

\bibitem[]{} Chokshi, A. \& Turner, E. L., 1992, MNRAS, 259, 421

\bibitem[]{} Cipollina, M., \& Bertin, G. 1994, A\&A, 288, 43

\bibitem[]{} Claussen, M.J., Hellingman, G.M., \& Lo, K.Y., 1984, Nature, 310, 298

\bibitem[]{} Claussen, M.J., \& Lo, K.Y., 1986, ApJ, 308, 592

\bibitem[]{} Collin-Souffrin, S., Dyson, J.E., McDowell, J.C., Perry, J.J. 1988, MNRAS, 232, 539

\bibitem[]{} Cowie, L.L., et al. 2003, ApJ, 584, L57

\bibitem[]{} Crane, P., \etal 1993, AJ, 106, 1371

\bibitem[]{} Crawford, M. K., Genzel, R., Harris, A. I., Jaffe, D. T., Lugten, J. B., Serabyn, E., Townes, C. H., \& Lacy, J. H.\ 1985, Nature, 315, 467

\bibitem[]{} Cretton, N., \& van den Bosch, F.C. 1999, ApJ, 514, 704

\bibitem[]{} Croom S. M., et al., 2003, astroph/0403040

\bibitem[]{} Curtis, H.D. 1918, Pub.Lick.Obs., 13, 31

\bibitem[]{} Dabrowski, Y., Fabian, A.C., Iwasawa, K., Lasenby, A.N., Reynolds, C.S. 1997, MNRAS, 288, L11

\bibitem[]{} Danzmann, K. 1995,  GEO600 - A 600-m Laser Interferometric Gravitational Wave Antenna,  in First Edoardo Amaldi Conference on gravitational wave experiments E. Coccia, G. Pizella and F. Ronga, eds., (World Scientific, Singapore), p.100-111 

\bibitem[]{} Danzmann, K. 2000, AdSpR, 25, 1129

\bibitem[]{} de Araujo, J.C.N., Miranda, O.D, \& Aguiar, O.D. 2001, ApJ, 550, 368

\bibitem[]{} Dettmar, R.-J., \& Koribalski, B. 1990, A\&A, 240, 15

\bibitem[]{} de Vaucouleurs, G., de Vaucouleurs, A., Corwin, H. G., Jr., Buta, R. J., Pasturel, G., \& Fouqué, P. 1991, Third Reference Catalogue of Bright Galaxies (New York: Springer)

\bibitem[]{} Devereux, N., Ford, H.C., Tsvetanov, Z., \& Jacoby, G. 2003, AJ, 125, 1226

\bibitem[]{} de Zeeuw, T. 2003, in "Hubble Science Legacy: Future Optical/Ultraviolet Astronomy from Space", eds. K.R.Sembach, J.C.Blades, G.D.Illingworth, \& R.C. Kennicutt, ASPC Conference Series, vol. 291, P. 205.

\bibitem[]{} Dietrich, M., Wagner, S.J., Courvoisier, T.J.-L., Bock, H., \& North, P. 1999, A\&A, 351, 31

\bibitem[]{} di Matteo, T., Fabian, A.C., Rees, M.J., Carilli, C.L., \& Ivison,
R.J. 1999, MNRAS, 305, 492

\bibitem[]{} Doeleman, S.S., et al. 2001, AJ, 121, 2610

\bibitem[]{} Done, C., \& Krolik, J.H. 1996, ApJ, 463, 144

\bibitem[]{} Done, C., Madejski, G.M., \& Zycki, P.T. 2000, ApJ, 536, 213

\bibitem[]{} dos Santos, P.M. and Lepine, J.R.D. 1979, Nature, 278, 34

\bibitem[]{} Ebisuzaki, T., Makino, J., \& Okumura, S. K. 1991, Nature,354, 212

\bibitem[]{} Eckart, A., Genzel, R., Hofmann, R., Sams, B. J., Tacconi-Garman, L. E. 1993, ApJ, 407, L77

\bibitem[]{} Eddington, A.S. 1930, `The internal Constitution of Stars', 
Cambridge University Press, Cambridge, England

\bibitem[]{} Edelson, R., \& Nandra, K. 1999, ApJ, 514, 682

\bibitem[]{} Einstein, A. 1915, {\it Sitzungsberichte der Deutschen Akademie der
Wissenschaften zu Berlin, Klasse fur Mathematik, Physik, und Technik},
844

\bibitem[]{} Eisenhauer, F., Sch\"odel, R., Genzel, R., Ott, T., Tecza, M., Abuter, R., Eckart, A., \& Alexander, T. 2003, ApJ, 597, L121

\bibitem[]{} Ekers, R. D., Goss, W. M., Schwarz, U. J., Downes, D., \& Rogstad, D. H. 1975, A\&A, 43, 159

\bibitem[]{} Elvis M., et al. 1994, ApJS, 95, 1

\bibitem[]{} Elvis, M., Risaliti, G., \& Zamorani, G. 2002, ApJ, 565, 75

\bibitem[]{} Emmering, R.T., Blandford, R.D., \& Schlosman, I. 1992, ApJ, 385, 460

\bibitem[]{} Emsellem, E., \etal 1999, MNRAS, 303, 495

\bibitem[]{} Faber, S. M., \etal 1989, ApJS, 69, 763

\bibitem[]{} Fabian, A.C., \& Canizares, C.A 1988, Nature, 333, 829

\bibitem[]{} Fabian, A.C., Rees, M.J., Stella, L., \& White, N. 1989, MNRAS, 238, 729

\bibitem[]{} Fabian, A.C., \etal 1995, MNRAS, 277, L11

\bibitem[]{} Fabian, A. C. \& Iwasawa, K. 1999, MNRAS, 303, L34, 121, 662

\bibitem[]{} Fabian, A.C., Iwasawa, K., Reynolds, C.S., \& Young, A.J. 2000, PASP, 112, 1145

\bibitem[]{} Fabian, A.C. 2003, Astron.Nach., 324, 4

\bibitem[]{} Falcke, H., Henkel, Chr., Peck, A. B., Hagiwara, Y., Almudena Prieto, M., \& Gallimore, J. F.\ 2000, A\&A, 358, L17

\bibitem[]{} Falomo, R., Kotilainen, J.K., Carangelo, N., \& Treves, A. 2003, ApJ, 595, 624

\bibitem[]{} Fan, X., \etal, AJ, 2001, 121, 54

\bibitem[]{} Ferland, G.J., Peterson, B.M., Horne, K., Welsh, W.F., \& Nahar, S.N. 1992, ApJ, 387, 95

\bibitem[]{} Ferrarese, L.. van den Bosch, F.C.. Ford, H.C., Jaffe, W. \& O'Connell, R.W. 1994, AJ, 108, 1598

\bibitem[]{} Ferrarese, L., Ford, H.C., \& Jaffe, W. 1996, ApJ, 470, 444	

\bibitem[]{} Ferrarese, L., \& Ford, H.C. 1999, ApJ, 515, 583

\bibitem[]{} Ferrarese, L. \& Merritt, D., 2000, ApJ, 539 L9

\bibitem[]{} Ferrarese, L., Pogge, R.W., Peterson, B.M., Merritt, D., Wandel, A. \& Joseph, C.M. 2001, ApJ, 555, L79

\bibitem[]{} Ferrarese, L. 2002a, in `Current high-energy emission around black
holes' Proceedings of the 2nd KIAS Astrophysics Workshop,  Eds
Chang-Hwan Lee and Heon-Young Chang. Singapore: World Scientific
Publishing,  p.3

\bibitem[]{} Ferrarese, L. 2002b, ApJ, 578, 90

\bibitem[]{} Ferrarese, L. 2003, in "Hubble Science Legacy: Future Optical/Ultraviolet Astronomy from Space", eds. K.R.Sembach, J.C.Blades, G.D.Illingworth, \& R.C. Kennicutt, ASPC Conference Series, vol. 291, P. 196.

\bibitem[]{} Field, G.B., 1964, ApJ, 140, 1434

\bibitem[]{} Filippenko, A.V., \& Ho, L.C. 2003, ApJ, 588, L13

\bibitem[]{} Ford, H.C., Dahari, O., Jacoby, G.H., Crane, P.C., \& Ciardullo, R. 1986, ApJ, 311, L7.

\bibitem[]{} Ford, H.C., \etal 1994, ApJ, 435, L27

\bibitem[]{} Fowler, R.H. 1926, MNRAS, 87, 114

\bibitem[]{} Franceschini, A., Vercellone, S., \& Fabian, A.C. 1998, MNRAS, 297, 817

\bibitem[]{} Fukugita, M., Hogan, C.J., \& Peebles, P.J.E. 1998, ApJ, 503, 518

\bibitem[]{} Gallimore, J.F., Baum, S.A., O'Dea, C.P.,  \& Pedlar, A. 1996a, ApJ, 458, 136 

\bibitem[]{} Gallimore, J.F., Baum, S.A., O'Dea, C.P., Brinks, E., \& Pedlar, A. 1996b, ApJ, 462, 740

\bibitem[]{} Gallimore,J.F., Baum, S.A., \& O'Dea, C.P. 1996c, ApJ, 464, 198 

\bibitem[]{} Gallimore,J.F., Baum, S.A., \& O'Dea, C.P. 1997, Ap\&SS, 248, 253 

\bibitem[]{} Gardner, F.F., \& Whiteoak, J.B. 1982, MNRAS, 201, 13

\bibitem[]{} Gebhardt, K., \etal 2000a, AJ, 119, 1157

\bibitem[]{} Gebhardt, K., \etal, 2000b, ApJ, 539, L13

\bibitem[]{} Gebhardt, K., \etal, 2000c, ApJ, 543, L5

\bibitem[]{} Gebhardt, K., \etal, 2003, ApJ, 583, 92

\bibitem[]{} Genzel, R., \& Townes, C. H.\ 1987, ARA\&A, 25, 377

\bibitem[]{} George, I.M., \& Fabian, A.C. 1991, MNRAS, 249, 352

\bibitem[]{} Gerhard, O. 1993, MNRAS, 265, 213

\bibitem[]{} Gerhard, O., Kronawitter, A., Saglia, R. P., \& Bender, R. 2001, AJ, 121, 1936

\bibitem[]{} Ghez, A. M., Klein, B. L., Morris, M., \& Becklin, E. E.\ 1998, ApJ, 509, 678

\bibitem[]{} Ghez, A. M., Morris, M., Becklin, E. E., Tanner, A., \& Kremenek, T.\ 2000, Nature, 407, 349

\bibitem[]{} Ghez, A.M., et al.\ 2003, ApJ, 586, L127

\bibitem[]{} Gilli, G.M., Salviati, M., \& Hasinger, G., 2001, A\&A, 366, 407

\bibitem[]{} Gondoin, P., Barr, P., Lumb, D., Oosterbroek, T., Orr, A., Parmar, A.N. 2001a, A\&A 378, 806

\bibitem[]{} Gondoin, P., Lumb, D., Siddiqui, H., Guainazzi, M., Schartel, N., 2001b, A\&A 373, 805

\bibitem[]{} Graham, A.W., Erwin, P., Caon, N., \& Trujillo, I. 2001, ApJ, 563, L11

\bibitem[]{} Greenhill, L.J., Jiang, D.R., Moran, J.M., Reid, M.J., Lo, K.Y., \& Claussen, M. J. 1995a, ApJ, 440, 619

\bibitem[]{} Greenhill, L.J., Henkel, C., Becker, R., Wilson, T.L., \& Wouterloot, J.G.A. 1995b, A\&A, 304, 21

\bibitem[]{} Greenhill, L.J., Gwinn, C.R., Antonucci, R., \& Barvainis, R. 1996, ApJ, 472, L21

\bibitem[]{} Greenhill, L.J. \& Gwinn, C.R. 1997, Astrophysics and Space Science, v. 248, 261

\bibitem[]{} Greenhill, L.J., Moran, J.M., \& Herrnstein, J.R. 1997, ApJ, 481, L23

\bibitem[]{} Greenhill, L., \etal 1997, ApJ, 486, L15

\bibitem[]{} Greenhill, L., \etal 2002, ApJ, 565, 836

\bibitem[]{} Greenhill, L., \etal 2003a, ApJ, 582, L11

\bibitem[]{} Greenhill, L., \etal 2003b, ApJ, 590, 162

\bibitem[]{} Greenstein, J.L., \& Schmidt, M. 1964, ApJ, 140, 1

\bibitem[]{} Haardt, F., \& Maraschi, L. 1991, ApJ, 380, L51

\bibitem[]{} Haardt, F., \& Maraschi, L. 1993, ApJ, 413, 507

\bibitem[]{} Haehnelt, M. G., Natarajan, P., \& Rees, M. J. 1998, MNRAS, 300, 817

\bibitem[]{} Haehnelt, M. G., \& Kauffmann, G. 2000, MNRAS, 318, L35

\bibitem[]{} Hagiwara, Y., Henkel, C., Menten, K.,M., \& Nakai, N. 2001a, ApJ, 560, L37

\bibitem[]{} Hagiwara, Y., Diamond, P.J., Nakai, N., \& Kawabe, R. 2001b, ApJ, 560, 119

\bibitem[]{} Hagiwara, Y., Diamond, P. J., \& Miyoshi, M.\ 2002, A\&A, 383, 65

\bibitem[]{} Harms, R.J., \etal 1994, ApJ, 435, L35

\bibitem[]{} Harrower, G.A. 1960, ApJ, 132, 22

\bibitem[]{} Harwit, M. 1998, ``Astrophysical concepts'', New York: Springer.

\bibitem[]{} Haschick, A.D., \& Baan, W.A. 1990, ApJ, 355, L23

\bibitem[]{} Haschick, A.D., Baan, W.A., \& Peng, E.W. 1994, ApJ, 437, L35

\bibitem[]{} Hasinger, G. 2002, in ``X-ray astronomy in the new millennium'', R. D. Blandford, A. C. Fabian and K. Pounds eds. Roy Soc of London Phil Tr A, vol. 360, Issue 1798, p.2077

\bibitem[]{} Hazard, MacKay \& Shimmins 1965, in `` Quasi-Stellar Sources and Gravitational Collapse'', Proceedings of the 1st Texas Symposium on Relativistic Astrophysics. Ivor Robinson, Alfred Schild and E.L. Schucking eds. Chicago: University of Chicago Press, p.448

\bibitem[]{} Henkel, C., Guesten, R., Downes, D., Thum, C., Wilson, T.L., \& Biermann, P. 1984, A\&A, 141, L1

\bibitem[]{} Henkel, C., Braatz, J.A., Greenhill, L.J., \& Wilson, A.S. 2002, A\&A, 394, L23

\bibitem[]{} Heraudean, Ph., Simien, F. 1998 A\&AS 133, 317

\bibitem[]{} Herrnstein, J.R., Greenhill, L.J., \& Moran, J.M. 1996, ApJ, 468, L17

\bibitem[]{} Herrnstein, J.R., \etal 1999, Nature, 400, 539

\bibitem[]{} Ho, L. C. 1999, in {\em Observational Evidence for Black Holes in the Universe}, eds. S. K. Chakrabarti (Dordrecht: Reidel), p. 157

\bibitem[]{} Ho, L.C. 2002, ApJ, 564, 120

\bibitem[]{} Hoyle, F., \& Fowler, W.A. 1963, MNRAS, 125, 169

\bibitem[]{} Hubeny, I., Agol, E., Blaes, O., \& Krolik, J.H. 2000,  ApJ, 533, 710

\bibitem[]{} Huchra, J. P., Geller, M. J., de Lapparent, V., \& Corwin, H. G., Jr.\ 1990, ApJS, 72, 433

\bibitem[]{} Illingworth, G. 1977, ApJ, 218, L43

\bibitem[]{} Ishihara, Y., Nakai, N., Iyomoto, N., Makishima, K., Diamond, P., \& Hall, P. 2001, PASJ 53, 215

\bibitem[]{} Iwasawa, K., \etal 1996, MNRAS, 281, L41

\bibitem[]{} Jaffe, W., Ford, H.C., Ferrarese, L., van den Bosch, F., \& O'Connell, R.W. 1993, Nature, 364, 213

\bibitem[]{} Jennison, R.C., \& Das Gupta M.K. 1953, Nature, 172, 996

\bibitem[]{} Jones, D.L., Wehrle, A.E., Meier, D.L., \& Piner, B.G. 2000, ApJ, 534, 165

\bibitem[]{} Jorgensen, I., Franx, M., \& Kjaergaard 1995, MNRAS, 276, 1371

\bibitem[]{} Kalogera, V., R. Narayan, R., Spergel, D.N. \& Taylor, J.H. 2001, Astrophys. J. 556, 340 

\bibitem[]{} Kaspi, S., Smith, P.S., Netzer, H., Maoz, D., Jannuzi, B.T., \& Giveon, U. 2000, ApJ, 533, 631

\bibitem[]{} Kollatschny, W. 2003, A\&A, 407, 461

\bibitem[]{} Komossa, S. \etal 2003, ApJ, 582, L15

\bibitem[]{} Koratkar, A. P., \& Gaskell, C. M. 1991, ApJ, 370, L71

\bibitem[]{} Kormendy, J., \etal 1988, ApJ, 335, 40

\bibitem[]{} Kormendy, J., \& Richstone, D. 1992, ApJ, 393, 559

\bibitem[]{} Kormendy, J., \& Richstone, D., 1995, ARA\&A, 581

\bibitem[]{} Kormendy, J., \& Gebhardt, K. 2001, in AIP Conf. Proc. 586, The 20th Texas Symposium on Relativistic Astrophysics, ed. J. C. Wheeler \& H. Martel (Melville: AIP), 363

\bibitem[]{} Krolik, J.H. 1998, in "Theory of Black Hole Accretion Disks", eds. Marek A. Abramowicz, Gunnlaugur Bjornsson, and James E. Pringle. Cambridge University Press, p.134

\bibitem[]{} Krolik, J.H. 2001, ApJ, 551, 72

\bibitem[]{} Kronawitter, A., Saglia, R. P., Gerhard, O., \& Bender, R. 2000, A\&AS, 144, 53

\bibitem[]{} Lacy, J. H., Townes, C. H., Geballe, T. R., \& Hollenbach, D. J.\ 1980, ApJ, 241, 132

\bibitem[]{} Lacy, J. H., Townes, C. H., \& Hollenbach, D. J.\ 1982, ApJ, 262, 120

\bibitem[]{} Lacy, M. \etal 2001, ApJ, 551, L17

\bibitem[]{} Landau, L. D.\ 1932, Phys. Z. Sowjetunion, 1, 285

\bibitem[]{} Laor, A. 1991, ApJ, 376, 90

\bibitem[]{} Laor, A. 2000, ApJ, 543, L111

\bibitem[]{} Lauer, T.R. 1995, AJ, 110, 2622

\bibitem[]{} Lee, M./H., \& Goodman, J. 1989, ApJ, 343, 594

\bibitem[]{} Lipunov, V.M., Postnov, K.A., \& Prokhorov, M.E. 1997, NewA., 2, 43 

\bibitem[]{} Liu, F. K., Wu, X.-B., \& Cao, S.L. 2003, MNRAS, 340, 411

\bibitem[]{} Loeb, A., \& Rasio, F. 1994, ApJ, 432, L52

\bibitem[]{} Lynden-Bell, D. 1969, Nature, 223, 690Mauder, H. 1973, A\&A, 28, 473

\bibitem[]{} Macchetto, D.F., \etal 1997, ApJ, 489, 579

\bibitem[]{} Maciejewski, W., \& Binney, J. 2001, MNRAS, 323, 831

\bibitem[]{} Magorrian, J., et al.\ 1998, AJ, 115, 2285

\bibitem[]{} Maiolino, R., \& Rieke, G.H. 1995, ApJ, 454, 95

\bibitem[]{} Malkan, M.A., \& Sargent, W.L.W. 1982, ApJ, 254, 22

\bibitem[]{} Malkan, M.A. 1983, ApJ, 268, 582

\bibitem[]{} Maoz, E. 1998, ApJL, 494, 181

\bibitem[]{} Marchesini, D., Ferrarese, L. \& Celotti, A., 2004, MNRAS, in press

\bibitem[]{} Marconi, A., \etal 2001, ApJ, 549, 915

\bibitem[]{} Marconi, A., \& Salvati, M. 2002, in "Issues in Unification of AGNs", R. Maiolino, A. Marconi and N. Nagar eds., ASP Conf. Series

\bibitem[]{} Marconi, A., \& Hunt, L.K. 2003, ApJ, 589, L21

\bibitem[]{} Marconi, A., \etal 2003, ApJ, 586, 868

\bibitem[]{} Marconi A., Risaliti G., Gilli R., Hunt L. K., Maiolino R., \& Salvati M., 2004, MNRAS, 351, 169

\bibitem[]{} Martocchia, A., \& Matt, G. 1996, MNRAS, 282, L53

\bibitem[]{} Marzke, R.O., \etal 1998, ApJ, 503, 617

\bibitem[]{} Matt, G., Perola, G.C., \& Piro, L. 1991, A\&A 247, 25

\bibitem[]{} Matt, G., Perola, G.C., Piro, L., \& Stella, L. 1992, A\&A 257, 63

\bibitem[]{} Mathur, S. 2000, NewAR, 44, 469

\bibitem[]{} Mauder, H. 1973, A\&A, 28, 473

\bibitem[]{} McHardy, I., 1988, Mem.Soc.Astr.It, 59, 239

\bibitem[]{} McLure, R.J., \& Dunlop, J.S. 2000, MNRAS, 317, 249

\bibitem[]{} McLure, R.J., \& Dunlop, J.S. 2001, MNRAS, 327, 199

\bibitem[]{} McLure, R.J., \& Jarvis, M.J. 2002, MNRAS, 337, 109

\bibitem[]{} McLure R. J., Dunlop. J. S., 2004, MNRAS, in press

\bibitem[]{} Melia, F., \& Falcke, H.\ 2001, ARA\&A, 39 309

\bibitem[]{} Merritt, D., \& Ferrarese, L. 2001a, MNRAS, 320, L30

\bibitem[]{} Merritt, D., \& Ferrarese, L. 2001b, ApJ, 547, 140

\bibitem[]{} Merritt, D., \& Ferrarese, L. 2001c, in ``The Central Kiloparsec of Starbursts and AGN: The La Palma Connection'', ASP Conference Proceedings Vol. 249. J. H. Knapen, J. E. Beckman, I. Shlosman, and T. J. Mahoney eds. San Francisco: Astronomical Society of the Pacific, p. 335.

\bibitem[]{} Merritt, D., \& Ekers, R.D. 2003, Science, 297, 1310

\bibitem[]{} Mezger, P. G., \& Wink, J. E.\ 1986, A\&A, 157, 252

\bibitem[]{} Miller, M.C., \& Colbert, E.J.M. 2004, Int.J.Mod.Phys., 13, 1

\bibitem[]{} Milosavljevic, M., \&  Merritt, D. 2001, ApJ, 563, 34

\bibitem[]{} Miyoshi, M., Moran, J., Herrnstein, J., Greenhill, L., Nakai, N., Diamond, P., \& Inoue, M. 1995, Nature, 373, 127

\bibitem[]{} Monaco, P., Salucci, P., \& Danese, L. 2000, MNRAS, 311, 279

\bibitem[]{} Moran, J., Greenhill, L., Herrnstein, J., Diamond, P., Miyoshi, M., Nakai, N., \& Inoue 1995, in "Quasars and AGN: High Resolution Imaging," Proceedings of the National Academy of Sciences, Volume 92, Issue 25, pp. 11427

\bibitem[]{} Moran, J., Greenhill, L., \& Herrnstein, J. 1999, J. Astroph. Astr., 20, 165

\bibitem[]{} Murray, N., Chiang, J., Grossman, S.A., \& Voit, G.M. 1995, ApJ, 451, 498

\bibitem[]{} Murray, N., \& Chiang, J. 1997, ApJ, 474, 91

\bibitem[]{} Mushotzky, R.F., Done, C., \& Pounds, K.A. 1993, ARA\&A, 31, 717

\bibitem[]{} Nakai, N., Inoue, M., \& Miyoshi, M., 1993, Nature, 361, 45

\bibitem[]{} Nakai, N., Inoue, M., Miyazawa, K., Miyoski, M., \& Hall, P. 1995, PASJ, 47, 771

\bibitem[]{} Nakano, T., \& Makino, J. 1999, ApJ, 510, 155

\bibitem[]{} Nandra, K., Pounds, K.A., Stewart, G.C., Fabian, A.C., \& Rees, M.J. 1989, MNRAS, 236, 39

\bibitem[]{} Nandra, K., George, I.M., Mushotzky, R.F., Turner, T.J., \& Yaqoob, T. 1997, ApJ, 477, 602

\bibitem[]{} Nandra, K., George, I. M., Mushotzky, R. F., Turner, T. J., \& Yaqoob, T.\ 1999, ApJ, 523, L17

\bibitem[]{} Narayan, R., \& Yi, I, 1994, ApJ, 428, L13

\bibitem[]{} Navarro, J. F., \& Steinmetz, M. 2000, ApJ, 538, 477

\bibitem[]{} Nelson, C. H., \& Whittle, M.\ 1996, ApJ, 465, 96

\bibitem[]{} Netzer, H. 2003, ApJ, 583, L5

\bibitem[]{} Neufeld, D.A., \& Melnick 1991, ApJ, 368, 215

\bibitem[]{} Neufeld, D.A., Maloney, P.R., \& Conger, 1994 ApJ, 436, L127

\bibitem[]{} Neufeld, D.A., \& Maloney, P.R. 1995 ApJ, 447, L17

\bibitem[]{} O'Brien, P. T., \etal 1998, ApJ, 509, 163

\bibitem[]{} Onken, C.A., \& Peterson, B.M. 2002, ApJ, 572, 746

\bibitem[]{} Onken, C.A., Peterson, B.M., Dietrich, M., Robinson, A., \& Salamanca, I.M., 2003, ApJ, 585, 121

\bibitem[]{} Onken, C.A., et al., 2004, ApJ,  in press

\bibitem[]{} Oppenheimer, J.R., \& Snyder, H. 1939, PhRv, 56, 455O

\bibitem[]{} Oppenheimer, J.R., \& Volkoff, G.M.\ 1938, Phys. Rev., 55, 374

\bibitem[]{} Oshlack, A.Y.K.N., Webster, R.L., \& Whiting, M.T. 2002, ApJ, 576, 81

\bibitem[]{} Padovani, P., Burg, R., \& Edelson, R.A., 1990, 353, 438

\bibitem[]{} Peebles, P. J. E. 1972, Gen. Relativ. Gravitation, 3, 63

\bibitem[]{} Pelletier, G., \& Pudritz, R.E. 1992, ApJ, 394, 117

\bibitem[]{} Peterson, B.M., 1993, PASP, 105, 247

\bibitem[]{} Peterson, B.M. 1999, in "Structure and Kinematics of Quasar Broad Line Regions", ASP Conference Series, Vol. 175. Eds. C. M. Gaskell, W. N. Brandt, M. Dietrich, D. Dultzin-Hacyan, and M. Eracleous. p.49

\bibitem[]{} Peterson, B.M., \& Wandel, A. 2000, ApJ, 540, L13

\bibitem[]{} Peterson, B.M., \etal 2000, ApJ, 542, 161

\bibitem[]{} Peterson, B.M., \etal 2002, ApJ, 581, 197

\bibitem[]{} Peterson, B.M., 2002, in "Advanced Lectures on the Starburst-AGN Connection", 2001 (Singapore:World Scientific), p.3

\bibitem[]{} Peterson, B.M., 2003 in "Future EUV/UV and Visible Space Astrophysics Missions and Instrumentation". Eds J. Chris Blades, Oswald H. W. Siegmund. Proceedings of the SPIE, Volume 4854, p. 311

\bibitem[]{} Peterson, B.M., \& Horne, K. 2003 in "Astrotomography, 25th meeting of the IAU". 

\bibitem[]{} Peterson, B.M., Polidan, R.S., \& Robinson, E.L. 2003, in `Future EUV/UV and Visible Space Astrophysics Missions and Instrumentation'. Eds J. Chris Blades, Oswald H. W. Siegmund. Vol. 4854, pp. 311-318 

\bibitem[]{} Peterson, B.M. 2004, in ``The Interplay Among Black Holes, Stars, and the ISM in Galactic Nuclei'', IAU Coll.222, eds. T. Storchi-Bergmann, L. Ho, and H.R. Schmitt, in press.

\bibitem[]{} Phinney, E.S., 2003, HEAD, 35, 2703

\bibitem[]{} Piro, L., Yamauchi, M., \& Matsuoka, M. 1990, ApJ, 360, L35

\bibitem[]{} Pogge, R.W. 1989, ApJS, 71, 433

\bibitem[]{} Press, W.H., \& Schechter, P. 1974, ApJ, 187, 425

\bibitem[]{} Quinlan, G. D., Hernquist, L., \& Sigurdsson, S. 1995, ApJ, 440, 554

\bibitem[]{} Quinlan, G. D., \& Hernquist, L. 1997, NewA, 2, 533

\bibitem[]{} Rees, M.J., Phinney, E.S., Begelman, M.C., \& Blandford, R.D. 1982,
Nature, 295, 17

\bibitem[]{} Rees, M.J. 1984, ARA\&A, 22, 471 

\bibitem[]{} Reid, M. J., Menten, K. M., Genzel, R., Ott, T., Sch\"odel, R., \& Eckart, A.\ 2003, ApJ, 587, 208

\bibitem[]{} Rest, A., \etal 2001, AJ, 121, 2431

\bibitem[]{} Reunanen, J., Kotilainen, J.K., \& Prieto, M.A. 2003, MNRAS, 343, 192

\bibitem[]{} Reynolds, C.S. 1996, PhD Thesis, University of Cambridge.

\bibitem[]{} Reynolds, C.S. 1997, MNRAS, 286, 513

\bibitem[]{} Reynolds, C.S., \& Nowak, M.A., 2003, Phys.Rept., 377, 389

\bibitem[]{} Rhoades, C. E., \& Ruffini, R.\ 1974, Phys. Rev. Lett., 32, 324

\bibitem[]{} Richstone, D.O., \& Schmidt, M. 1980, ApJ, 235, 361

\bibitem[]{} Richstone, D.O., \& Tremaine, S. 1985, ApJ, 286, 370

\bibitem[]{} Richstone, D. O. \etal 1998, Nature, 395, A14

\bibitem[]{} Robinson, I., Schild, A., \& Schucking, E.L., Proceedings of the 1st
Texas Symposium on Relativistic Astrophysics, Chicago: University of
Chicago Press, 1965

\bibitem[]{} Rosati, P., et al. 2002, ApJ, 566, 667

\bibitem[]{} Ross, R.R., Fabian, A.C., \& Mineshige, S. 1992, MNRAS, 258, 189

\bibitem[]{} Salpeter, E.E. 1964, ApJ, 140, 796

\bibitem[]{} Salucci, P. \etal 1999, MNRAS, 307, 637

\bibitem[]{} Sandage, A. 1964, ApJ, 139, 416

\bibitem[]{} Sargent, W.L.W., \etal 1978, ApJ, 221, 731

\bibitem[]{} Sarzi, M., \etal 2001, ApJ, 550, 65

\bibitem[]{} Schlegel, D. J., Finkbeiner, D. P., \& Davis, M. 1998, ApJ, 500, 525

\bibitem[]{} Sch\"{o}del, R., Ott, T., Genzel, R., Eckart, A., Mouawad, N., \& Alexander, T.\ 2003, ApJ, 596, 1015

\bibitem[]{} Schwarzschild, K. (1916a), 1916, {\it Sitzungsberichte der Deutschen
Akademie der Wissenschaften zu Berlin, Klasse fur Mathematik, Physik,
und Technik}, 189

\bibitem[]{} Schwarzschild, K. (1916b), 1916, {\it Sitzungsberichte der Deutschen
Akademie der Wissenschaften zu Berlin, Klasse fur Mathematik, Physik,
und Technik}, 424

\bibitem[]{} Schwarzschild, M. 1979 ApJ, 232, 236

\bibitem[]{} Scorza, C., \& Bender, R. 1995, A\&A, 293, 20

\bibitem[]{} Serabyn, E., \& Lacy, J. H.\ 1985, ApJ, 293, 445

\bibitem[]{} Sergeev, S.G., Pronik, V.I., Peterson, B.M., Sergeeva, E.A., \& Zheng, W. 2002, ApJ, 576, 660

\bibitem[]{} Seyfert, C.K. 1943, ApJ, 97, 28

\bibitem[]{} Shankar, F., Salucci, P., Granato, G.L., De Zotti, G., \& Danese, L. 2004, MNRAS, in press

\bibitem[]{} Sheth, R.K., et al. 2003, ApJ, 594, 225

\bibitem[]{} Shields, G.A. 1978, Nature, 272, 706

\bibitem[]{} Shields, G.A.  \etal 2003, ApJ, 583, 124

\bibitem[]{} Shklovski, I.S. 1954, Doklady Akad.Nauk., 98, 353

\bibitem[]{} Sigurdsson, S., Hernquist, L., \& Quinlan, G. D. 1995, ApJ, 446, 75

\bibitem[]{} Siemiginowska, A., Kuhn, O., Elvis, M., Fiore, F., McDowell, J., \& Wilkes, B.J. 1995, ApJ, 454, 77

\bibitem[]{} Silk, J., \& Rees, M. J. 1998, A\&A, 331, L1

\bibitem[]{} Simien, F., \& de Vaucouleurs, G. 1986, ApJ, 302, 564

\bibitem[]{} Sincell, M.W., \& Krolik, J.H. 1998, ApJ, 496, 737

\bibitem[]{} Smith, H.J., \& Hoffleit, D. 1963, AJ, 68, 292

\bibitem[]{} Smith, M.D., \& Raine, D.J. 1985, MNRAS, 212, 425

\bibitem[]{} Smith, M. D., \& Raine, D. J.\ 1988, MNRAS, 234, 297

\bibitem[]{} Small, T., \& Blandford, R.D. 1992, MNRAS, 259, 725

\bibitem[]{} Snellen, I.A.G., Lehnert, M.D., Bremer, M.N., \& Schilizzi, R.T. 2003, MNRAS, 342, 889

\bibitem[]{} Sofue, Y., \& Irwin, J.A. 1992, PASJ, 44, 353

\bibitem[]{} Soltan, A. 1982, MNRAS, 200, 115

\bibitem[]{} Steinmetz, M., \& Muller, E. 1995, MNRAS, 276, 549

\bibitem[]{} Stern, B.E., Poutanen, J., Svensson, R., Sikora, M., \& Begelman, M.C. 1995, ApJ, 449, L13

\bibitem[]{} Stirpe, G.M., \etal 1994, A\&A, 285, 857

\bibitem[]{} Sudou, H., Iguchi, S., Murata, Y., \& Taniguchi, Y. 2003, Science, 300, 1263
  
\bibitem[]{} Sun, W.-H, \& Malkan, M.A. 1989, ApJ, 346, 68

\bibitem[]{} Sunyaev, R.A., \& Truemper, J.1979, Nature, 279, 506

\bibitem[]{} Szuszkiewicz, E., Malkan, M.A., \& Abramowicz, M.A. 1996, ApJ, 458, 474

\bibitem[]{} Tadhunter, C., Marconi, A., Axon, D., Wills, K., Robinson, T. G., \&  Jackson, N. 2003, MNRAS, 342, 861

\bibitem[]{} Takahashi, K., Inoue, H., \& Dotani, T. 2002, PASJ, 54, 373

\bibitem[]{} Tanaka, Y., \etal 1995, Nature, 375, 659

\bibitem[]{} Thatte, N., Quirrenbach, A., Genzel, R., Maiolino, R., \& Tecza 1997, ApJ, 490, 238

\bibitem[]{} Thorne, K.S., \& Price, R. H. 1975, ApJ, 195, 101

\bibitem[]{} Thorne, K.S. 1994, {\it Black Holes and Time Warps, Einstein's
Outrageous Legacy}, W W Norton \& Company, New York and London

\bibitem[]{} Tonry, J.L. \etal 2001, ApJ, 546, 681

\bibitem[]{} Tran, H.D., \etal 2001, AJ, 121, 2928

\bibitem[]{} Tremaine, S., \etal 2002, ApJ, 574, 740

\bibitem[]{} Trotter, A.S., \etal 1998, ApJ, 495, 740

\bibitem[]{} Turler, M., et al. 1999, A\&AS, 134, 89

\bibitem[]{} Turner, J.L., \& Ho, P.T.P. 1994, ApJ, 421, 122

\bibitem[]{} Ulrich, M.-H., \& Horne, K. 1996, MNRAS, 283, 748

\bibitem[]{} Urry, C.M., \& Padovani, P. 1995, PASP, 107, 803

\bibitem[]{} Valluri, M., Merritt, D., \& Emsellem, E. 2004, ApJ, 602, 66

\bibitem[]{} van de Ven, G., Hunter, C., Verolme, E.K., \& de Zeeuw, P.T. 2003, MNRAS, 342, 1056

\bibitem[]{} van den Bosch, F. C., \& de Zeeuw, P. T. 1996, MNRAS, 283, 381

\bibitem[]{} van den Bosch, F. C., Jaffe, W., \& van der Marel, R. P. 1998, MNRAS, 293, 343

\bibitem[]{} van der Kruit, P.C.,  Oort, J.H., Mathewson, D.S. 1972, A\&A, 21, 169

\bibitem[]{} van der Marel, R.P., Franx, M.1993, ApJ, 407, 525

\bibitem[]{} van der Marel, R.P. 1994, MNRAS, 270, 271

\bibitem[]{} van der Marel, R.P., \& van den Bosch, F.C. 1998, AJ, 116, 2220

\bibitem[]{} van der Marel, R.P. 1999, AJ, 117, 744

\bibitem[]{} Verolme, E.K., \etal 2002, MNRAS, 335, 517

\bibitem[]{} Vestergaard, M., Wilkes, B.J., \& Barthel, P.D. 2000, ApJ, 538, L103

\bibitem[]{} Vestergaard, M. 2002, ApJ, 571, 733

\bibitem[]{} Vestergaard M. 2004, ApJ, 601, 676

\bibitem[]{} Vila-Vilaro, B. 2000, PASPJ, 52, 305

\bibitem[]{} Walker, R.C., Matsakis, D.N., and Garcia-Barreto, J.A. 1982, ApJ, 255, 128.

\bibitem[]{} Wandel A., 1999, ApJ, 519, 39

\bibitem[]{} Wandel, A., Peterson, B.M., \& Malkan, M.A. 1999, ApJ, 526, 579

\bibitem[]{} Wandel, A. 2002, ApJ, 565, 762

\bibitem[]{} Wanders, I., \& Horne, K. 1994, A\&A, 289, 76 

\bibitem[]{} Wanders, I., \etal 1995, ApJ, 453, L87

\bibitem[]{} Wanders, I., \& Peterson, B. M. 1996, ApJ, 466, 174

\bibitem[]{} Wanders, I., \etal 1997, ApJS, 113, 69

\bibitem[]{} Wang, J.-M., Szuszkiewicz, E., Lu, F.-J., \&  Zhou, Y.-Y. 1999, ApJ, 522, 839

\bibitem[]{} Wang, J.-X., Wang, T.-G., \& Zhou,  Y.-Y. 2001, ApJ, 549, 891

\bibitem[]{} Watanabe, M., et al. 2003, ApJ, 591, 714

\bibitem[]{} Watson, W.D., \& Wallin, B.K. 1994, ApJ, 432, L35

\bibitem[]{} Wyithe, J.S.B., \& Loeb, A. 2002, ApJ, 581, 886

\bibitem[]{} Wyithe, J.S.B., \& Loeb, A. 2003, ApJ, 595, 614

\bibitem[]{} Willke, B., et al. 2002, CQGra, 19, 1377

\bibitem[]{} Wilms, \etal 2001 MNRAS, 328, L27

\bibitem[]{} Wollman, E. R., Geballe, T. R., Lacy, J. H., Townes, C. H., \& Rank, D. M.\ 1977, ApJ, 218, L103

\bibitem[]{} Woo, J.-H., \& Urry, C. M.\ 2002a, ApJ, 579, 530

\bibitem[]{} Woo, J.-H., Urry, C. M.\ 2002b, ApJ, 581, L5

\bibitem[]{} Wu, X.-B., \& Han, J.L. 2001, A\&A, 380, 31

\bibitem[]{} Young, P. J. 1980, ApJ, 242, 1232

\bibitem[]{} Young, J.S., Claussen, M.J., \& Scoville, N.Z. 1988, ApJ, 324, 115

\bibitem[]{} Young, A.J., \& Reynolds, C.S. 2000, ApJ, 529, 101

\bibitem[]{} Yu, Q. 2002,  MNRAS, 331, 935

\bibitem[]{} Yu, Q., \& Tremaine, S. 2002, MNRAS, 335, 965

\bibitem[]{} Zdziarski, A.A., \etal 1994, MNRAS, 269, L55

\bibitem[]{} Zel'dovich, Y. B., \& Novikov, I. D.\ 1964, Sov. Phys. Dokl., 158, 811 

\end{thebibliography}
\end{document}